\documentclass{aastex61}
\newcommand\pcc{\;{\rm cm}^{-3}}
\newcommand\Msun{\; {\rm M}_{\odot}}
\newcommand\kms{\; {\rm km}\;{\rm s}^{-1}}
\newcommand\ergs{\; {\rm erg}\;{\rm s}^{-1}}
\newcommand\erg{\; {\rm erg}}

\newcommand\cm{\;{\rm cm}}
\newcommand\yr{\; {\rm yr}}
\newcommand\Myr{\;{\rm Myr}}

\newcommand\pc{\;{\rm pc}}
\newcommand\kpc{\;{\rm kpc}}
\newcommand\sfrunit{\Msun \kpc^{-2} \yr^{-1}}
\newcommand\momunit{\Msun \kms}
\newcommand\Punit{\pcc\,{\rm K}}
\newcommand\Surf{\Msun\;{\rm pc^{-2}}}
\newcommand\rhounit{\Msun\;{\rm pc^{-3}}}
\newcommand\Kel{\;{\rm K}}
\newcommand\eV{\;{\rm eV}}

\newcommand\simgt{\lower.5ex\hbox{$\; \buildrel > \over \sim \;$}}
\newcommand\simlt{\lower.5ex\hbox{$\; \buildrel < \over \sim \;$}}
\newcommand\pderiv[2]{\frac{\partial {#1}}{\partial {#2}}}

\newcommand\rbrackets[1]{\left({#1}\right)}
\newcommand\sbrackets[1]{\left[{#1}\right]}

\newcommand\divergence[2][\rbrackets]{\nabla \cdot #1{#2}}
\newcommand\curl[2][\rbrackets]{\nabla \times #1{#2}}

\newcommand\vel{\mathbf{v}}
\newcommand\Bvec{\mathbf{B}}

\newcommand\xhat{\hat{\mathbf{x}} }
\newcommand\yhat{\hat{\mathbf{y}} }

\newcommand\torb{t_{\rm orb}}

\newcommand\tage{t_{\rm m}}
\newcommand\snrate{\xi_{\rm SN}}
\newcommand\sncrate{\Xi_{\rm SN}}
\newcommand\model[1]{{\tt MHD-{#1}pc}}
\newcommand\vaturb{\delta v_{\rm A}}
\newcommand\vamean{\overline{v}_{\rm A}}
\shorttitle{TIGRESS: Algorithms, Fiducial Model, and Convergence}
\shortauthors{Kim \& Ostriker}

\begin{document}

\title{Three-phase Interstellar medium in Galaxies
  Resolving Evolution
  with Star formation and
  Supernova feedback (TIGRESS): Algorithms, Fiducial
  Model, and Convergence}
\author[0000-0003-2896-3725]{Chang-Goo Kim}
\affiliation{Department of Astrophysical Sciences, Princeton University, Princeton, NJ 08544, USA}
\author[0000-0002-0509-9113]{Eve C. Ostriker}
\affiliation{Department of Astrophysical Sciences, Princeton University, Princeton, NJ 08544, USA}
\email{cgkim@astro.princeton.edu, eco@astro.princeton.edu}

\begin{abstract}
We introduce TIGRESS, a novel framework for multi-physics
numerical simulations of the star-forming interstellar medium (ISM)
implemented in the {\tt Athena} MHD code.  The algorithms of TIGRESS
are designed to spatially and temporally resolve key physical
features, including: (1) the gravitational collapse and ongoing
accretion of gas that leads to star formation in clusters, (2) the
explosions of supernovae (SNe) both near their progenitor birth sites
and from runaway OB stars, with time delays relative to star formation
determined by population synthesis, (3) explicit evolution of SN
remnants prior to the onset of cooling, which leads to the creation of
the hot ISM, (4) photoelectric heating of the warm and cold phases of
the ISM that tracks the time-dependent ambient FUV field from the
young cluster population, (5) large-scale galactic differential
rotation, which leads to epicyclic motion and shears out overdense
structures, limiting large-scale gravitational collapse, (6) accurate
evolution of magnetic fields, which can be important for vertical
support of the ISM disk as well as angular momentum transport.  We
present tests of the newly-implemented physics modules, and
demonstrate application of TIGRESS in a fiducial model representing
the Solar neighborhood environment.  We use a resolution study to
demonstrate convergence and evaluate the minimum resolution $\Delta x$ required
to correctly recover several ISM properties, including the star
formation rate, wind mass-loss rate, disk scale height,
turbulent and Alfv\'enic velocity dispersions, and volume fractions of
warm and hot phases.  For the Solar neighborhood model, 
all these ISM properties are 
converged at $\Delta x \le 8\pc$.
\end{abstract}


\section{Introduction}

Feedback from massive young stars plays a crucial role in regulating
star formation and the properties of the interstellar medium (ISM) at
all scales \citep[e.g.,][]{2007ARA&A..45..565M,2014prpl.conf..243K}.
Stellar winds and ionizing+non-ionizing radiation from massive stars
profoundly affect their birth environment, and the large-scale
outflows that these processes produce impose an upper limit on the
lifetime star formation efficiency of the parent giant molecular cloud
(GMC)
\citep[e.g.,][]{2014ApJ...795..121L,2015NewAR..68....1D,2016ApJ...829..130R}.
Protostellar outflows and jets driven in the course of low mass star
formation may also contribute to supporting the parent molecular
cloud, and extending its life
\citep[e.g.,][]{2010ApJ...709...27W,2014prpl.conf..451F}.  When a
massive stars dies, the most extreme feedback event occurs, with the
instantaneous release of $\sim10^{51}\erg$ in the form of high-velocity
supernova (SN) ejecta
\citep[e.g.,][]{1999ApJS..123....3L,2003ApJ...591..288H}.  Among the
many forms of stellar feedback, SNe are believed to be the most
important for driving turbulence of the warm neutral
medium and cold neutral medium of the ISM (WNM and CNM, respectively)
\citep[e.g.,][]{2004RvMP...76..125M} and for regulating galactic star
formation rates (SFRs)
\citep[e.g.,][]{2010ApJ...721..975O,2011ApJ...731...41O,2011ApJ...743...25K},
as well as for creating the hot component of the ISM
\citep[e.g.,][]{1974ApJ...189L.105C,1977ApJ...218..148M}.
The collective effects of many (spatially and temporally) correlated
SNe create superbubbles that expand away from the disk midplane and
contribute to driving galactic fountains and winds
\citep[e.g.,][and references therein]{2017ApJ...834...25K}.  Expulsion of baryons in
winds driven by star
formation feedback are believed to play a critical role in shaping the galaxy
stellar mass function, particularly in low-mass haloes
\citep[e.g.,][]{2015ARA&A..53...51S}.

Although the importance of SN feedback to the ISM
has long been recognized \citep[e.g.,][]{1977ApJ...218..148M},
proper implementation in numerical simulations is challenging 
due to the requirements of very high spatial and temporal resolution.
A SN releases prodigious energy that is extremely
concentrated in both space and time.  This produces high velocity
shock waves propagating into the surrounding gas, and the shocked
ambient medium that comprises the 
interior of a SN remnant is extremely hot.  
At later stages, after the denser portion of the hot gas is able to
cool, both individual SN remnants and superbubbles
are bounded by shells of cooled, dense gas, while their interiors remain
hot.   Since the initial momentum of the SN ejecta is boosted
(more than an order of magnitude) during the
energy-conserving stage by the work of the expanding hot bubble on the
surrounding ISM, it is crucial to resolve this evolutionary stage 
\citep[e.g.,][]{1988ApJ...334..252C,1998ApJ...500...95T,2015ApJ...802...99K}.
For an isolated SN propagating into gas of number density $n=100-0.1\pcc$,
the energy-conserving Sedov-Taylor stage ends at the shell formation
time $t_{\rm sf} \sim 10^3 - 10^5\yr$, when the SN remnant radius is
$r_{\rm sf}\sim 1 - 10\pc$.
In order to obtain numerically converged results -- in
particular for the history of hot gas mass and the final radial
momentum -- these temporal and spatial scales must be resolved.  
\citet{2015ApJ...802...99K} found that convergence requires $r_{\rm sf}$
to be resolved by at least three grid zones,
which imposes an upper limit on the initial size of the SN feedback region and
the grid resolution $\Delta x$.  

In cosmological galaxy formation simulations using
current computational resources (where the best resolution is several tens
of pc), it is generally
not possible to resolve the Sedov-Taylor stage of individual SN remnants.
Thermal energy dumped into a scale larger than the expected
$r_{\rm sf}$ (or shared with a larger mass than this radius would enclose) 
immediately cools away, and this feedback has no impact on subsequent
star formation 
\citep[e.g.,][]{1992ApJ...391..502K}. ``Over-cooling'' from unresolved
SN feedback leads to overly efficient conversion of gas to stars,
and galaxies which are too massive.
To deal with the ``over-cooling'' problem in galaxy formation simulations, 
a wide variety of sub-grid models for treating feedback has been developed  
\citep[e.g.,][]{2012MNRAS.426..140D,2013ApJ...770...25A,2013MNRAS.429.3068T,2014ApJ...788..121K,2014MNRAS.445..581H}.
Many sub-grid models have been calibrated to reproduce 
basic observables such as 
the stellar mass to halo mass relation 
\citep[e.g.,][]{2013ApJ...770...57B,2013MNRAS.428.3121M},
the Kennicutt-Schmidt relation \citep{1998ApJ...498..541K}, 
star formation history, and other properties 
\citep[e.g.,][]{2014MNRAS.445..581H,2014Natur.509..177V,2015MNRAS.446..521S,2015MNRAS.450.1937C,2016MNRAS.462.3265D,2016ApJ...824...79A,2016MNRAS.463.1431K}.

While recent efforts have been successful in matching stellar abundances
and SFRs, detailed gas properties in the ISM and
also circumgalactic and intergalactic media (CGM and IGM, respectively)
can differ substantially depending on
the feedback treatment adopted.  For example, in
\citet{2017MNRAS.466...11R}, delayed cooling 
and kinetic feedback approaches unphysically enhance the ratio of gas outflow rate
to SFR (a.k.a. the mass loading factor),
altering the volume filling factors and compositions of the CGM and IGM. 
Approaches that accumulate enough energy from SNe that the temperature
of the feedback region is high are able to limit cooling.
However, from higher resolution simulations, 
it is known that the spatial and temporal correlations (or de-correlations) of
SNe with the gas can strongly affect the proportions of ISM mass and volume
in different phases \citep[e.g.,][]{2014A&A...570A..81H,2015MNRAS.449.1057G,2015ApJ...814....4L,2017ApJ...834...25K}. ``Accumulation'' of feedback energy based on
the resolution of simulation will not in general recover the true space-time
correlations of SN with each other and with the ISM gas.

In intermediate scale simulations, it is possible to incorporate
detailed ISM physics and to resolve SN feedback.
Utilizing vertically-stratified boxes that
represent local patches in galactic disks,
the roles of SNe in driving turbulence and
shaping the multiphase ISM in galactic disks have been extensively investigated
\citep[e.g.,][]{1999ApJ...514L..99K,2000MNRAS.315..479D,
  2004A&A...425..899D,2006ApJ...653.1266J,2012ApJ...750..104H,
  2013MNRAS.432.1396G,2015MNRAS.454..238W,2016MNRAS.456.3432G,
  2016arXiv161008971L}.
Most studies to date have simply imposed a SN rate based on an empirical
relationship between gas surface density and the surface density of
star formation \citep[e.g., the Kennicutt-Schmidt relation of][]{1998ApJ...498..541K},
and evolve the ISM subject to this fixed rate, setting off SNe either at
random locations or at loci restricted by their density.
However, imposed feedback of this kind can lead to ISM states
that are quite unphysical, e.g. with the hot medium occupying either a
negligible volume or almost the entire domain.  Instead, for self-consistency,
the rate of star formation and SN explosions should be self-regulated,
since SN feedback provides a strong negative feedback 
\citep[e.g.,][]{2010ApJ...721..975O,2011ApJ...731...41O,2011ApJ...743...25K,2013ApJ...776....1K}.  

Only a few recent simulations have studied the effects of
resolved SN feedback within models in which the SN rate in the three-phase ISM
is self-consistently determined by collapse under self-gravity
\citep[][]{2014A&A...570A..81H,2017MNRAS.466.1903G,2017MNRAS.466.3293P}.
Furthermore,
these simulations have had relatively brief duration ($\le 100\Myr$),
and do not appear to have reached a quasi-steady state in which
the rates of star formation and feedback and the ISM properties are
independent of initial  conditions.  
These simulations have also neglected galactic differential rotation,
which is responsible for limiting 
gravitational collapse at large scales \citep[e.g.,][]{2002ApJ...581.1080K}
and creating and maintaining magnetic fields via galactic dynamos 
\citep[e.g.,][]{2015ApJ...815...67K}.

In this paper, we present a comprehensive framework for modeling the
turbulent, magnetized, multiphase, self-gravitating ISM in a local,
vertically-stratified box with sheared rotation, including star
formation and SN feedback.  The TIGRESS (Three-phase 
  Interstellar medium in Galaxies Resolving
  Evolution with Star formation and Supernova feedback)
framework we have developed is designed to be applied in a wide
variety of galactic environments.  The investigation underway will produce 
detailed theoretical representations of the ISM and star formation
properties for comparison to observations of the Milky Way and nearby
galaxies.  In addition, resolved TIGRESS simulations may be used to
develop and calibrate sub-grid models for galaxy formation simulations
in which direct resolution of the ISM, star formation, and feedback
effects -- including driving winds -- is not possible.

To achieve the highest level of realism and predictive power in
simulating the star-forming ISM, standard algorithms to evolve the
equations of magnetohydrodynamics (MHD), solve for self-gravity, and
apply heating and cooling must be coupled to a number of more
specialized methods that represent key physical elements that are
particular to the problem at hand.  The specialized TIGRESS algorithms
include treatment of local gravitational
collapse and accretion onto sink/star particles
representing clusters, and treatment of direct feedback to the ISM
from sink/star particles in the form of SNe and photoelectric heating,
based on a stellar population synthesis model.
In addition to OB stars in clusters represented by sink/star particles,
we model runaway OB stars ejected from these sites; both produce SNe.
TIGRESS allows for
three different types of SN feedback, representing different
evolutionary stages of SN remnants, and the type applied depends on
the ambient density surrounding the explosion site and the local grid
resolution.  In addition to explicitly following the energy-conserving
stage of SN remnants (which is almost always possible at the typical grid
resolution), TIGRESS allows for two other treatments of SNe
that follow the momentum-conserving or the free-expansion stages
of SN remnants (SNRs) when
the ambient density is either very high or very low, respectively.
In the former
case this captures the correct momentum injection by a SN to the surrounding
ISM when the pre-radiative SNR 
evolution is unresolved,
 and in the latter case this captures early evolution more correctly
 before thermalizing the interior of the SNR. In this paper, we
describe the specialized TIGRESS algorithms (and tests of these
methods) fully so that the simulations we conduct will be reproducible
by other groups.

To demonstrate the application of TIGRESS, we present results from a
fiducial simulation with parameters similar to those of the Solar
neighborhood.
This model runs long enough (three galactic orbits, or $\sim 700\Myr$) that
a fully self-consistent quasi-steady state is reached, with 
more than ten self-regulation cycles of 
star formation.
We show time histories of
star formation and feedback loops, as well as basic gas properties
including phase balance, turbulent velocities, magnetic field strengths,
and disk scale heights.  
The resulting ISM is highly realistic, with cold, warm, and hot
phases; 
velocity dispersions and magnetic field strengths are also realistic.  

In addition to simple tests of individual numerical modules, it is important
to test the overall behavior of the complex, highly-coupled nonlinear
model system. An essential ``system test'' that is required in any
complex application, and particularly any application involving
turbulence, is the study of convergence with respect to numerical resolution.
Previously, there have been
  resolution studies mainly focused on convergence in
  the mass fractions of the three
thermal phases mediated by SN feedback (at fixed rates) in different models  
\citep[e.g.,][]{2004A&A...425..899D,2012ApJ...750..104H,2013MNRAS.432.1396G}. 
In the present work, a significant advance is that the 
SN feedback rate is self-regulated (by the response of star formation to
changes in the ISM induced by feedback), and the temporal/spatial correlations
of SNe are realistic with respect to the various ISM phases.
Systematic exploration of resolution effects and convergence criteria
for the self-consistent TIGRESS implementation is required.
We present a resolution study for the fiducial simulation, varying
resolution from $\Delta x =2\pc$ to $64\pc$.  We analyze convergence of
several properties including the SFR,
wind mass-loss rate, disk scale height, turbulent and Alfv\'enic velocity
dispersions, and volume fractions of warm and hot phases.

The remainder of this paper is organized as follows.
In Section~\ref{sec:method},
we describe our numerical algorithms. Section \ref{sec:sp}
describes treatment of sink/star particles (including creation, accretion,
aging, and integration of motion). 
Section \ref{sec:feedback} describes feedback algorithms
 including our population synthesis treatment,
 treatment of FUV radiation/photoelectric heating (Section \ref{sec:sp_fuv}),
and treatment of SN rates and energy injection prescriptions 
(Sections \ref{sec:snr} - \ref{sec:explosion}).
In Section~\ref{sec:Fiducial}, we present results from
our fiducial Solar neighborhood model 
with spatial resolution of $\Delta x=4\pc$.  
In Section~\ref{sec:conv}, we present a convergence study of the fiducial
run with varying spatial resolution.   
Section~\ref{sec:summary} summarizes the paper.

\section{Numerical Methods}\label{sec:method}
\subsection{Magnetohydrodynamic Equations and Additional Physics}
We solve the ideal MHD equations in a local, shearing box
\citep[e.g.,][]{2010ApJS..189..142S}.
The transformation maps from global cylindrical $(R,\phi)$ coordinates 
to local Cartesian coordinates as $(x,y)=(R-R_0, R_0[\phi-\Omega t])$,
where $R_0$ is the galactocentric distance of the box center
and $\Omega$ is the angular velocity of galactic rotation at $R_0$. 
The vertical coordinate $z$ remains global.
Assuming galactic differential rotation
with the shear parameter $q\equiv -d\ln\Omega/d\ln R|_{R_0}=1$ 
for a flat rotation curve, we have background shear velocity of 
$\vel_s=-q\Omega x\yhat$.
In this local, rotating frame, inertial forces emerge
in the form of the Coriolis force, $-2\mathbf{\Omega}\times \vel$, and 
the tidal potential, $\Phi_{\rm tidal}=-q\Omega^2 x^2$.
Including gaseous and (young) stellar self-gravity, a fixed ``external''
gravitational potential to represent the old stellar disk
and dark matter halo, and optically thin cooling and heating,
the governing equations are given by
\begin{equation}\label{eq:cont}
\pderiv{\rho}{t}+\divergence{\rho\vel}=0,
\end{equation}
\begin{equation}\label{eq:mom}
\pderiv{(\rho\vel)}{t}+\divergence{\rho\vel\vel + P + \frac{B^2}{8\pi}-
\frac{\Bvec\Bvec}{4\pi}}=
-2\mathbf{\Omega}\times(\rho\vel)-\rho\nabla\Phi_{\rm tot},
\end{equation}
\begin{equation}\label{eq:energy}
\pderiv{}{t}\rbrackets{\frac{1}{2}\rho v^2+\frac{P}{\gamma-1}+\frac{B^2}{8\pi}}+
\divergence[\sbrackets]{\rbrackets{\frac{1}{2}\rho v^2 +
\frac{\gamma}{\gamma-1}P+\rho\Phi_{\rm tot}}\vel
+\frac{(\Bvec\times\vel)\times\Bvec}{4\pi}}=
-\rho\mathcal{L},
\end{equation}
\begin{equation}\label{eq:induction}
\pderiv{\mathbf{B}}{t}=\curl{\vel\times\mathbf{B}},
\end{equation}
\begin{equation}\label{eq:poisson}
\nabla^2\Phi=4\pi G(\rho+\rho_{\rm sp}).
\end{equation}
Here, $\rho$ is the gas density, $\vel$ is the gas
velocity, $P$ is the gas thermal pressure,
$\Bvec$ is the magnetic field, 
$\Phi_{\rm tot}=\Phi+\Phi_{\rm ext} +\Phi_{\rm tidal}$ is the total gravitational
potential,
$\rho\mathcal{L}=n_H[n_H\Lambda(T)-\Gamma]$ is the net cooling function,
$n_H=\rho/(\mu_H m_H)$ is the number density of hydrogen nuclei,
and all other symbols have their usual meaning. The quantity
$\rho_{\rm sp}$ is the density of
sink/star particles (representing stellar clusters
formed by gravitational collapse) mapped onto the nearest $3^3$ grid cells using
the triangular-shaped-cloud scheme \citep{1981csup.book.....H}.

\begin{figure}
\plotone{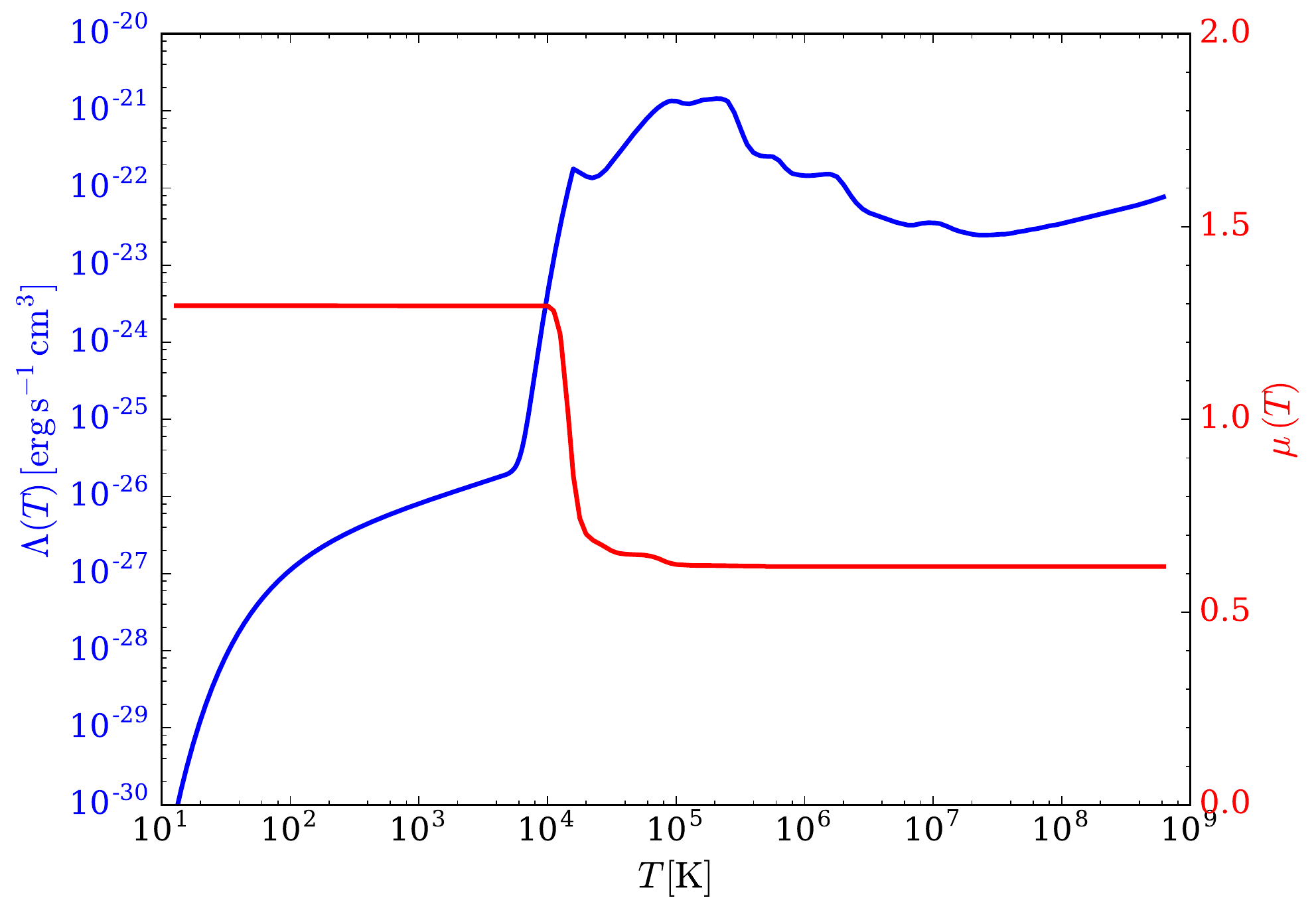}
\caption{Left, blue: adopted cooling coefficient as a function of
  temperature, combining a fitting formula for the neutral ISM from
  \citet{2002ApJ...564L..97K} at $T<10^{4.2}$, and CIE cooling from
  \citet{1993ApJS...88..253S} at $T>10^{4.2}$.  Right, green: adopted
  mean molecular weight as a function of temperature. We smoothly
  interpolate $\mu$ between $\mu_{\rm ato}=1.295$ and $\mu_{\rm
    ion}=0.618$ following the shape presented in
  \citet{1993ApJS...88..253S}.  }
\label{fig:cool}
\end{figure}

To follow radiative heating and cooling over the full range of temperatures
and densities that occur in the three-phase ISM, a variety of approaches can
be adopted.  For most accurate treatment, time-dependent chemistry 
is required; although in general this is computationally quite expensive, it
is also possible to design a reduced chemical network that retains just the
most essential reactions for key coolants
\citep[e.g.,][]{2016arXiv161009023G}.
For present purposes, we take the simpler approach of
adopting a cooling coefficient $\Lambda(T)$ using the 
fitting formula in \citet[][see \citealt{2008ApJ...681.1148K} 
  for form with correction of typographical error]{2002ApJ...564L..97K}
for $T<10^{4.2}$,
and collisional ionization equilibrium (CIE) cooling with solar metalicity
from \citet{1993ApJS...88..253S} for $T>10^{4.2}$. In order to obtain gas 
temperature from pressure and density, in principle we need to know
the gas composition (or, in other words, the mean molecular weight, $\mu$). 
However, here we do not follow
detailed chemistry of molecule formation/dissociation and
ionization/recombination.
Instead, we simply use a tabulated mean molecular weight for the adopted
CIE cooling at Solar metallicity of \citet{1993ApJS...88..253S}. 
As temperature increases, $\mu(T)$ varies from $\mu_{\rm ato}=1.295$ for 
neutral gas (here we do not include molecular gas) 
to $\mu_{\rm ion}=0.618$ for fully ionized gas. 
At intermediate temperatures, $\mu$ is calculated iteratively 
for a given pair of $P$ and $\rho$. The shape of the cooling function and mean
molecular weight as a function of $T$ are shown in Figure~\ref{fig:cool}.
For neutral gas, we include heating due to the photoelectric effect on grains;
to exclude the ionized component we apply a normalization 
whose shape follows $\mu(T)$ (see Equation (\ref{eq:heat})).
The heating rate scales with the instantaneous FUV luminosity from sink/star
particles (see \S~\ref{sec:sp_fuv} for details).

The external gravity in the vertical direction is modelled with
a fixed potential. We use the Kuijken \& Gilmore form
\citep{1989MNRAS.239..571K} with a modification for the
dark matter potential at large $|z|$.
The combined (old) stellar disk and dark matter halo potential is given by
\begin{equation}\label{eq:phi_ext}
\Phi_{\rm ext}\equiv 2\pi G \Sigma_* z_*\sbrackets{\rbrackets{1+\frac{z^2}
{z_*^2}}^{1/2}-1} + 2\pi G \rho_{\rm dm} R_0^2\ln\rbrackets{1+\frac{z^2}{R_0^2}}.
\end{equation} 
Adopted parameters are given for the fiducial Solar neighborhood model in 
Section \ref{sec:Fiducial}.  Alternative
values of $R_0$, $\Sigma_*$, $z_*$, and $\rho_{\rm dm}$ can be adopted
to represent other local environments, whether within the Milky Way or
within other disk galaxies.

Our physics modules are implemented within
the {\tt Athena} code, which employs 
directionally unsplit Godunov methods to solve the ideal MHD equations, 
including
the constrained-transport algorithm to preserve $\nabla \cdot \Bvec =0$
\citep{2008ApJS..178..137S}.
In this paper, we use {\tt Athena}'s predictor-corrector type integrator 
\citep{2009NewA...14..139S}, adopting piecewise linear spatial reconstruction,
and employing Roe's Riemann solver. In addition, we apply ``H-correction'' 
\citep{1998JCoPh.145..511S} when the difference in signal speed
is larger than 
$20\kms$ 
to suppress the carbuncle instability
that can arise from strong blastwaves produced by SN explosions.  We also apply
 ``first-order-flux-correction'' to cells with negative
pressure and/or density after the second-order update;
this situation can be produced by strong rarefaction waves
in highly turbulent medium \citep{2009ApJ...691.1092L}. 

We solve Poisson's equation (Eq.~(\ref{eq:poisson}))
using the FFT method with shearing-periodic boundary conditions in 
the horizontal directions \citep[e.g.,][]{2001ApJ...553..174G}
and vacuum boundary conditions in the vertical direction 
\citep{2009ApJ...693.1316K}. The net cooling source term is solved fully 
implicitly in an operator-split manner
 before the integrator step.
Since the cooling time $t_{\rm cool}\equiv |P/[(\gamma-1)\rho\mathcal{L}]|$
in cold, dense gas is usually much shorter than the MHD time step, 
$\Delta t$,
the cooling solver is sub-cycled if $t_{\rm cool}<\Delta t$.  For sub-cycling
the cooling, we use an adaptive time step 
corresponding to the instantaneous cooling time.
We neglect explicit thermal conduction, viscosity, and Ohmic resistivity.
Other key features in the TIGRESS implementation are sink/star particles
and feedback in the form of stellar heating and SN explosions.
We delineate the details of these below.

\subsection{Sink/Star Particles}\label{sec:sp}

We use sink/star particles to trace formation and evolution of 
star clusters and to apply appropriate feedback from massive stars 
based on a population synthesis model. 
The original implementation of our sink particle method is described in 
\citet[][GO13]{2013ApJS..204....8G}.
We modify this to accommodate a non-isothermal equation of state for
sink creation, and to allow for non-accreting star particles (representing
older clusters).  We also implement a symplectic particle integration method for
the shearing-box rotating frame, and introduce a particle age attribute
needed to handle feedback from massive stars (the feedback itself is
described in Section \ref{sec:feedback}).

\subsubsection{Sink Creation}\label{sec:sp_creation}

We create a sink particle when the gas in a cell and its
surroundings satisfies
three conditions:
\begin{itemize}
\item[(1)]{ the density of the cell exceeds a threshold},
\item[(2)]{ the cell is at the local potential minimum}, 
\item[(3)]{  the flow is converging}.
\end{itemize}
  For the first condition, we use the Larson-Penston
\citep[][LP]{1969MNRAS.145..271L,1969MNRAS.144..425P} density threshold 
suggested by GO13,
\begin{equation}\label{eq:rhoLP}
\rho_{\rm thr}\equiv\rho_{\rm LP}(\Delta x/2)=\frac{8.86}{\pi} \frac{c_s^2}{G \Delta x^2},
\end{equation}
where $c_s\equiv (P/\rho)^{1/2}$ is the sound speed of the cell. Note
that the density threshold suggested by \citet{1997ApJ...489L.179T}
has the same parameter dependence, but with a coefficient of $\pi/16$
instead of $8.86/\pi$.  As discussed by GO13, the motivations for
using the LP density threshold are (1) the LP solution
$\rho_{\rm LP} (r) \rightarrow 8.86 c_s^2/(4 \pi G r^2)$
for $r\rightarrow 0$ is an
``attractor'' of gravitational collapse not just for the 
spherical case, but for arbitrary turbulent flows; and (2) the LP
asymptotic solution guarantees supersonic collapse ($v_r \rightarrow
-3.28c_s$ for $r\rightarrow 0$). For an isothermal equation of
state, numerical simulations have validated that these two conditions 
hold quite generally in collapsing regions 
\citep[e.g.,][and references therein]{1993ApJ...416..303F,2005MNRAS.360..675V,2011ApJ...729..120G,2015ApJ...806...31G}.  
From the first condition, if the density reaches
the threshold of Equation (\ref{eq:rhoLP}), it may be a signature that
runaway gravitational collapse is occurring; the singularity implies that
this collapse is inherently unresolved on a computational grid.
The second condition is important because introduction of a sink particle
could potentially alter the fluid variables in the surrounding volume if
the inflow were subsonic; supersonic inflow ensures that the sink region
is causally disconnected from its surroundings.

In order to check the validity of the LP approximation for a non-isothermal
collapse, we have run collapsing simulations 
that include time-dependent heating and cooling.
To set this up, we start with a spherical profile in which 
instantaneous thermal and dynamical equilibrium hold at each radius.
This initial equilibrium profile can be uniquely determined by 
solving the static momentum equation, Poisson's equation,
and thermal equilibrium for the CNM branch, 
with given central temperature and edge pressure. Here, we use
$T_{\rm center}=20\Kel$ and $P_{\rm edge}/k_B=2000\Punit$. 
To initiate collapse, we multiply the density and pressure by a factor of 2.
Due to the short cooling time, the temperature immediately readjusts
to restore thermal equilibrium, but gravity is not balanced by pressure
and the sphere undergoes collapse similar to that of an isothermal
sphere.

\begin{figure*}
\plotone{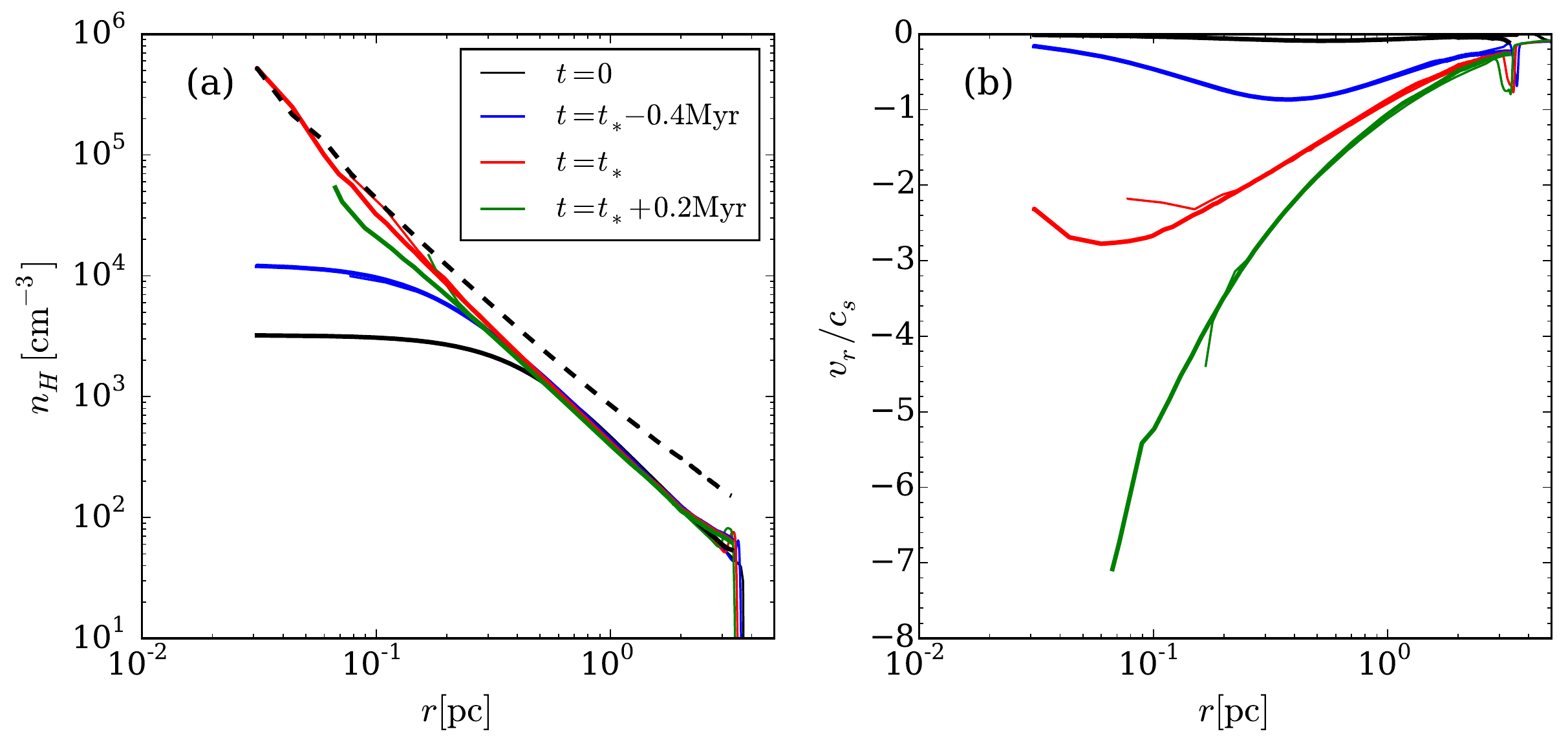}
\caption{Collapse of non-isothermal sphere.
Radial profiles of 
(a) number density and (b) radial Mach number $v_r/c_s$
are shown for epochs of $t=0$ (initial condition; black),
$t=t_*-0.4\Myr$ (before  singularity formation; blue) $t=t_*$ 
(at the time of  singularity formation; red), and $t=t_*+0.2\Myr$
(after  singularity formation; green).
Thick and thin lines show high ($\Delta x = 0.031\pc$) and low resolution
($\Delta x = 0.078\pc$) runs, respectively.
The dashed line in (a) denotes the LP density profile with
$c_s(r)$ set by local values at the time of singularity formation.
}
\label{fig:besphere}
\end{figure*}

Figure~\ref{fig:besphere} plots
(a) number density and (b) radial Mach number $v_r/c_s$ as a function of 
radius for several different times.  The initial condition is shown in
black; profiles  before (blue), at (red), and after
 singularity formation (green) are also shown.
The dashed line in (a) denotes the LP density profile,
$\rho_{\rm LP}=8.86c_s^2(r)/(4\pi G r^2)$ with
sound speed at each radius set by local values at the time of singularity.
Because $c_s$ is not a constant but increases outward,
this density profile is slightly shallower than that in the isothermal case
$\rho_{\rm LP}\propto r^{-2}$.  
The collapse starts from large radius and propagates inward
(i.e. the location of the velocity extremum moves inward; 
e.g., \citealt{2009ApJ...699..230G}).
Eventually, the radial collapsing velocity exceeds the local sound speed,
and the density profile of the inner part approaches the dashed line.
When the central density reaches the LP density threshold
(Eq. (\ref{eq:rhoLP}) using the local $c_s$),
we create a sink particle, and continue to follow the
late-time accretion. At the time of singularity formation,
$v_r\sim -3c_s$ for the high resolution run.
If we do not create a sink particle when the central density
reaches the LP threshold, unresolved collapse 
causes unphysical overshoots of density and out-of-equilibrium
temperature; this failure is why sink particles are required.
Figure~\ref{fig:besphere} shows that when we apply the LP density  
threshold as the creation criterion for sink
particles, unphysical behavior does not arise and 
both low and high resolution runs give essentially the same density
and velocity profiles before, at, and after singularity formation.

Since cold gas where collapse occurs is generally quite close to
thermal equilibrium,
we can estimate typical values of the LP threshold density for given $\Delta
x$ and given cooling/heating functions.
For the CNM ($T<200\Kel$), the cooling function can be
approximated by $\Lambda(T)\approx
2.8\times10^{-28}\sqrt{T}\exp(-92/T)\ergs{\rm cm}^{3}$, where $T$ is
the temperature in Kelvins. Then, the thermal equilibrium condition 
($n_H\Lambda(T)-\Gamma=0$) can be solved for the equilibrium density at a given
temperature. By equating the thermal equilibrium density to 
$n_{\rm thr}=\rho_{\rm thr}/(\mu_H m_H)\propto T/(\Delta x)^2$,
the threshold temperature ($T_{\rm thr}$) is given by
the nonlinear equation
\begin{equation}\label{eq:TLP}
T_{\rm thr}^{3/2}\exp\rbrackets{-\frac{92}{T_{\rm thr}}}=0.6 \rbrackets{\frac{\Gamma}{\Gamma_0}}
\rbrackets{\frac{\Delta x}{{\rm pc}}}^2,
\end{equation}
where $\Gamma_0=2\times10^{-26}\ergs$ is adopted for the Solar neighborhood
\citep[see][]{2002ApJ...564L..97K}.
Given the solution $T_{\rm thr}$, the sink particle density threshold using
the LP condition is then
obtained from Equation (\ref{eq:rhoLP}).  The corresponding density
threshold from the Truelove condition would use a coefficient of $\pi/16$
instead of $8.86/\pi$.

\begin{figure}
\plotone{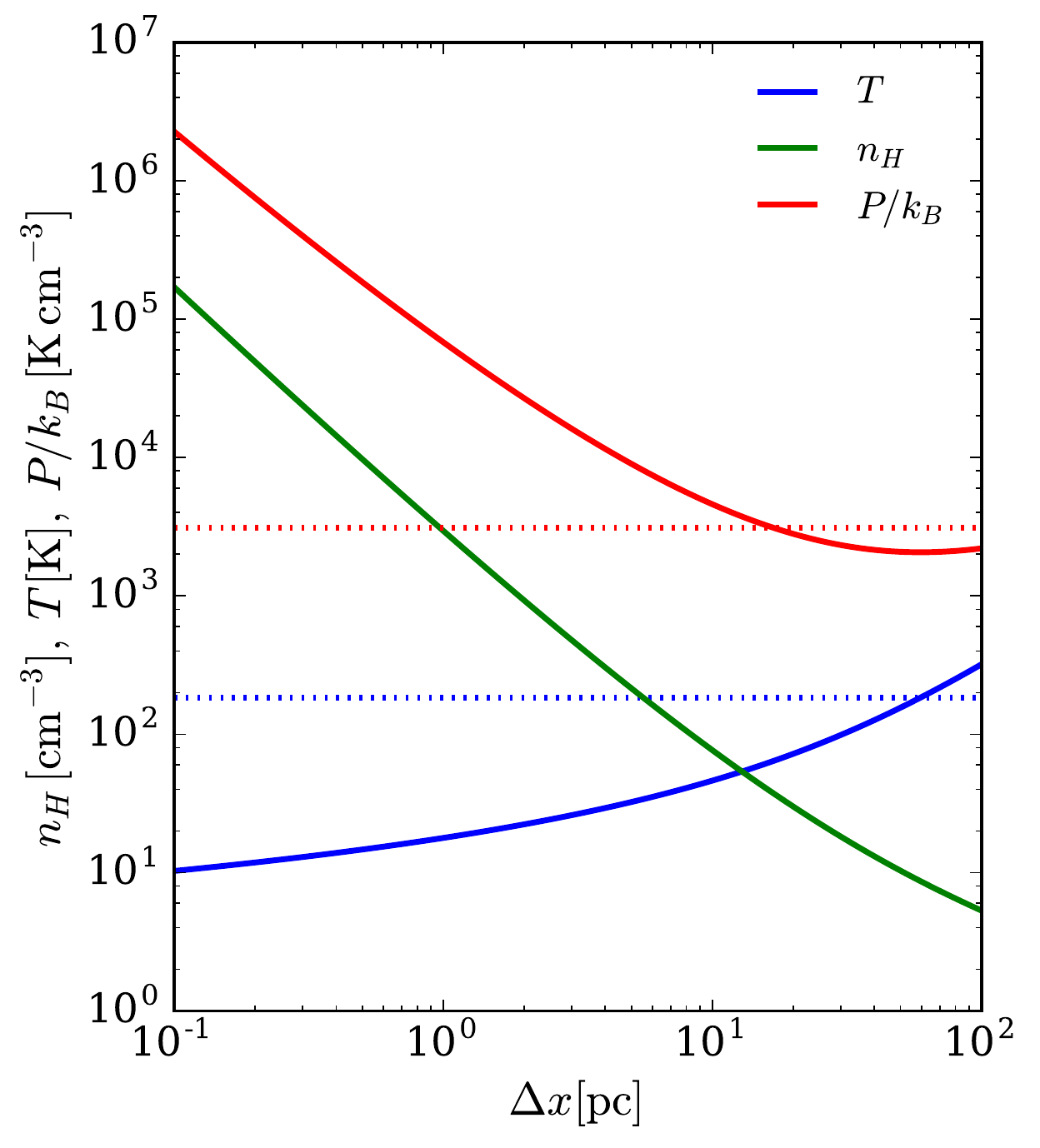}
\caption{Estimated threshold density, temperature, and pressure
for sink particle creation as a function
of spatial resolution $\Delta x$. Solid lines show the threshold conditions
defined by simultaneous satisfaction of thermal equilibrium and the
LP density threshold condition (Equation \ref{eq:rhoLP}).
The horizontal dotted lines
denote the maximum equilibrium temperature of the CNM ($=184\Kel$, blue)
and the two-phase pressure 
($P_{\rm two}=3110k_B\Punit (\Gamma/\Gamma_0)$, red). At high resolution, the
threshold pressure is well above $P_{\rm two}$.  
}
\label{fig:threshold}
\end{figure}

Figure~\ref{fig:threshold} shows the threshold density, temperature,
and pressure that satisfy both the LP density
threshold and thermal equilibrium, as a function of $\Delta x$
for $\Gamma=\Gamma_0$.
We overplot as dotted lines the maximum equilibrium temperature of the CNM 
($184\Kel$, blue)
and the two-phase pressure 
($P_{\rm two}=3110k_B\Punit (\Gamma/\Gamma_0)$, red) 
defined by the harmonic mean of the maximum and minimum equilibrium 
pressures of the WNM and CNM, respectively.  
  $P_{\rm two}$ is characteristic of the midplane thermal pressure 
in the warm/cold atomic ISM when star formation is in a 
self-regulated state 
\citep{2010ApJ...721..975O,2011ApJ...743...25K,2013ApJ...776....1K}.
Note that these particular values
of $T_{\rm CNM, max}$ and $P_{\rm two}$ depend on the adopted cooling curve
and on the value of $\Gamma$.

For self-gravitating collapse to be numerically captured under typical
ISM conditions, the central pressure of a collapsing cloud at the local 
simulation resolution must be
sufficiently high compared to the typical ambient ISM pressure;
i.e. $P_{\rm thr}$ must be large enough compared to $P_{\rm two}$.
If the resolution is too low, collapse cannot
be captured.   For example,
Figure~\ref{fig:threshold} shows that resolution $\Delta x < 20\pc$ would be
needed for the expected central pressure of a collapsing cloud
to exceed that of the environment under 
Solar-neighborhood conditions ($\Gamma=\Gamma_0$).
We note that the resolution needed to obtain $P_{\rm thr} \gg P_{\rm two}$
and capture collapse is easily accessible for
local simulations, but that even the marginal resolution
needed for $P_{\rm thr} \sim P_{\rm two}$ is inaccessible
for current cosmological galaxy formation simulations.

After identifying any location where the threshold density is exceeded,
we check the second two criteria for sink particle formation.  In
particular, we first test whether the gravitational potential
in a candidate cell is the minimum in the control volume of the
$3^3$ surrounding cells.  
Finally, we check flow convergence in all directions,
$dv_x/dx<0$, $dv_y/dy<0$, and $dv_z/dz<0$ (not simply
$\nabla\cdot\vel <0$).  We note that
if the resolution is high enough, the additional criteria are essentially
always satisfied (see GO13).  The control volume has an effective radius of 
$r_{\rm ctrl}\equiv 1.5\Delta x$, and we check whether 
there are any existing sink particles within $2r_{\rm ctrl}$. 
If there are, a new sink particle will not be created.

For a cell that passes all the above conditions,
we create a sink particle at the cell center.
When a sink particle is created, the control volume
is treated as ghost zones for the purpose of resetting the fluid variables.
For example, the cells at faces, sides, and corners 
of the cubic control volume use averages of (respectively)
one, two, and three contact cells in the surrounding grid 
to compute the density, momentum, and pressure.
The cell containing the sink particle
uses the average of six contact cells to compute its density, momentum,
and pressure.  
In the shearing box, we need to carefully subtract and add
the background velocity for corresponding cells,
before taking and after applying averages, respectively.
Note that we do not reset magnetic fields in the control volume.
After the control volume
density and momentum are reset as ghost zones, the sink particle
mass and velocity are assigned such that the total mass and momentum of
sink particle + control volume are equal to the value integrated over
all $3^3$ zones prior to sink particle creation.  

Finally, we note that we immediately treat a sink particle as a star
  cluster. This implicitly assumes that the ``star formation time scale'' --
  representing the interval from the beginning of gravitational collapse to
  star cluster formation -- is negligible compared to other relevant
  timescales.   In particular, our treatment is valid provided that the
  ``feedback time scale'' -- the interval between star cluster
formation and the first SN event ($\sim 4\Myr$ for {\tt STARBURST99}) -- is
longer than ``the star formation time scale.''  Physically,
it is reasonable to assume that massive stars would form within a free-fall
time at the threshold density.
In this light, instantaneous introduction of a
sink particle would be valid up to resolution of 
$\Delta x=8\pc$, where $n_{\rm thr}\simgt 100\pcc$ and $t_{\rm ff}\sim4\Myr$.
In low resolution models with lower $n_{\rm thr}$, the free-fall
time would exceed the true time before the onset of feedback.  However,
we do not explicitly introduce a time delay to account for this.  Also, low
resolution tends to lead to very massive sink particles, whereas in a higher
resolution simulation there would have been several smaller sinks that form
at different times.  Taken together, these effects tend to (unphysically)
increase the space-time correlation of star formation and feedback at 
low resolution, as we shall show later.

\subsubsection{Sink Accretion and Merging}\label{sec:sp_acc}
In contrast to the original GO13 implementation,
subsequent to particle creation we reset control volume fluid variables
and add mass and momentum to sink particles only when gas is converging
to the control volume in all three directions. 
This modification is
desirable because the current simulations develop very strong turbulence,
such that there can be large relative velocities of particles and the
surrounding gas.
To check for flow convergence, we define the right/left face-averaged 
fluxes in the sink particle's reference frame as
\begin{eqnarray}
\overline{\mathcal{F}}_{r/l}\equiv
\sum_{j,k}\mathcal{F}_{\rho,i+2/i-1,j,k}
\rbrackets{1 - v_{\rm x, sp}\frac{\mathcal{F}_{\rho,i+2/i-1,j,k}}{\mathcal{F}_{\rho v_x,i+2/i-1,j,k}}},\\
\overline{\mathcal{G}}_{r/l}\equiv
\sum_{i,k}\mathcal{G}_{\rho,i,j+2/j-1,k}
\rbrackets{1 - v_{\rm y, sp}\frac{\mathcal{G}_{\rho,i,j+2/j-1,k}}{\mathcal{G}_{\rho v_y,i,j+2/j-1,k}}},\\
\overline{\mathcal{H}}_{r/l}\equiv
\sum_{i,j}\mathcal{H}_{\rho,i,j,k+2/k-1}
\rbrackets{1 - v_{\rm z, sp}\frac{\mathcal{H}_{\rho,i,j,k+2/k-1}}{\mathcal{H}_{\rho v_z,i,j,k+2/k-1}}},
\end{eqnarray}
where $\mathcal{F}$, $\mathcal{G}$, and $\mathcal{H}$
stand for fluxes in the $x$-, $y$-, and $z$-directions
defined at cell faces, respectively,
and $i$, $j$, and $k$ in the summation run from $i-1$ to $i+1$, $j-1$ to $j+1$, and $k-1$ to $k+1$, respectively,
for a particle in a cell at $(i,j,k)$.
We check the converging flow condition using
$\overline{\mathcal{F}}_r<0$ and $\overline{\mathcal{F}}_l>0$,
or equivalently, $\overline{\mathcal{F}}_r-\overline{\mathcal{F}}_l<0$
and $\overline{\mathcal{F}}_r\overline{\mathcal{F}}_l<0$,
and similarly for $\overline{\mathcal{G}}$ and $\overline{\mathcal{H}}$.
Note that these fluxes are calculated at $t^{n+1/2}$ 
so that we need to know particles' velocity also at this time,
which is what the particle integrator returns (see \S\ref{sec:sp_move}).

When the converging-flow
condition is satisfied, the particle accretes mass and momentum
as in GO13.
The accretion rates of mass and momentum to each sink particle
are calculated based on the fluxes returned by the Riemann solver
at the control volume boundary,
combined with the mass and momentum differences
within the control volume 
between the new, step $(n+1)$, and old, step $n$, control volumes
if sink particles move across grid zones (see GO13).
In order to calculate the accretion rate due to particles' movement,
we need to advance sink particles' positions before the MHD integrator
step, where we calculate the accretion rate based on the fluxes.
This is important in preserving Galilean invariance of particle accretion
(see test problem in \S3.3 of GO13).
Note that accretion can occur for a given  particle 
before it hosts its first SN event (see \S~\ref{sec:feedback});
after this time, sink particles become star particles and cease to accrete.

When the distance between two sink particles
is smaller than $2r_{\rm ctrl}$, we merge 
them by creating a new sink particle at the center of mass
of the two. All properties for the new sink particle 
are set by mass-weighted averages.

\subsubsection{Sink/Star Particle Types}\label{sec:sp_type}

We separate sink/star particles into three categories based on its (mean) age $\tage$.
\begin{itemize}
\item {\bf ``growing'' particles}: $0<\tage<t_{\rm SN}$.  This group
  consists of particles between their birth and their first SN event.
  The mean SN onset time is
  $t_{\rm SN}\sim 4\Myr$, but for any sink/star particle individual 
SN events are determined stochastically 
based on the rate from {\tt STARBURST99} (see Figure~\ref{fig:syn}).
``Growing'' particles are treated as sinks, and as described above can accrete
gas and merge with other ``growing'' particles.
They exert gravity and contribute to the total mean FUV radiation.
\item {\bf ``feedback'' particles}: $t_{\rm SN}<\tage<t_{\rm life}$. This
  group consists of 
  star particles between their first SN event and the adopted
  feedback lifetime $t_{\rm life}\equiv 40 \Myr$.  These particles do not accrete
  and merge; their motion is simply integrated as in Section \ref{sec:sp_move}.
  These particles exert gravity by contributing to $\rho_{\rm sp}$, and
  contribute total mean FUV radiation (Section \ref{sec:sp_fuv}) and
  SN feedback (Section \ref{sec:snr}).  
\item {\bf ``passive'' particles}: $t_{\rm life}<\tage$.  This group of particles
  is no longer active, and affects the gas and other particles only through
  the gravity they exert.
\end{itemize}

In addition, we use ``runaway'' particles to follow the position of runaways
(see \S~\ref{sec:snr}). These particles are massless ``passive'' particles
and have no effect on the gravity and the gas. They are ejected from
sink/star particles and host
one SN event after an assigned delay time $\tau_{\rm run}$.

\subsubsection{Sink/Star Aging}\label{sec:sp_age}

After their formation, sink/star particles must age in time, in order
for feedback to be appropriately applied based on the stellar
population of the clusters they represent.  To represent birth of
a young stellar population from newly accreted gas, we assign 
individual sink/star particles a mass-weighted mean age:
\begin{equation}
\tage = \frac{m_{\rm sp}(\tage+\Delta t)+
\Delta m \Delta t}{m_{\rm sp}+\Delta m},
\end{equation}
where $\Delta m$ is the mass increment during a hydrodynamic time step
$\Delta t$.  Note that simple aging $\tage = \tage + \Delta t$ is
recovered as $\Delta m \rightarrow 0$.  When two particles merge, the
age is also calculated by taking a mass-weighted mean.

\subsubsection{Particle Motion Integration}\label{sec:sp_move}

We integrate particles' positions and velocities
from the equation of motion in a shearing box with gravity
from both gas and particles, as well as the fixed gravitational potential
from the old stellar disk.
Since self-gravity is evaluated at steps $n$ and $n+1$, not at
$n+1/2$,
it is advantageous to use a 
``Kick-Drift-Kick (KDK)'' form of a leap-frog integrator.
The general form of a KDK integrator can be written as
\begin{eqnarray}\label{eq:kdk}
\vel^{n+1/2}&=&\vel^{n-1/2}+ \mathbf{a}(\mathbf{x}^n,\vel)\Delta t,\\
\mathbf{x}^{n+1}&=&\mathbf{x}^n +  \vel^{n+1/2}\Delta t.
\end{eqnarray}
The acceleration $\mathbf{a}$ is the same as in the
right-hand side of Equation (\ref{eq:mom}) divided by $\rho$.  It 
consists of the effective gravity, $-\nabla \Phi_{\rm tot}$,
which depends on position,
and the Coriolis force, $-2\mathbf{\Omega}\times\vel$,
which depends on velocity.
Depending on the choice of velocity in the acceleration term, 
one can write down KDK integrators in explicit $\vel\equiv\vel^{n-1/2}$,
implicit $\vel\equiv\vel^{n+1/2}$, 
and semi-implicit $\vel\equiv(\vel^{n-1/2}+\vel^{n+1/2})/2$ forms
(see \citealt{2010ApJS..190..297B} for DKD integrators).
However, those integrators are not symplectic in a shearing box.

Here, we adopt a symplectic
integrator for Hill's equation suggested by \citet[][Q10]{2010AJ....139..803Q}
with a generalization for an arbitrary rotation profile with
shear parameter $q\equiv -d\ln\Omega/d\ln R$.
The full set of equations to advance velocity and position of a particle from
$t^n$ to $t^{n+1}$ is as follows.
\begin{itemize}
\item First Kick:
  \begin{eqnarray}
    \label{eq:kick1x}
v_x^{n+1/2}& =& v_x^{n}+hf_x^{n}+2h\Omega (P_y^n-2\Omega x^{n})\\
    \label{eq:kick1y}
v_y^{n+1/2}& =& P_y^{n}-2\Omega x^n-2h\Omega v_x^{n+1/2}\\
    \label{eq:kick1z}
v_z^{n+1/2}& =& v_z^{n}+hf_z^{n},
\end{eqnarray}
where $h\equiv\Delta t/2$, $P_y^n\equiv v_y^n+hf_y^n+2\Omega x^n$,
and $\mathbf{f}=-\nabla \Phi_{\rm tot}^n $;
\item Full Drift:
\begin{eqnarray}
\mathbf{x}^{n+1} = \mathbf{x}^n + \mathbf{v}^{n+1/2}\Delta t;
\end{eqnarray}
\item Second Kick:
  \begin{eqnarray}
    \label{eq:kick2x}
v_x^{n+1}& =& v_x^{n+1/2}+hf_x^{n+1}+2h\Omega (P_y^n-2\Omega x^{n+1})\\
\label{eq:kick2y}
v_y^{n+1}& =& P_y^{n}-2\Omega x^{n+1}+hf_y^{n+1}\\
\label{eq:kick2z}
v_z^{n+1}& =& v_z^{n+1/2}+hf_z^{n+1}.
\end{eqnarray}
\end{itemize}
In practice, the self-gravity that enters in the second kick step
(through $\mathbf{f}^{n+1}$) is not available until time $t^{n+1}$, because
it requires an update of the gas density by the MHD integrator.
Thus, while the position can be fully drifted to $t^{n+1}$, the velocity
can be kicked only to $t^{n+1/2}$.
This means that $v_x^n$ and $v_y^n$ appearing in Equations (\ref{eq:kick1x})
and (\ref{eq:kick1y}) are not immediately available.
We therefore must begin each step
by applying the ``second kick'' to the velocity associated with the
previous timestep, i.e. for step $n'$ we would apply Equations
(\ref{eq:kick2x})-(\ref{eq:kick2z})
with $n=n'-1$, which make use of $\mathbf{f}^{n'}$.  
Having particle velocities at $t^{n+1/2}$ is also useful for
computing mass fluxes for evaluating the accretion onto sink particles.

We test our particle integrator using epicyclic orbits. 
With only the gravitational force that produces the background rotation
curve of the galaxy ($\Phi_{\rm tot}=\Phi_{\rm tidal}$),
a particle's motion in a shearing box follows a planar epicyclic orbit
described by 
\begin{equation}\label{eq:epi_xy}
x(t)=A \cos (\kappa t),\quad y(t)=\frac{2\Omega}{\kappa}A\sin(\kappa t),
\end{equation}
where $\kappa\equiv\sqrt{2(2-q)}\Omega$ is the epicyclic frequency,
and $A$ is the (arbitrary) amplitude of the orbit.
For this test, we adopt $(A,q,\Omega)\equiv(0.4,1,1)$,
which gives the total energy of the orbit,
\begin{equation}
E=\frac{1}{2}(\dot{x}^2+\dot{y}^2)+\Phi_{\rm tidal}= (2-q)\Omega^2 A^2=0.16.
\end{equation}

\begin{figure*}
\plotone{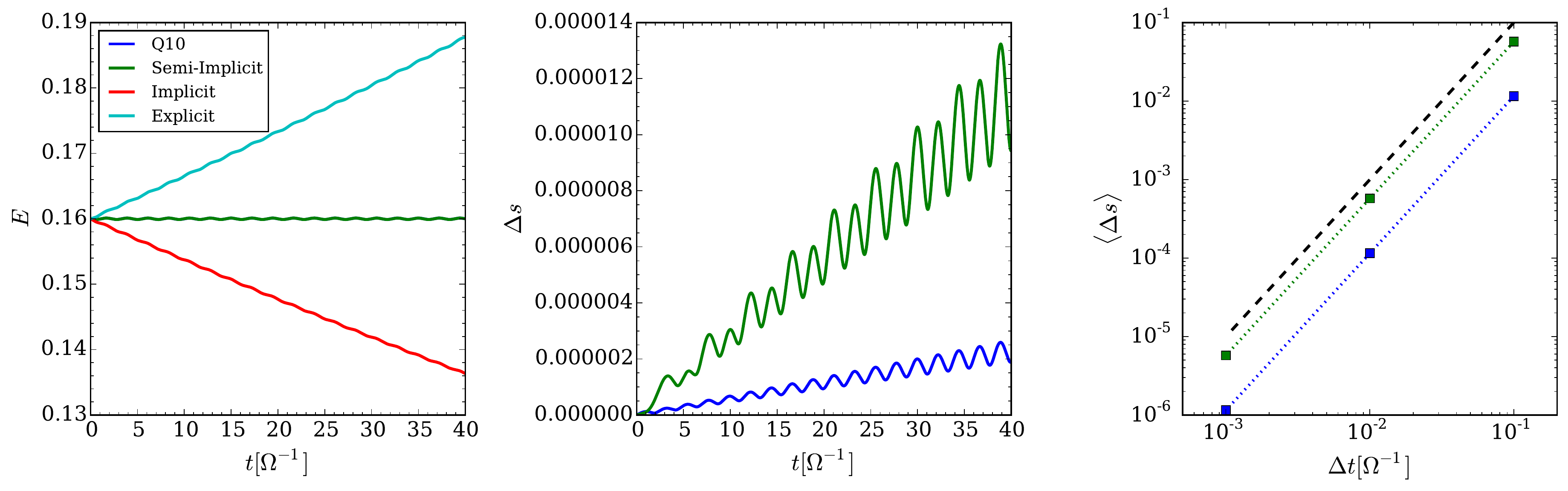}
\caption{Results of epicyclic motion tests for
  our adopted symplectic integrator (Q10; shown in
  all panels in blue), in comparison to 
  explicit (cyan; left panel), semi-implicit (green; all panels), and
  implicit (red; left panel) KDK integrators.
Left and middle panels respectively show 
time evolution of energy ($E$) and  
position offset ($\Delta s$) of the orbit for 
$\Delta t=10^{-3}\Omega$.  Right panel shows 
convergence of the mean $\Delta s$ as a function of time step.
The dashed line in the right panel shows second order convergence. 
}
\label{fig:epicycle}
\end{figure*}

Figure~\ref{fig:epicycle} shows energy (left) and position offset (middle)
of the orbit for $\Delta t=10^{-3}/\Omega$, and convergence of position
offset as a function of time step (right). 
For comparison, we also present results from explicit,
semi-implicit, and implicit KDK integrators.
Both the semi-implicit and the Q10 integrator show good energy conservation,
without the secular energy increase of the explicit integrator
or decrease of the implicit integrator.
Position offsets (consisting of a phase shift, but not a change in the
semi-major axis) increase over time for both semi-implicit
and Q10 methods, with the Q10 integrator giving better results and more
regular fluctuation.
The mean phase errors over $t\Omega=40$ (right panel) 
show second order convergence as the time step gets smaller.
For our fiducial model with $\Delta x= 4\pc$, typical
timesteps have $\Omega \Delta t \sim 10^{-4}$-$10^{-5}$.

\subsection{Star Formation Feedback}\label{sec:feedback}

\subsubsection{FUV radiation and photoelectric heating}\label{sec:sp_fuv}

\begin{figure}
\plotone{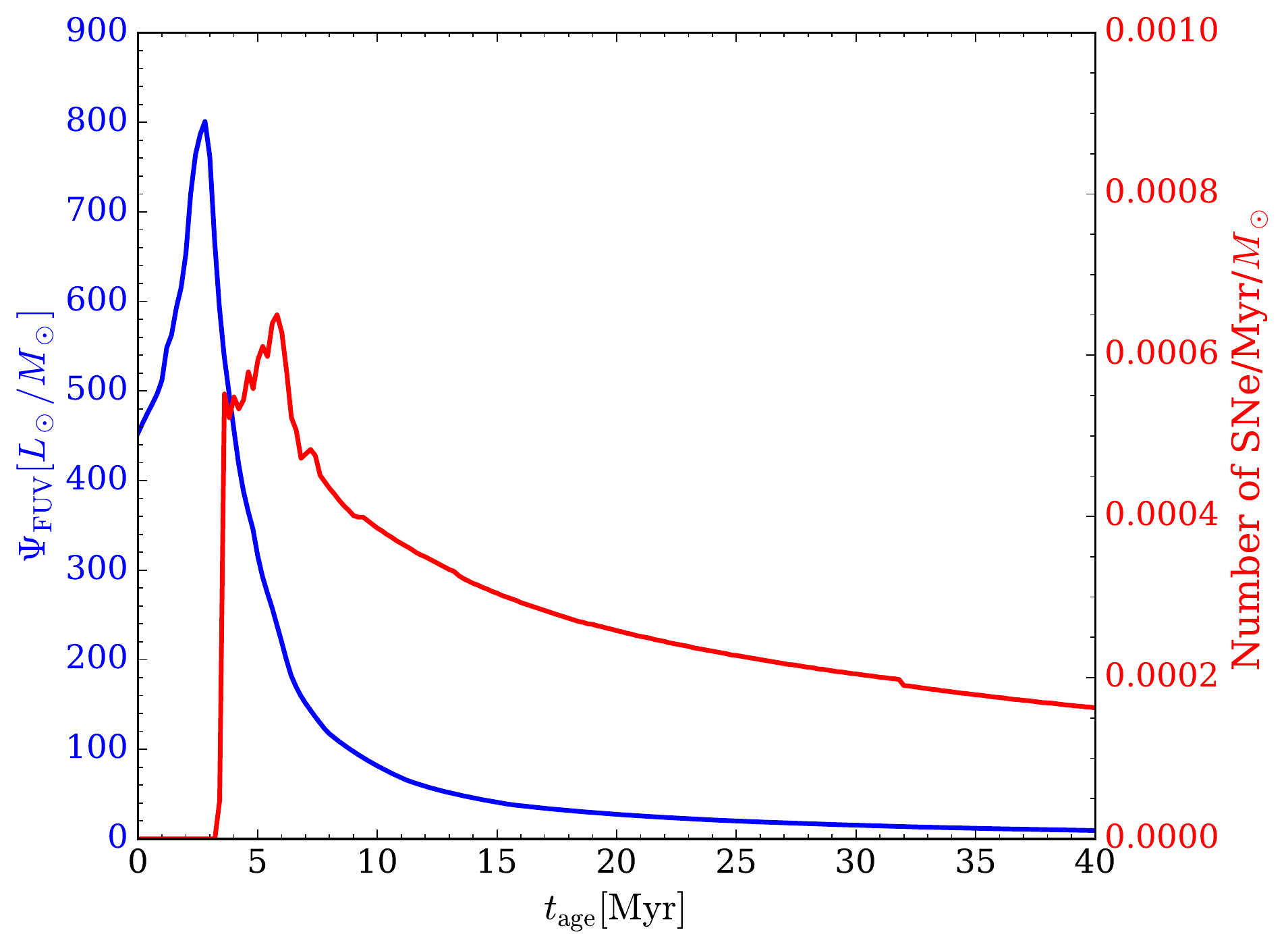}
\caption{
{\it Left axis, blue}: Specific FUV luminosity of star clusters 
($\Psi_{\rm FUV}\equiv L_{\rm FUV}/M_{\rm cl}$)
as a function of cluster age, assuming a fully sampled
Kroupa IMF. 
{\it Right axis, red}: Specific SN rate of star clusters
($\snrate\equiv d(\mathcal{N}_{\rm SN}/M_{\rm cl})/dt$) 
as a function of cluster age.
}
\label{fig:syn}
\end{figure}

For warm and cold gas, we implement photoelectric heating on
grains. In reality, the heating rate at any location in the ISM
depends on the dust density (which we assume is simply proportional to
the gas density) and on the angle-averaged FUV intensity $J_{\rm
  FUV}$.  In the present implementation, we do not attempt direct
radiative transfer from the active sink/star particles to compute $J_{\rm
  FUV}$ at every spatial location (see note below),
but we do vary the heating rate
temporally based on the mean FUV radiation that the massive young
stellar population would produce at any time.  For a sink/star particle
with mass $m_{\rm sp}$ and mass-weighted mean age $\tage$,
the FUV luminosity is 
$L_{\rm FUV,sp}=\Psi_{\rm FUV}(\tage)m_{\rm sp}$.
Here, we use a tabulated luminosity-to-mass ratio,
$\Psi_{\rm FUV}(t)$,
for FUV radiation ($6\eV < h\nu<13.6\eV$) 
from {\tt STARBURST99} with 
a fully sampled Kroupa IMF (\citealt{2001MNRAS.322..231K}; 
see blue solid line and left axis in Figure~\ref{fig:syn}).

We calculate total FUV luminosity $L_{\rm FUV}$ by summing $L_{\rm FUV,sp}$
from all sink/star particles.
We further assume that
$4 \pi J_{\rm FUV} \propto \Sigma_{\rm FUV}\equiv L_{\rm FUV}/(L_xL_y)$, where
for the optically thin case and a uniform source distribution at the
midplane, equality would hold.\footnote{
The mean FUV intensity, $J_{\rm FUV}$, should be obtained 
by full radiation transfer.
Although computationally expensive, time-dependent transfer
can be incorporated with ray-tracing methods provided that the number of
active sources is not too large; implementation of this is underway.
For present purposes, we note \citep{2010ApJ...721..975O} that simple slab
geometry of gas with UV optical depth of $\tau_\perp = \kappa\Sigma$ 
gives $4\pi J_{\rm FUV}=\Sigma_{\rm FUV} (1 - E_2(\tau_\perp/2))/\tau_\perp$,
where $E_2$ is the second exponential integral. For $\tau_\perp\sim 0.1-1$,
the correction term varies from $\sim 1.7$ to $0.7$. Therefore, 
$4\pi J_{\rm FUV}\approx \Sigma_{\rm FUV}$ is a valid assumption 
to zeroth order.
In forthcoming work, we will investigate the distribution
of FUV radiation for simulation snapshots of gas and star particle 
source distributions.}
We compute the contribution to the photoelectric heating rate
from the local disk assuming $\Gamma \propto J_{\rm FUV}$,
where 
$\Sigma_{\rm FUV,0}=6.9 L_\odot/{\rm pc^2}$, corresponding to 
$J_{\rm FUV,0} = 2.2\times10^{-4}\ergs\cm^{-2}{\rm \; sr^{-1}}$
(or $G_0=1.7$ in Habing units; \citealt{1978ApJS...36..595D}), 
yields $\Gamma_0=2\times10^{-26}\ergs$.

Including the temperature dependence of the mean molecular weight 
(to turn off photoelectric heating at high temperature)
and adding a heating floor due to the metagalactic FUV 
($4\pi J_{\rm FUV,meta}=6.7\times10^{-6}\ergs\cm^{-2}$
or $J_{\rm FUV,meta}/J_{\rm FUV,0}=0.0024$, \citealt{2002ApJS..143..419S})
the adopted heating rate becomes
\begin{equation}\label{eq:heat}
\Gamma=\Gamma_0 \rbrackets{
\frac{\mu(T)- \mu_{\rm ion}}{\mu_{\rm ato}-\mu_{\rm ion}}}
\rbrackets{\frac{\Sigma_{\rm FUV}}{\Sigma_{\rm FUV,0}}+
0.0024}.
\end{equation}

Both ``growing'' and ``feedback'' sink/star particles contribute 
to the FUV luminosity.

\subsubsection{Supernova Rates and Runaways}\label{sec:snr}

Let $\snrate(\tage)\equiv d(\mathcal{N}_{\rm SN}/M_{\rm cl})/dt$
be the specific SN rate, defined as
the number of SNe per unit time per star cluster mass 
for a cluster of mean age in the interval $(\tage, \tage+dt)$.
We adopt results for $\snrate$
tabulated from {\tt STARBURST99} \citep{1999ApJS..123....3L}
for a fully sampled Kroupa IMF 
(see red solid line and right axis in Figure~\ref{fig:syn}). 
This gives a total mass of new stars per SN, $m_*$, 
\begin{equation}
m_* \equiv\rbrackets{\int_0^{t_{\rm life}}\snrate  dt}^{-1}=95.5\Msun.
\end{equation}

For a sink/star particle with mass $m_{\rm sp}$ and 
mass-weighted mean age $\tage$, 
 the expected number of SNe during the MHD time step $\Delta t$ is
 $\mathcal{N}_{\rm SN}=m_{\rm sp} \snrate(\tage)\Delta t$.
 In the simulation, we sample a
 random number $\mathcal{U}_{\rm SN}$ in (0,1) and a SN
event occurs if $\mathcal{N}_{\rm SN} > \mathcal{U}_{\rm SN}$. 
The mean value of $\snrate$ over the lifetime of a sink/star particle is
$\langle \snrate\rangle =(m_* t_{\rm life})^{-1}$ and the SN rate varies by
only a factor $\sim 3$, so typically 
$\mathcal{N}_{\rm SN} \sim m_{\rm sp} \Delta t/(m_* t_{\rm life})$.
Note that if $\mathcal{N}_{\rm SN}>1$, 
more than one SN would be expected in a single epoch. 
However, considering our typical timestep,
$\Delta t\sim 10^{-3}-10^{-5}\Myr$, and sink/star particle mass, $m_{\rm
sp}=10^3-10^5\Msun$, this never happens in practice.

We include runaway OB stars to represent SN events that occur 
far from star clusters.\footnote{Note that in this work we
neglect Type Ia SNe, whose rates are typically 10\% to 20\% of the SNe rate
from massive stars
\citep[e.g.,][]{1994ApJS...92..487T}. However, runaway SNe
capture the most important effect of Type Ia SNe, namely a broad 
spatial distribution of SN events far from young star clusters.}
 Runaways may occur because of  
(1) dynamical interactions during birth of a star cluster \citep[e.g.,][]{1967BOTT....4...86P,2011Sci...334.1380F,2015ApJ...805...92O} and 
(2) SNe in OB binary systems \citep[e.g.,][]{1961BAN....15..265B,2000ApJ...544..437P,2011MNRAS.414.3501E}.
In the current TIGRESS implementation, our treatment of
runaways is based on the second mechanism.
Whenever there is a SN event 
($\mathcal{N}_{\rm SN} > \mathcal{U}_{\rm SN}$), 
we additionally check whether this event occurs in a single star
or a binary system for a given binary fraction of OB stars $f_{\rm bin}$ 
(2/3 is our standard choice).
Here, this binary fraction is defined as  
\begin{equation}
f_{\rm bin}\equiv\frac{\textrm{Number of OB stars in binaries}}
{\textrm{Number of total OB stars}}
\equiv\frac{N_{\rm bin}}{N_{\rm tot}}=
\frac{N_{\rm bin,p}+N_{\rm bin,s}}{N_{\rm single}+N_{\rm bin}},
\end{equation}
where the number of primaries and secondaries is
$N_{\rm bin,p}=N_{\rm bin,s}=N_{\rm bin}/2$.  
If the event
is not in a binary, we simply assign an SN event at the sink/star particle
position following \S~\ref{sec:explosion}.

If the event is defined as a SN in a binary, 
we allow for both an {\it in situ}
primary SN explosion immediately
and a runaway secondary that produces a SN explosion after a time delay. 
Runaways are  massless star particles ejected isotropically
with initial velocity distribution 
consistent with a binary population synthesis model 
(Fig. 2 of \citealt{2011MNRAS.414.3501E}). 
Each runaway is a time bomb
with a delay time $\tau_{\rm run}$.
In order to have the overall SN rate consistent with $\snrate m_{\rm sp}$,
we set the delay time using the probability integral transform,
$\tau_{\rm run} = \sncrate^{\rm inv}(\mathcal{U}_{\rm run})$,
where $\mathcal{U}_{\rm run}$ is another uniform random number in (0,1),
the normalized cumulative distribution of SNe is 
\begin{equation}
\sncrate(t)  \equiv\frac{ \int_{0}^{t} \snrate dt}
{\int_0^{t_{\rm life}}\snrate dt}=\int_{0}^{t} \snrate m_* dt,
\end{equation}
and $\sncrate^{\rm inv}$ is the inverse function of $\sncrate$.
We tabulate $\sncrate$ from the tabulated $\snrate$.  
Once we obtain $\tau_{\rm run}$, we compare it with $\tage$. 
When $\tau_{\rm run}>\tage$, a massless
star particle is ejected and will be exploded after $\tau_{\rm run}$. 
We also explode a SN at the
original sink/star particle position, representing the explosion of the primary.
Otherwise (if $\tau_{\rm run}<\tage$), 
we simply do nothing for this unphysical situation (causality violation).

\subsubsection{Supernova Feedback Treatment}\label{sec:explosion}

For any SN event, we first determine the feedback prescription that
is applied.  We use three types of feedback in TIGRESS: {\tt EJ} (Ejecta), {\tt ST} (Sedov-Taylor), and {\tt MC} (Momentum Conserving).
Our goals in defining and selecting from among these prescriptions  
are to assign thermal energy and momentum appropriate for the
resolution of the simulation and local
 properties of the ambient environment,
avoiding both numerical ``overcooling'' and a subsequent time step
that is very small.
In the {\tt ST} and {\tt EJ} feedback prescriptions, 
we reset density, momenta, and thermal energy
within a sphere of radius $R_{\rm snr}$.
In the {\tt MC} feedback prescription, the thermal energy and density are
unchanged but velocities are reset within $R_{\rm snr}$.
Magnetic fields remain unchanged.

To choose which prescription is applied, we calculate the
mean gas properties in cells surrounding the SN with $d_{ijk}<R_{\rm snr}$, 
where $d_{ijk}$ is the
distance of a cell from the SN. We vary $R_{\rm snr}$ starting from $R_{\rm
  snr,min}=3 \Delta x$ to $R_{\rm snr,max}$
(the choice of $R_{\rm snr,max}$ is model-dependent)
with an increment of 
$\Delta R=\Delta x/2$.
For every $R_{\rm snr}$, we first calculate the total mass within
the initial ``SN remnant,''
$M_{\rm snr}\equiv \sum\rho_{ijk}(d_{ijk}<R_{\rm snr}) \Delta V +M_{\rm ej}$, and the
mean density $\rho_{\rm snr}\equiv M_{\rm snr}/\sum_{d_{ijk}<R_{\rm snr}} \Delta V$,
where $M_{\rm ej}=10\Msun$ represents the mass of ejecta plus 
circumstellar medium.  We then calculate the ratio of the initial remnant mass 
to the expected shell formation mass, $\mathcal{R}_M\equiv M_{\rm snr}/M_{\rm sf}$,
where $M_{\rm sf}=1679\Msun (n_H/\pcc)^{-0.26}$ 
  can be estimated from numerical simulations of individual expanding
  SNR with ambient density $n_H$ that include cooling 
  \citep[e.g.,][KO15a]{2015ApJ...802...99K}. 
Here, we use the mean molecular weight per hydrogen atom $\mu_H\equiv 1.427$
to obtain the hydrogen number density $n_H = \rho_{\rm snr}/(\mu_H m_H)$.
Shell formation 
occurs when post-shock gas at the forward shock of
an expanding SN remnant first becomes strongly radiative.  

If $\mathcal{R}_M>1$ for $R_{\rm snr,min}$, 
the Sedov-Taylor stage of this SN is unresolved.  This case
occurs if resolution is low and/or local density is high, and the SN
remnant would become radiative at a scale smaller than that is locally resolved.
In this case, our treatment of feedback just injects momentum to the grid using
the final radial momentum  
$p_{\rm snr}=2.8\times10^5\Msun\kms (n_H/\pcc)^{-0.17}$
computed from resolved numerical simulations 
of a single SN in a two-phase medium (see KO15a); other
resolved simulations for an inhomogeneous medium find similar final momenta
\citep{2015MNRAS.450..504M,2015A&A...576A..95I,2015MNRAS.451.2757W}. 
We refer to this type of feedback as {\tt MC},
as momentum is assumed to be conserved from the unresolved scale at
which the remnant would cool to the resolved scale at which momentum is
injected to the grid.

If $0.027<\mathcal{R}_M<1$ for given $R_{\rm snr}$, 
we assign total SN energy $E_{\rm SN}=10^{51}\erg$
in both thermal and kinetic forms to the
gas within the feedback region, $d_{ijk} < R_{\rm snr}$. 
The energy ratio
is appropriate for the energy conserving phase 
($\sim 72\%$ in thermal, and $\sim 28\%$ in kinetic).
We refer to this type of feedback as {\tt ST}, as it represents a remnant
that is in the Sedov-Taylor stage.

Note that as long as $\mathcal{R}_M<1$ and the ambient medium is
  uniform, KO15a showed that SNRs are
  sufficiently resolved that either the {\tt ST} prescription
  or pure thermal energy prescription (dumping the total SN energy to thermal
  energy)   will
  provide correct feedback.
  However, $\mathcal{R}_M\sim1$ (or $\Delta x/r_{\rm sf}\sim 1/3$) is marginal.
Thus, if the density were {\it not} uniform
within $R_{\rm snr}$, the gas at higher-than-average density would cool faster.
Equivalently, the value of $M_{\rm sf}(n_H)$ evaluated using $n_H$ from
the overdense portions of the feedback region would be lower than if the
mean density were used, and the
corresponding $\mathcal{R}_M$ would exceed unity.  Therefore, if a  thermal
dump were applied to {\it inhomogeneous} gas with $\mathcal{R}_M\sim1$
it would lead to ``overcooling,'' and the SNR evolution would not be properly
resolved. We conclude that
it is unsafe to use a criterion $\mathcal{R}_M\sim1$ unless
the material within the feedback region is reset to a uniform density (see
below).  More generally, to avoid vulnerability to overcooling, we use as
small a feedback region as possible so as to make $\mathcal{R}_M$ as close as possible to 0.027.  This stricter criterion corresponds to
$\Delta x / r_{\rm sf}=1/10$ for $R_{\rm snr,min}=3\Delta x$, the
``consistent convergence condition'' of KO15a.
If a SN event occurs in a very
rarefied medium so that $\mathcal{R}_M<0.027$ for $R_{\rm snr,min}$, we
increase $R_{\rm snr}$ and recalculate $\mathcal{R}_M$ until
$\mathcal{R}_M>0.027$, and assign {\tt ST} type feedback as above.
Increasing $M_{\rm snr}$ from extremely small values
reduces the initial temperture from extremely high values, which would
otherwise lead to extremely short time steps.

Finally, in the case that the surrounding medium has density so low that
$\mathcal{R}_M < 0.027$ even for $R_{\rm SNR}=R_{\rm snr,max}$, 
we assign pure kinetic energy within the remnant.
We refer to this type of feedback as {\tt EJ},
as it represents the effects of ejecta in the free expansion stage.

For {\tt ST} and {\tt EJ} feedback,
the mass density, momentum density, and internal energy of the gas within a SN
feedback region are initially set to constant values 
using the mean density, the mass-weighted mean velocity,
and the mean internal energy, respectively, within $R_{\rm snr}$.  
We then assign additional momentum with 
a radial velocity profile of $v_r\propto r^2$ 
and uniform thermal energy within $R_{\rm snr}$.
Note that the ejecta mass is already added in prior to
calculation of the mean density.

For {\tt MC} feedback,
we do not reset the gas properties in the feedback region
prior to applying the momentum associated with the SN event.
When we assign velocity fields, we add corresponding radial momentum 
to each gas parcel's momentum in the star particle's rest frame.
In low resolution simulations (used for our convergence test),
more than one SN event can be assigned at a given
time and position. We then inject momentum and energy additively from multiple SNe.

\begin{figure}
\plotone{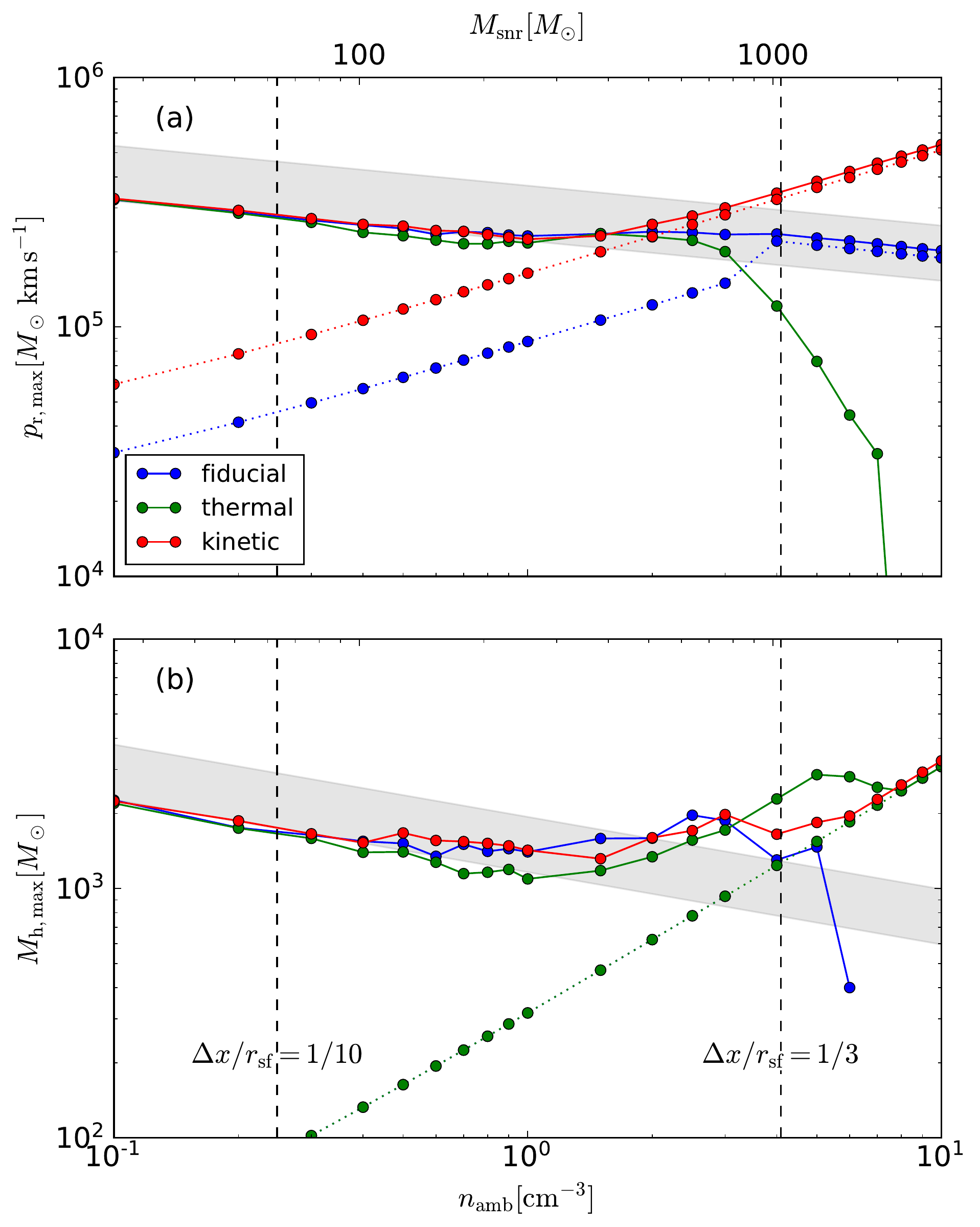}
\caption{
Tests of SN feedback prescriptions, based on
simulations of radiative SNR evolution in a uniform medium 
with spatial resolution of $\Delta x=4\pc$:
(a) the
final radial momentum and (b) the maximum mass of
hot gas ($T>10^5\Kel$), both  
as a function of the number density of the ambient medium ($n_{\rm amb}$). 
Blue symbols connected with solid lines show results based on our
fiducial feedback prescription, as described in the text.
Green and red symbols connected with solid lines show results of
purely thermal and purely kinetic energy feedback methods, respectively.  
Symbols connected with dotted lines 
indicate the initial values of radial momentum and hot gas mass for
different feedback methods.
The dashed vertical lines demarcate loci of different 
resolution for the initial SNR of radius
$r_{\rm sf}=22.6\pc (n_{\rm amb}/\pcc)^{-0.42}$ (KO15a), with 
$\Delta x/r_{\rm sf}=1/10$ (left) and $\Delta x/r_{\rm sf}=1/3$ (right).
The grey shaded regions in (a) and (b) respectively 
enclose fitting results from KO15a within $\pm25\%$
for the final momentum 
$p_{\rm final}=2.95\times10^5 (n_{\rm amb}/\pcc)^{-0.16}\momunit$
and the hot gas mass at shell formation 
$M_{\rm sf}=1.55\times10^3 (n_{\rm amb}/\pcc)^{-0.29}\Msun$.}
\label{fig:sntest}
\end{figure}

To demonstrate the results of our SN feedback method, 
we run test simulations of radiative SNR evolution
in a uniform medium with fixed spatial resolution of $\Delta x=4\pc$.
Figure~\ref{fig:sntest} plots (a)
final radial momentum and (b) the maximum
mass of hot gas ($T>10^5\Kel$), both as a function of the number density of 
the ambient medium ($n_{\rm amb}$; bottom axis)
and the enclosed mass of the feedback region ($M_{\rm snr}$; top axis).   
Results using our fiducial prescription as described above are shown in blue.
For comparison, we also show the result of using purely thermal energy
injection (green) and purely kinetic energy injection (red) 
within a radius of $R_{\rm snr,min}=3\Delta x$.
For reference, 
symbols connected with dotted lines indicate the corresponding
initial values of (a) radial momentum and (b) hot gas mass,
as applied by the different feedback prescriptions.
In (b), the dotted
blue and green lines coincide for $\Delta x/r_{\rm sf} < 1/3$.
The vertical dashed lines denote $\Delta x/r_{\rm sf}=1/10$ and $1/3$,
where $r_{\rm sf}=22.6\pc (n_{\rm amb}/\pcc)^{-0.42}$ is the shell formation
radius at the corresponding ambient density (KO15a). The vertical
line at $\Delta x/r_{\rm sf}=1/3$ corresponds to the transition from
the {\tt ST} (at lower density) to the {\tt MC} (at higher density)
feedback type in our prescription.
Note that the transition to {\tt EJ} type feedback would occur for
$n_{\rm amb}<9\times10^{-4}\pcc(R_{\rm snr,max}/128\pc)^{-2.38}$,
beyond the domain of this figure.

In both panels, if the Sedov-Taylor stage is fully resolved
(according to the assessment of KO15a) with $\Delta x/r_{\rm sf}<1/10$,
all feedback prescriptions give the same result for both final energy
and maximum hot gas mass.  As the Sedov-Taylor stage is 
marginally resolved with $\Delta x/r_{\rm sf}=1/3$, the different
prescriptions lead to a factor of a few differences. 
In particular, Figure~\ref{fig:sntest}(a) indicates that the final
momentum is significantly  under- and over-estimated for unresolved
cases ($\Delta x/r_{\rm sf}>1/3$) when feedback is implemented via
purely thermal and kinetic energy, respectively.
Our fiducial method injects momentum to the medium
within 25\% (grey shaded region) of that found by KO15a for all values of
the density.
We note that the cooling curve and the treatment of 
the mean molecular weight are slightly different from those of KO15a,
so that agreement in the overall trend is more important than absolute values.
If $\Delta x/r_{\rm sf} <1/3$,
the hot gas mass peaks at around the shell formation time
and remains within $\sim 25\%$ of KO15a.
If $\Delta x/r_{\rm sf}>1/3$, as the initial SNR size
$R_{\rm snr}=R_{\rm sir,min}=3\Delta x$ is larger than the shell formation radius,
the hot gas mass peaks initially and drops abruptly for all cases.  For
these unresolved cases, the hot gas mass is initially overestimated,
but at later times is severely underestimated relative to resolved
simulations (with smaller $\Delta x$) at the same $n_{\rm amb}$.

Careful inspection of the final momentum
  for the case where $\Delta x/ r_{\rm sf}=1/3$ and $\mathcal{R}_M=1$
  (i.e. $M_{\rm snr}=1000\Msun$, near the right vertical dashed
line in Figure~\ref{fig:sntest}(a)), shows that both kinetic
and thermal prescriptions lead to deviations from the correct solution.
Thus, if one wanted to adopt a pure-thermal feedback for resolved
regions (i.e. low $M_{\rm snr}$) and {\tt MC} feedback for unresolved
regions, it would be necessary to choose $M_{\rm snr}\sim 400\Msun$ rather
than $M_{\rm snr}=1000\Msun$ as the transition point.  In a clumpy medium, even
the lower value of $M_{\rm snr}$ might be risky.  We note that 
\citet[][see also
  \citealt{2015MNRAS.449.1057G}]{2017MNRAS.466.1903G} adopt
for their resolved treatment (applied when $\mathcal{R}_M<1$)
a pure-thermal prescription with a feedback region size
chosen such that $M_{\rm snr}=1000\Msun$, which based
on our analysis would be marginal.  Also, since Gatto et al
did not redistribute
mass to make the density in the feedback region uniform in the thermal-feedback
case, it would make their treatment even more vulnerable to overcooling,
as described above.  
Overcooling may
potentially explain why their SN-only model results in
quite high SFRs, while our SN-only SFRs 
are much lower (and consistent with observations).  
Note that a feedback prescription that transitions from {\tt EJ}
type (red line in Figure~\ref{fig:sntest}) to {\tt MC} type feedback
at $M_{\rm snr}\sim 400\Msun$ 
can also inject the correct momentum
(and produce hot gas if resolved).
The SN feedback prescriptions described in \citet{2014ApJ...788..121K} and
\citet{2014MNRAS.445..581H} use an {\tt EJ}-transitioning-to-{\tt MC}
approach.

\subsection{Gas Phases}\label{sec:phase}

We distinguish the gas in simulations by defining five components: 
cold ($c$) $T<184\Kel$; unstable ($u$)
$184\Kel<T<5050\Kel$; warm ($w$) $5050\Kel<T<2\times10^4\Kel$; ionized ($i$)
$2\times10^4\Kel<T<5\times10^5\Kel$; hot ($h$) $T>5\times10^5\Kel$.  
We use the
Heaviside step function $\Theta(C)$ to filter each component, 
where $C=c,u,w,i,h$.
Sometimes we refer to $C=c+u+w$ as warm-cold medium ($wc$) 
and to $C=i+h$ as hot-ionized medium ($hi$).
In the analysis below, we define summation of the quantity
$Q$ over each component as
$\sum_{C} Q\equiv \sum Q\Theta(C)$.

\section{Fiducial Solar Neighborhood Model}\label{sec:Fiducial}

In this section, we begin by describing the evolution of
a solar neighborhood model with standard resolution, $\Delta x=4\pc$.
The simulation domain size is $L_x=L_y=1024\pc$ and $L_z=4096\pc$.
We limit the maximum SN feedback region radius to $R_{\rm snr,max}=128\pc$.
For the external gravity
(Equation \ref{eq:phi_ext}), we use the parameter set
determined by \citet{2013ApJ...772..108Z}, $\Sigma_*=42\Surf$,
$z_*=245\pc$, $\rho_{\rm dm}=0.0064\rhounit$, and $R_0=8\kpc$.  In the
limit of $|z|\ll z_*$, the external gravity can be approximated by a
linear profile $g_{\rm ext}\approx -4\pi G \rho_{\rm sd} z$. In our
previous work (e.g., \citealt{2013ApJ...776....1K,2015ApJ...815...67K}, 
which had smaller vertical domain
than is required to follow the hot ISM and wind launching), we used
$\rho_{\rm sd}=0.05\rhounit$ for the volume density of stars and dark
matter at the midplane in the Solar neighborhood.  The current
parameter set gives a larger value of $\rho_{\rm
  sd}=\Sigma_*/(2z_*)+\rho_{\rm dm}=0.092\rhounit$.  
We adopt $\Omega = 28\kms\kpc^{-1}$, which yields $\torb=220\Myr$,
and binary OB fraction $f_{\rm bin}=2/3$.

For the initial vertical gas density profile, we use a 
double exponential 
\begin{equation}
\rho(z)=\rho_1(z)+\rho_2(z)=\rho_{10}\exp(-\Phi_{\rm 0,tot}(z)/\sigma_1^2)+
\rho_{20}\exp(-\Phi_{\rm 0,tot}(z)/\sigma_2^2),
\end{equation}
where $\Phi_{\rm 0,tot}(z)=\Phi_{\rm ext}(z) + 2\pi G \Sigma |z|$ is the 
total gravitational potential 
using a thin-disk approximation for gaseous
self-gravity. The adopted initial effective sound speeds of two phases 
are set to $\sigma_1=7\kms$ and
$\sigma_2=10\sigma_1$, respectively, representing warm and hot media.
The midplane density of the warm medium is $\rho_{10}=2.85 m_H\pcc$, while
for the hot medium $\rho_{20}=10^{-5}\rho_{10}$ such that it has 
negligible mass contribution.  The total initial gas surface density is 
$\Sigma=13\Surf$.  The initial pressure profile is set to 
$P=\rho_1\sigma_1^2+\rho_2\sigma_2^2$. The initial
magnetic fields have only azimuthal ($\yhat$) components 
with constant plasma beta, $\beta\equiv 8\pi P/B^2=10$, everywhere.
This yields an initial midplane magnetic field strength of
$B = 2.6\mu G$.
The initial heating rate is set to $\Gamma_0$.

Our initial conditions are not in thermal equilibrium, and the density
near the midplane is higher than the maximum density of the WNM 
for given cooling and heating rates. 
As soon as the simulation begins, the gas immediately cools 
and looses vertical support from thermal pressure. 
Thermal instability develops rapidly and 
separates the gas near the midplane into 
the CNM and WNM. As a consequence, if the system is initiated without
any turbulent support,  
the CNM falls to the midplane and forms a very thin and dense slab 
\citep[e.g.,][]{2010ApJ...720.1454K}.
This leads to very bursty star formation, making the overall evolution
converge slowly. 

In order to reduce the artificial initial burst of star formation
(and subsequent ``ringing''),  
for an initial period 
we drive turbulence that provides overall vertical support.
At the same time, compressions seeded by these initial velocity perturbations
enhance thermal instability and promote star formation. 
We use one-dimensional velocity power spectrum 
$\mathcal{P}_k\propto k^{-2}$ for $1<kL_x/2\pi<64$, with driving rate
$\dot{E}_{\rm turb}=10^4L_\odot$.  
Turbulence is driven with this full strength for $50\Myr$, and
then slowly turned off from $50\Myr$ to $100\Myr$.
During the driving phase, we place a minimum on the heating rate of 
$\Gamma_0$ for the first $50\Myr$, and 
reduce this minimum slowly as the turbulent driving decreases.  
As stars form, turbulence and heating driven by SN and FUV feedback
begin to exceed the imposed turbulent driving and minimum
heating rate. All of our quantitative assessments are made after
the artificial driving/heating phase has ended, and a quasi-steady
saturated state has been reached.  
We have confirmed that evolution and properties in the saturated state
are insensitive to the exact form of initial turbulent driving and heating.
The exact initial density and pressure profiles are also not
  important.

\subsection{Overall Evolution and Star Formation Cycles}
\begin{figure}
\plotone{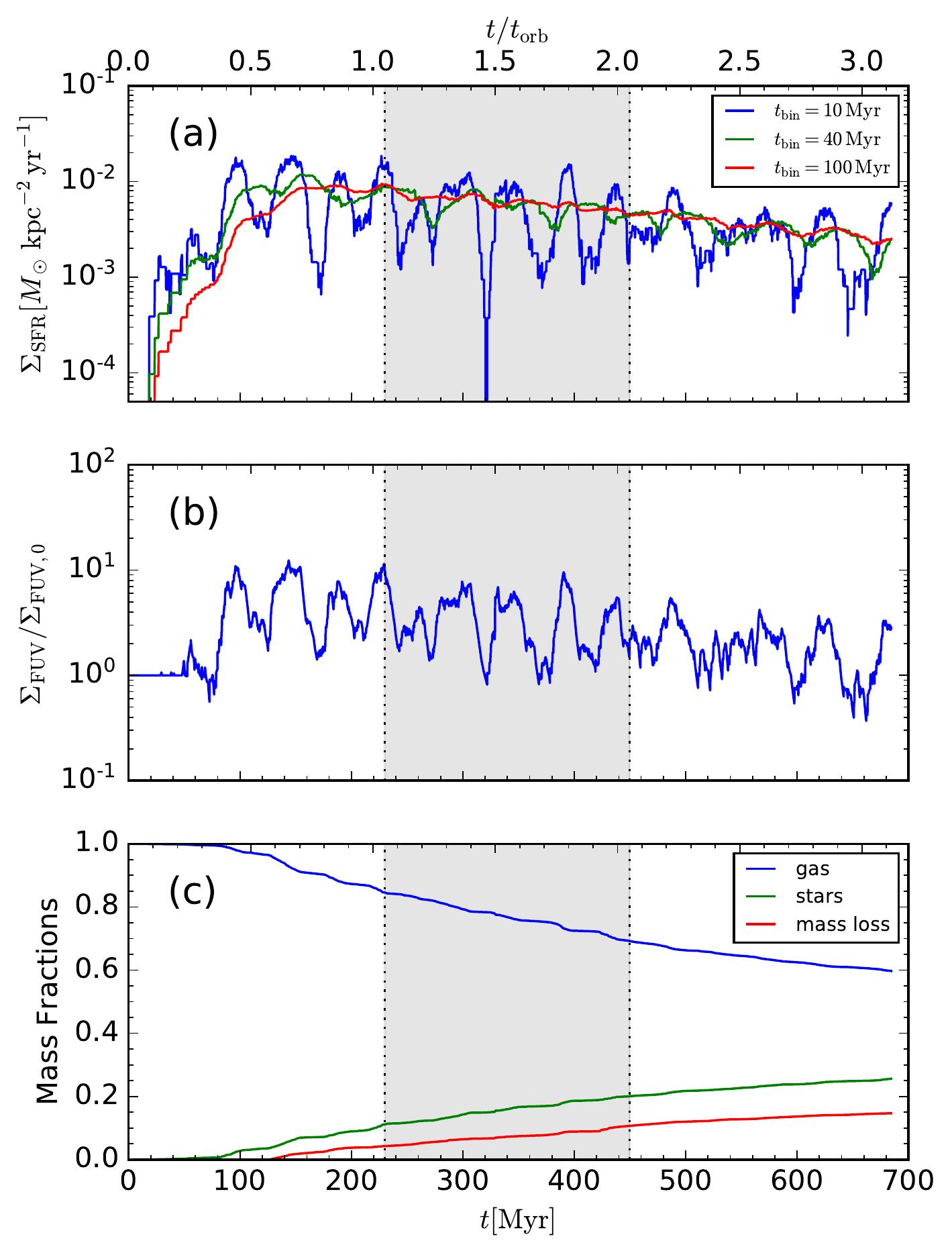}
\caption{Time evolution of (a) surface density of SFR, $\Sigma_{\rm SFR}$,
for star clusters younger than 10 (blue), 40 (green), and 100 (red) Myr; 
(b) surface density of FUV luminosity, $\Sigma_{\rm FUV}$, in units
of $\Sigma_{\rm FUV,0}=6.9 L_\odot/{\rm pc^2}$; and (c)
mass fractions of gas (blue), new stars (green), and 
gas mass loss through the vertical boundaries (red) 
compared to initial total gas mass. 
The shaded region between vertical dotted lines
denotes the time range for $11>\Sigma/(\Surf)>9$ over which 
statistical properties are calculated.
}
\label{fig:tevol1}
\end{figure}

Figure~\ref{fig:tevol1}(a) shows time evolution of
the recent SFR surface density calculated using the total mass of young stars;
\begin{equation}\label{eq:sfr}
\Sigma_{\rm SFR} (\Delta t = t_{\rm bin})\equiv\frac{\sum m_{\rm sp} (\tage<t_{\rm bin})}{L_x L_y t_{\rm bin}}
\end{equation}
We choose three different bins, $t_{\rm bin}=10$, 40, and 100 Myr, 
to indicate the way of SFR surface density would vary when
traced by different diagnostics of young stars   
(e.g. the simple burst model adopted in \citealt{2012AJ....144....3L} would have
95\% of H$\alpha$ and FUV emitted within 4.7Myr and 65 Myr, respectively).
As the early driven turbulence effectively limits the bursty behavior in 
the initial star formation, 
$\Sigma_{\rm SFR}$ increases with only a factor of a few variation
up to $t\sim 100\Myr$.
After this early ``imposed driving'' phase,
star formation feedback self-consistently offsets cooling and drives
turbulence.  Time evolution after $t=100\Myr$ 
reaches a quasi-steady state with self-regulation cycles 
involving large amplitude temporal fluctuations in
$\Sigma_{\rm SFR}$.  In each cycle, gas falls to 
the midplane and is collected by self-gravity and large-scale turbulent flows
into giant clouds, where collapse occurs in the highest density regions.
The massive stars in the newly born clusters that form strongly 
increase the FUV radiation field and SN rate, which disperses
the dense gas and enhances heating and turbulent driving throughout the ISM.
Star formation shuts off as the gas disk puffs up and becomes warmer.
With the corresponding reduction in star formation feedback, gas can settle
back to the midplane and once again collect into large clouds where star
formation occurs.
The mean duty cycle is 
 $\sim 45 \Myr$ for the simulation shown, and
there is an order of magnitude variation in
$\Sigma_{\rm SFR}(\Delta t = 10 \Myr)$.  For $\Sigma_{\rm SFR}(\Delta t = 40 \Myr)$
there is less than a factor of two variation, and for 
$\Sigma_{\rm SFR}(\Delta t = 100 \Myr)$ it is only tens of percent.  It should be
kept in mind, however, that the amplitude of variations depend on the
horizontal box size because this determines the number of independent
star-forming patches.  Larger boxes that contain a larger number of independent
patches would have reduced variation in $\Sigma_{\rm SFR}(\Delta t = 10 \Myr)$.

Figure~\ref{fig:tevol1}(b) plots time evolution of 
the surface density of FUV luminosity $\Sigma_{\rm FUV}=L_{\rm FUV}/L_xL_y$
normalized by $\Sigma_{\rm FUV,0}$ to show the time evolution
of the heating rate. As seen in
Figure~\ref{fig:syn}, most of FUV comes from very massive stars 
in star clusters younger than $\sim 10\Myr$. Therefore, the temporal
variation of the heating rate is very similar to 
$\Sigma_{\rm SFR} (\Delta t = 10\Myr)$. 
Similar to $\Sigma_{\rm SFR} (\Delta t = 10\Myr)$,
there is an order of magnitude temporal fluctuation in the amplitude of
$\Sigma_{\rm FUV}$.

Figure Figure~\ref{fig:tevol1}(a)
shows a secular decrease of $\Sigma_{\rm SFR}$ over time, which is
clearest in $40\Myr$ and $100\Myr$ averages.
This reduction is mainly because the total gas mass decreases, 
as seen in Figure~\ref{fig:tevol1}(c); there we show
time evolution of gas, star, and outflow mass fractions compared
to the initial gas mass. In this simulation, 
we convert gas into sink/star particles, but we do not replenish it. 
Also, a significant amount of the gas flows out through vertical boundaries
in the form of hot winds and warm fountains 
driven by the clustered SN feedback 
(and SNe from runaways at high-$|z|$).
\footnote{The driving and properties of winds and fountains are
  an important aspect of our simulations (as well as real galaxies), and
  will be carefully analyzed in a forthcoming companion paper.}
Considering the secular evolution of the simulation,
we limit our analysis of statistical properties to the time range
between $t_{11}=1.05\torb=231\Myr$ and $t_9=2.05\torb=450\Myr$ (shaded region
in Figure~\ref{fig:tevol1}), corresponding to the times
when the gas surface density is $11\Surf$ and $9\Surf$,
respectively. This is reduced from an initial gas surface density
of $13\Surf$.

\begin{figure*}
\plotone{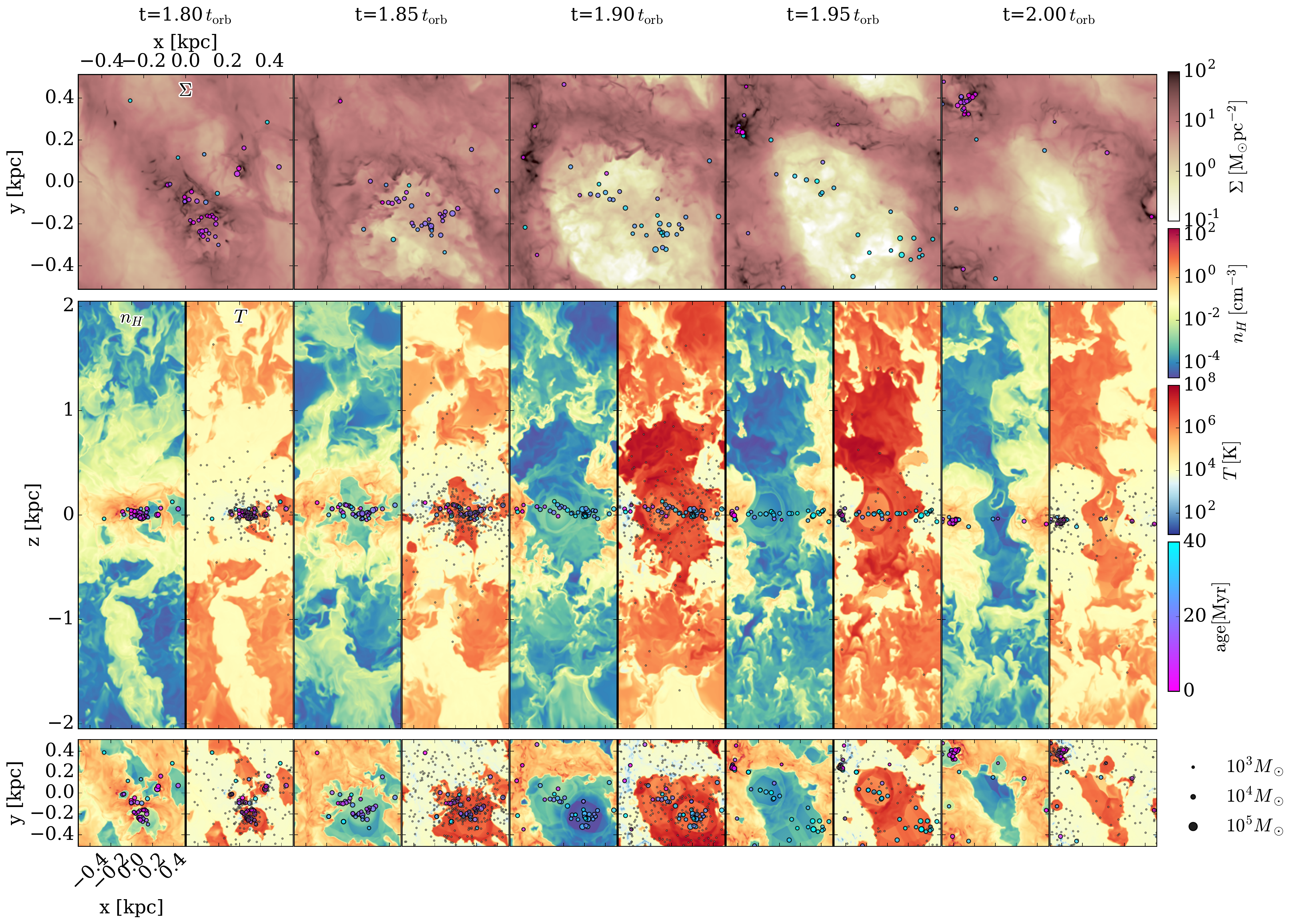}
\caption{
  Time evolution of the ISM and young star population
  in the Solar neighborhood model, shown at intervals of
  $\Delta t= 0.05\torb \approx 11 \Myr$, 
  from $t=1.8\torb=395\Myr$ to $t=2\torb=439\Myr$.
  Top row: gas surface density $\Sigma$ projected onto the
XY ($\xhat$-$\yhat$) plane.
Middle row: paired vertical slices (through $y=0$) of 
number density $n_H$ (left) and gas temperature $T$ (right). 
Bottom row: paired midplane slices (through $z=0$) of 
$n_H$  (left) and $T$ (right).
In all panels, colored circles denote locations of all sink and star particles 
younger than 40 Myr (see the colorbar) projected onto each plane.
The symbol size of sink/star particles denotes their mass (see legend).
Runaway OB stars are shown as black dots only in the temperature panels
for visual clarity.
}
\label{fig:MHD_4pc}
\end{figure*}

\begin{figure*}
\plotone{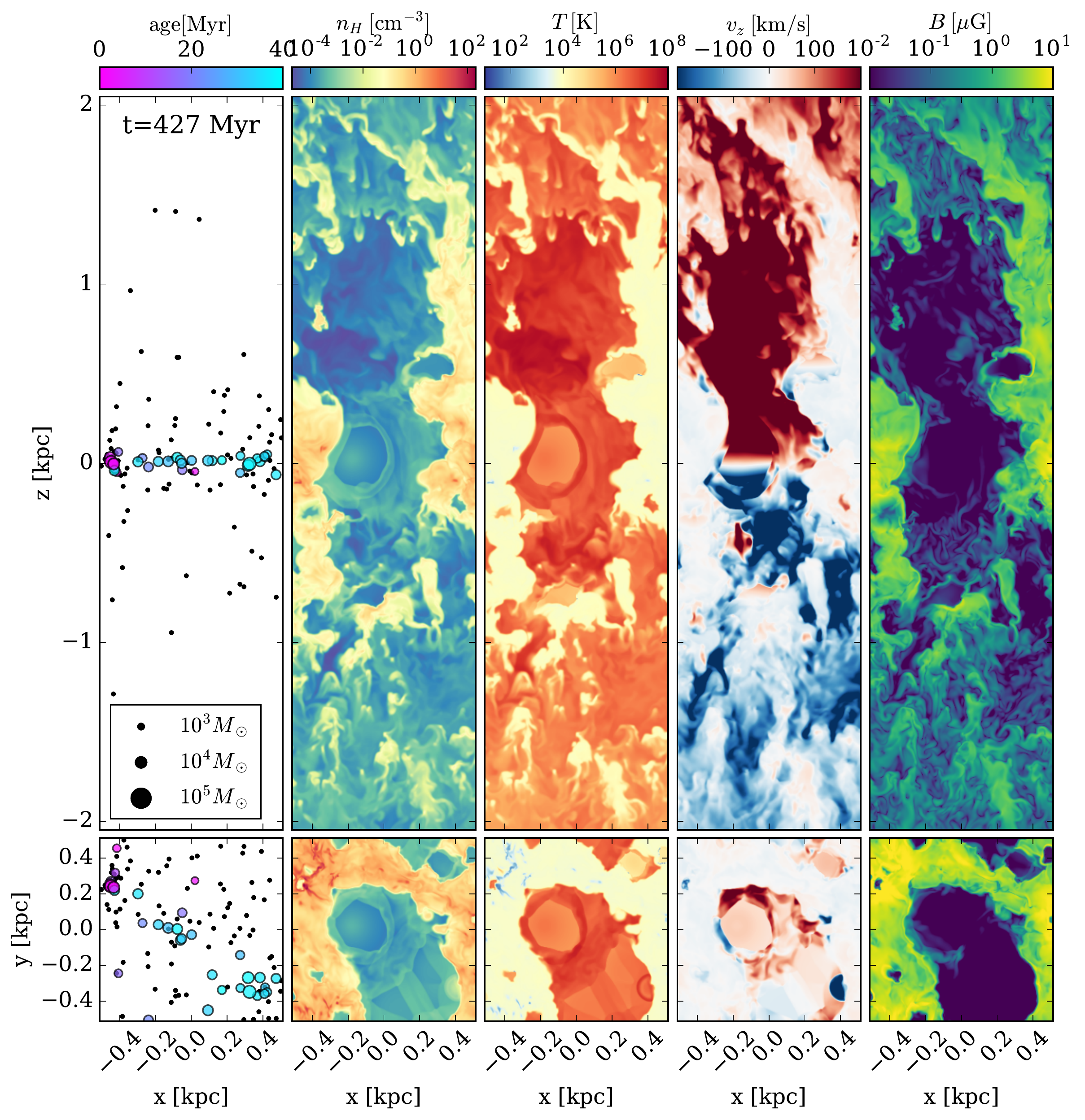}
\caption{
Snapshots of (a) sink/star particles, (b) number density, (c)
temperature, (d) vertical velocity, and (e) magnetic field strength at
$t=1.95\torb=428\Myr$. At this time, a strong outflow is driven by SN-heated
gas.  In
columns (b)-(e), we present XZ (through $y=0$; top) and XY (through
$z=0$; bottom) slices of ISM properties.  In column (a) positions of
sink/star particles with age younger than 40 Myr (colored circles) and runaway
stars (black dots) are projected onto XZ (top) and XY (bottom) planes.
\label{fig:MHD_4pc_slices}}
\end{figure*}

Figure~\ref{fig:MHD_4pc} displays a series of 5-panel snapshots 
showing surface density projections and number density and temperature slices
at an interval of 
$0.05t_{\rm orb}\sim11\Myr$, starting at $t=1.8\torb=395\Myr$. Sink and star
particles are also shown (see caption for details).
For one selected time
$t=1.95\torb=428\Myr$
when a strong outflow driven by SNe is prominent,
Figure~\ref{fig:MHD_4pc_slices} shows the distribution of sink/star particles, 
number density, temperature, vertical velocity, and magnetic field strength.

The series of snapshots in Figure~\ref{fig:MHD_4pc} clearly shows the
characteristic evolutionary cycle of self-regulated star formation.
In the first snapshot (at $t=1.8\torb$), a collection of star clusters
has formed in a dense cloud complex slightly below the center of the
XY plane. These young clusters heat up the gas and evaporate most of
the CNM. As a result, the WNM predominates in the midplane (see
temperature slice), although the hot gas still fills $\sim 20\%$ of
the volume near the midplane.  These sink/star particles produce a large
number of SN explosions (from $t=1.8\torb$ to $t=1.9\torb$), 
and as a result a superbubble forms, clearly
evident in the low density and high temperature regions of the XY panels of
the $t=1.9\torb$ snapshots.   At this epoch, the sink/star particles
from this burst have aged (see color scale) and emit less FUV radiation.
Plenty of the CNM can now form within the WNM, as is evident in the XY temperature
panel at $t=1.9\torb$.  As the hot bubble occupies more than
the half of the volume near the midplane, the CNM/WNM 
is pushed aside and aggregated within a smaller volume outside of the bubble.
Within this favorable environment, a massive dense cloud assembles (upper
left of XY panel) and
promotes a second round of star formation (see snapshots at $t=1.95$ and $2\torb$). 
At $t=2\torb$ (last set of snapshots), 
the first hot bubble has mostly merged with surrounding gas as
its interior has cooled off and turbulence redistributes material into
the former bubble volume.  Meanwhile, the result of new star formation
is evident in the upper-left corner of the midplane.
The feedback from this collection of sink/star particles will produce another superbubble.  
This cycle of massive cloud assembly  
leading to a burst of star formation, followed by a burst of feedback leading
to star formation quenching, continues 
throughout the run. 

Breakout of superbubbles from the warm-cold midplane layer drives hot gas to 
high-$|z|$; e.g. the XZ slices at $t=1.9$ and $1.95\torb$ in
Figure~\ref{fig:MHD_4pc} show this clearly.
 As shown in Figure~\ref{fig:MHD_4pc_slices},
 the hot gas has high outward vertical velocity, i.e. large $|v_z|$ at large
 $|z|$, with $sign(v_z) = sign(z)$.
We anticipate that this rapidly expanding hot outflow would 
escape to the CGM and/or IGM as a galactic wind. 
From Figures~\ref{fig:MHD_4pc} and \ref{fig:MHD_4pc_slices},
substantial WNM is also pushed to high-$|z|$  by
superbubble expansion. However, outward vertical velocities of this warm
material
are insufficient to  escape the gravitational potential of the galaxy 
($|v_z|>200\kms$ would be required).
The majority of the WNM falls back towards the midplane (when feedback
is at a low state), creating the ``return flow'' of a galactic fountain.
Detailed analysis of the multiphase high-$|z|$ wind and fountain flows 
is deferred to the companion paper.

\subsection{Multiphase Structures and Turbulence Properties of the ISM}\label{sec:4pc_phase}

\begin{figure*}
\plotone{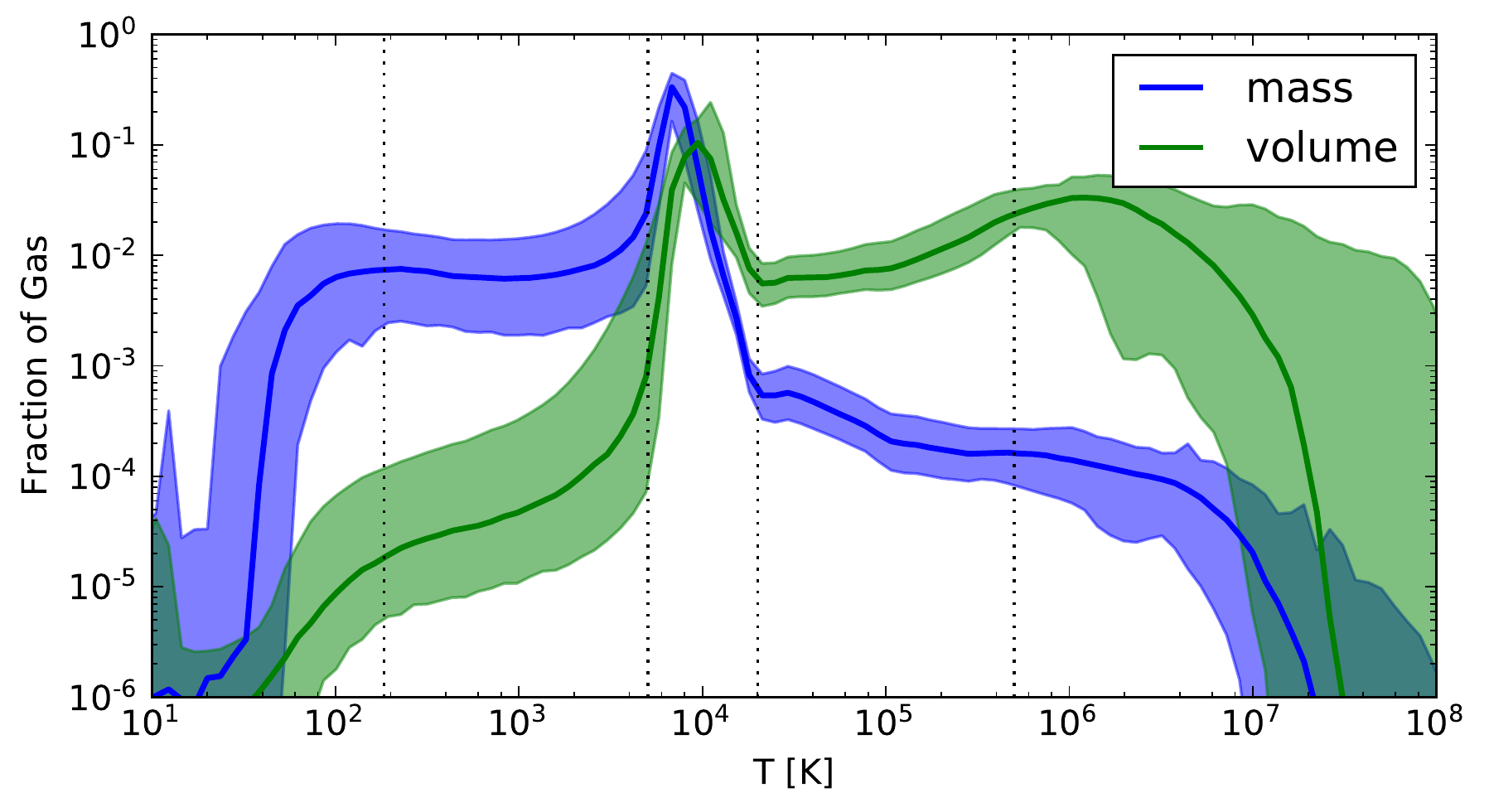}
\caption{Temperature PDFs weighted by mass (blue)
and volume (green). The solid lines
denote the median value of each bin over the time
span of $t_{11}$-$t_9$, and
the shaded area envelopes the minimum and maximum values over this time
span.  The vertical dotted lines indicate
$T=184\Kel$, 5050\,K, $2\times10^4 \Kel$, and $5\times10^5\Kel$
separating the gas into the five components we define (from left to right:
CNM, UNM, WNM, ionized, and hot).
}
\label{fig:T-pdf}
\end{figure*}

The ISM material in our simulation populates a wide range of temperature,
with three distinct phases.
Figure~\ref{fig:T-pdf} plots probability density
distributions (PDFs) of 
gas temperature weighted by mass (blue) and volume (green).
Thermal instability and rapid cooling tend to reduce the
amount of gas in the unstable (UNM) and ionized components.
However, strong turbulence, large temporal fluctuations of the heating rate,
and expanding superbubbles in our simulation continuously
repopulate these regimes.
In addition, since our numerical resolution
is not high enough to spatially resolve the transition
layer between the WNM and CNM, 
the CNM (UNM) mass fraction may be numerically
reduced (enhanced).\footnote{With our adopted numerical resolution,
we cannot fully resolve the Field length
for realistic thermal conductivity.
However, \citet{2008ApJ...681.1148K} showed that 
numerical diffusion caused by translational
motion produces an effective ``numerical conductivity,''
which is much larger than the physical thermal conductivity.
The phase transition layer is thicker 
and the UNM mass fraction increases
as the numerical resolution gets poorer 
(increasing ``numerical conductivity'').
As a consequence, the CNM mass fraction decreases at low resolution, while
the WNM mass fraction remains the same.}
Therefore, for studying numerical convergence
we will consider the mass fraction of the CNM plus UNM, rather than
their individual fractions.  
We obtain mean mass fractions between $t_{11}$ and $t_9$
of $f_{c+u}=24\%$ and $f_w=75\%$.

\begin{figure*}
\plotone{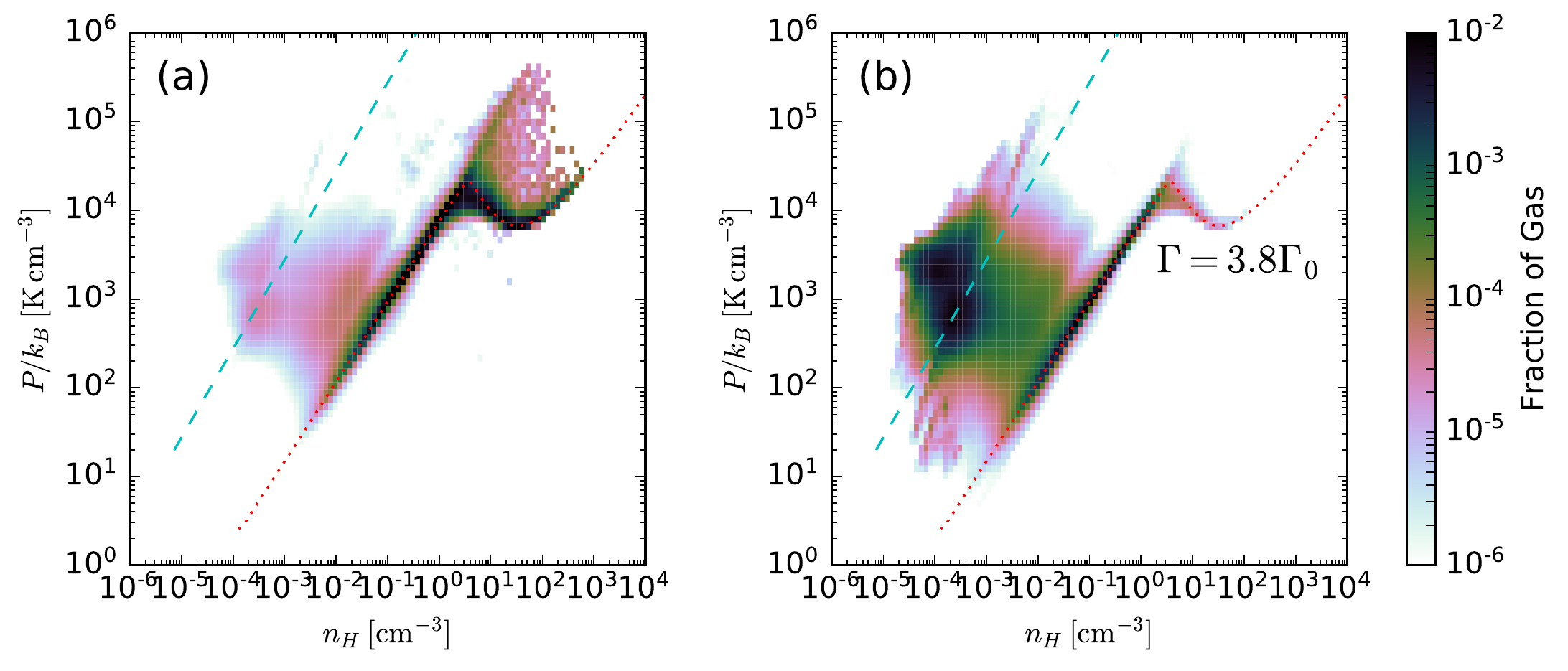}
\plotone{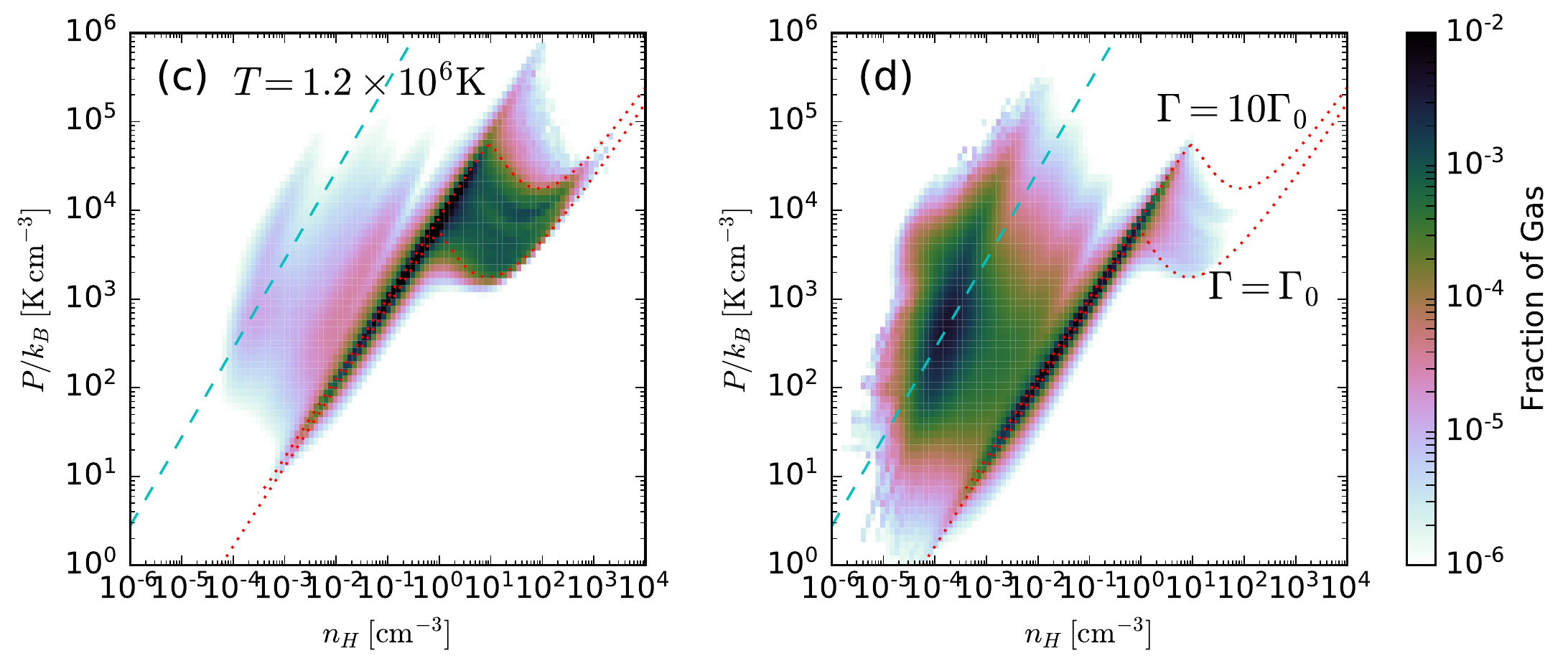}
\caption{Instantaneous ($t=1.95\torb=427\Myr$; (a) and (b))
and time averaged ($t_{11}$-$t_9$ or $1.05\torb$-$2.05\torb$; (c) and (d))
gas distribution in the number density ($n_H$)
and pressure ($P/k_B$) phase plane.
Color represents the mass (left column)
and volume (right column) fractions in logarithmic bins.
Dotted red
lines denote the thermal equilibrium curves 
for instantaneous heating rate $3.8\Gamma_0$ in (a) and (b)
and representative heating rates $\Gamma_0$ (lower) and $10\Gamma_0$ (upper)
in (c) and (d).
The dashed cyan line denotes the typical temperature of the hot-ionized medium, 
$T=1.2\times10^6\Kel$.
}
\label{fig:nP}
\end{figure*}

Figure~\ref{fig:nP} displays
gas PDFs at $t=1.95\torb=428\Myr$ (top row)
and averaged over time ranges between 
$t_{11}=1.05\torb=231\Myr$ and $t_9=2.05\torb=231\Myr$ (bottom row)
weighted by mass (left column) and volume (right column)
in the $n_H$-$P/k_B$ phase plane.
We draw as a dashed line the locus $T=1.2\times10^6\Kel$,
representing a typical hot gas temperature defined by 
the peak of the volume PDF (Figure~\ref{fig:T-pdf}) for the hot-ionized medium.
The dotted line in (a) and (b) shows the instantaneous thermal equilibrium curve
for $\Gamma=3.8\Gamma_0$.
Two dotted lines in (c) and (d) show the thermal equilibrium curves for $\Gamma=\Gamma_0$
and $\Gamma=10\Gamma_0$, which approximately brackets
the variation in the
heating rate from the varying $\Sigma_{\rm FUV}$ shown
in Figure~\ref{fig:tevol1}(b).
The mass PDF shows that the CNM, UNM, and WNM components are most populated,
while the volume PDF is dominated by the WNM, ionized, and hot components.

The warm-cold medium ($T<2\times10^{4}\Kel$) 
tends to evolve to a two-phase states with a short
cooling time, and the majority of 
the warm-cold medium indeed follows the instantaneous
thermal equilibrium curve and is within the envelope of
the two thermal equilibrium curves in Figure~\ref{fig:nP}.   
However,
due to strong turbulence and the large time variation in the heating rate,
the distinction between CNM and UNM is not as clear as expected
from the classical theory \citep{1965ApJ...142..531F,1969ApJ...155L.149F}.

In the volume PDF, the WNM is strongly concentrated around 
thermal equilibrium, while the hot-ionized medium shows a broader distribution.
At early stages of expansion for individual SN remnants or superbubbles,
when the temperature of shock heated gas
is high enough ($T>10^6\Kel$), 
adiabatically expanding hot interiors of bubbles forms a sequence
in the phase plane with a slope of $5/3$.
As shown in Figure~\ref{fig:T-pdf},
the typical hot-ionized medium temperature ranges between
$10^6\Kel$ and $10^7\Kel$
\citep{2017ApJ...834...25K}.
As the shock expansion speed drops, the temperature of post-shock gas is
lower, and due to the strong cooling 
peak at $T\sim 10^5\Kel$ (see Figure~\ref{fig:cool})
this material quickly cools down to join the WNM.  The centers of superbubbles 
remain hot until turbulence causes them to merge with the surrounding ISM. 
Thus, the volume PDF shows distinct hot and warm phases 
occupying the majority of the simulation volume.
Continuous SN explosions and mixing maintains a non-negligible fraction
of gas at intermediate temperature, between $10^4$-$10^6\Kel$.

\begin{figure}
\plotone{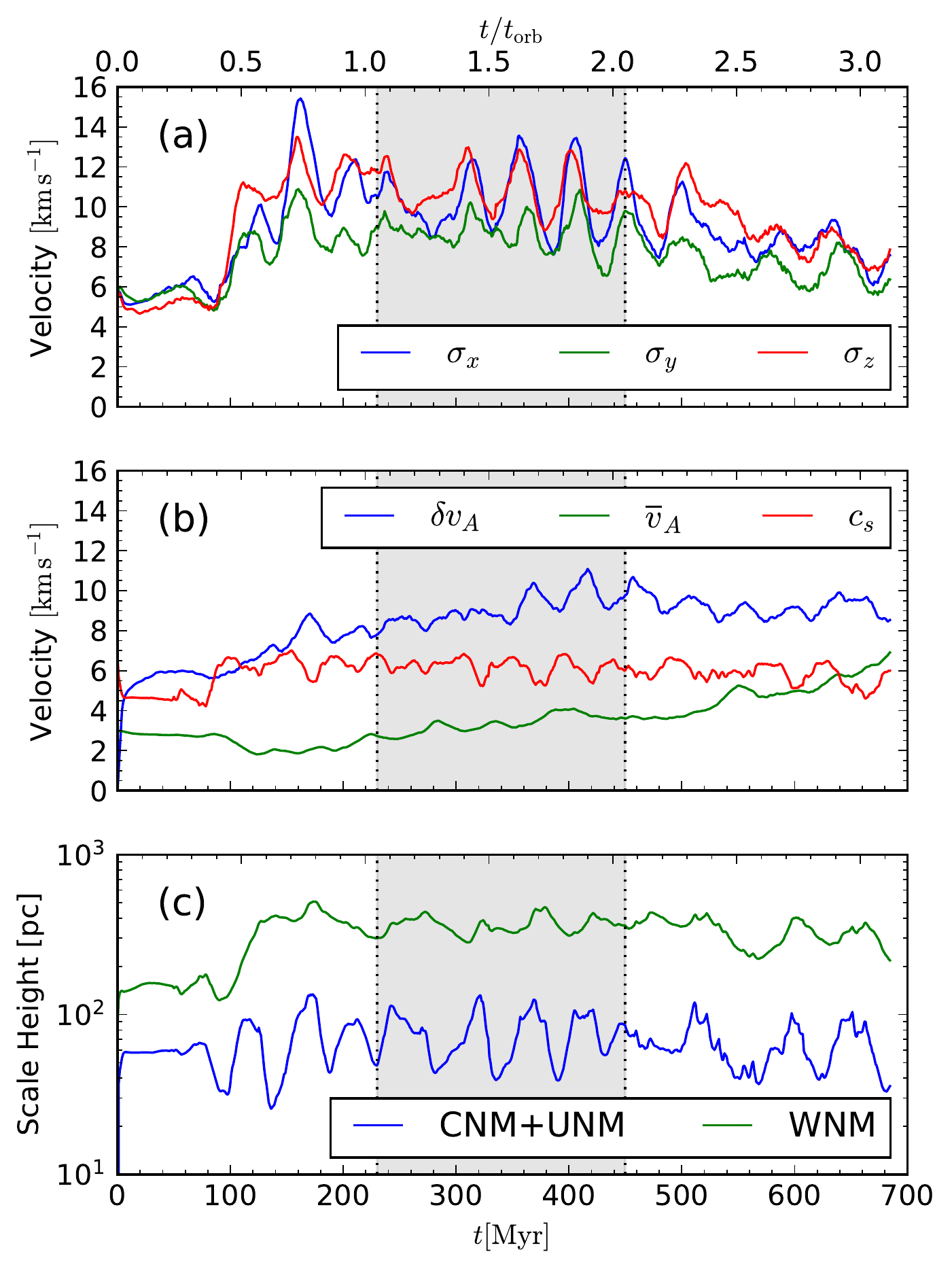}
\caption{Time evolution of (a) turbulent velocity dispersions,
(b) turbulent and mean Alfv\'en velocities and sound speed,
  and (c) scale heights.  All quantities are averaged over
  the warm-cold medium. See text for definitions.
}
\label{fig:tevol2}
\end{figure}

Figures~\ref{fig:tevol2}(a) and (b) plot time evolution
averaged over the warm-cold medium  ($T<2\times10^4\Kel$) of 
various mass-weighted velocities defined by
\begin{equation}\label{eq:veld}
\sigma_i\equiv
\rbrackets{\frac{\sum_{wc} \rho \delta v_i^2 \Delta V}{\sum_{wc} \rho \Delta V}}^{1/2},
c_s \equiv \rbrackets{\frac{\sum_{wc} P \Delta V}{\sum_{wc} \rho \Delta V}}^{1/2},
\vaturb \equiv \rbrackets{
\frac{\sum_{wc} \delta\Bvec\cdot\delta\Bvec \Delta V}
{4\pi\sum_{wc} \rho \Delta V}}^{1/2},
\vamean \equiv \rbrackets{
\frac{\sum_{wc} \overline{\Bvec}\cdot\overline{\Bvec}\Delta V}
{4\pi\sum_{wc} \rho \Delta V}}^{1/2}.
\end{equation}
Here, $\sigma_i$ is the turbulent velocity dispersion in each direction,
$c_s$ is the sound speed,
$\vaturb$ is the turbulent Alfv\'en speed,
and $\vamean$ is the mean Alfv\'en speed.
The perturbed velocity (removing the background shear flow) is defined by 
$\delta \vel\equiv \vel+q\Omega x\yhat$,
the mean magnetic field $\overline{\Bvec}$ is
calculated based on a horizontal average at each $z$, and the
turbulent magnetic field is defined by 
$\delta{\Bvec}\equiv\Bvec-\overline{\Bvec}$.

Turbulent velocity dispersions of the warm-cold medium are about $11\kms$
($\sigma_z\sim5\kms$ for the CNM and $12\kms$ for the WNM),
a factor of 1.5 to 2 higher than those in our previous simulations 
\citep[e.g.,][]{2013ApJ...776....1K,2015ApJ...815...67K}, 
where we found $\sigma\sim 5-7\kms$.
A number of effects could contribute to this increase in $\sigma$,
including the correlation of supernovae in superbubbles and
the cooling of hot gas that has expanded to large $|z|$.
Note that the turbulent velocity dispersions fluctuate with a similar
period to, but much lower amplitude than, $\Sigma_{\rm SFR}$.   
Turbulent magnetic fields are generated very quickly
by a turbulent dynamo \citep[e.g.,][]{2015ApJ...815...67K}, 
so that the turbulent Alfv\'en velocity is expected
to quickly saturate, and indeed this rapid growth and saturation is seen.  
The turbulent magnetic field strength depends on both the turbulent
kinetic energy and mean magnetic energy. 
As the mean field keeps growing for $t<400\Myr$, the saturation level of the
turbulent magnetic energy also gradually increases in
time. For $t>400\Myr$, the turbulent magnetic energy stays constant
as the mean magnetic field strength
reaches a certain level.\footnote{
\citet{2011PhRvL.107k4504F,2012ApJ...754...99S} have shown that
the saturation level of the turbulent magnetic energy
is much lower than the turbulent kinetic energy (less than a percent)
for a dynamo driven by compressible turbulence, when the initial
mean fields were almost negligible.
Here, we instead find that turbulent magnetic energy saturates
at a level similar to the turbulent kinetic energy.  This may in part
owe to our the larger initial mean magnetic fields, and in part to
the presence of background sheared rotation.  Understanding
galactic dynamo behavior for realistic ISM turbulence combined with 
realistic shear, rotation, and vertical stratification
is a very interesting question.  
However, this will require carefully controlled numerical
studies and analyses, which we do not attempt here.}
At this stage, the ratio between turbulent kinetic and magnetic energies
is about 7:3 as in \citet{2015ApJ...815...67K}.
Growth of the mean magnetic field
is slow, with the time scale similar to the orbit time,
so that mean Alfv\'en velocity increases throughout the simulation.

In Figure~\ref{fig:tevol2}(c), we show the scale heights of the
CNM+UNM and WNM
($C=c+u$, and $w$, respectively) defined by 
\begin{equation}\label{eq:scaleH}
H_C\equiv\rbrackets{\frac{\sum_C \rho z^2 \Delta V}
{\sum_C \rho \Delta V}}^{1/2}.
\end{equation}
Note that the vertical box size is not large enough to define
a meaningful scale height for the ionized and hot components,
so we omit these in this Figure.
The mean values of the scale heights averaged over $t_{11}$-$t_9$
are $H_{c+u}=76\pc$ and $H_w=363\pc$, giving
a scale height of the warm-cold medium of $H_{wc}=317\pc$.

\section{Numerical Convergence}\label{sec:conv}

\begin{figure*}
\plotone{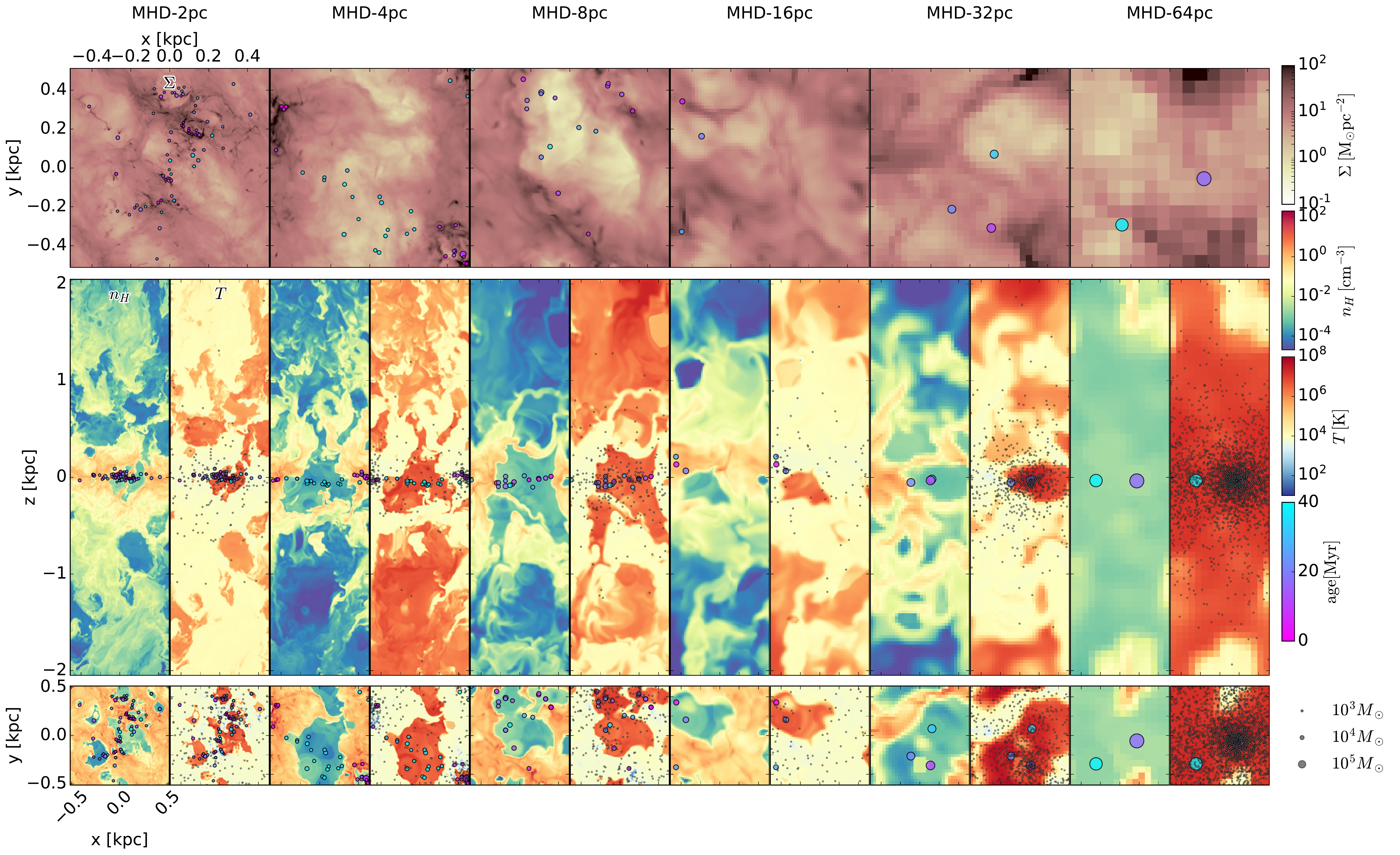}
\caption{
  Comparison of structure at varying simulation resolution, for the same
  model parameters, 
when the gas surface density is $\Sigma=10\Surf$,
after a quasi-steady state has been reached.
  Each row shows the same properties as in Figure~\ref{fig:MHD_4pc}.
  Numerical resolution is poorer by a factor of two in each set
  from left to right, as labeled 
  with $\Delta x= 2\pc$, $4\pc$, $8\pc$, $16\pc$, $32\pc$, and
  $64\pc$.
}
\label{fig:conv}
\end{figure*}

In this section, we compare statistical properties of the
same solar neighborhood model run at different numerical grid
resolutions, varying from $\Delta x=2\pc$ 
to $64\pc$. Hereafter, each model is referred as {\tt MHD-$\Delta x$}.
 Since all simulations secularly evolve due to the decline in gas mass, 
 for fair intercomparison 
 we obtain statistical properties over the
range of times ($t_{11}$,$t_{9}$) in each simulation when the gas
   is in the same gas surface density range
of $11>\Sigma/\Surf>9$.
Before diving into the detailed quantitative comparison, 
in Figure~\ref{fig:conv} we display $\Sigma$, $n_H$, and $T$
snapshots at the respective times when $\Sigma=10\Surf$
(after saturation) for all models.
As expected, overdense structures and small-scale thermal variations
are more smeared out as the resolution gets poorer. A discernible 
qualitative change only emerges with Model \model{64},
where we observe thermal runaway with highly correlated SNe
that is not seen in higher-resolution models. But, as we shall show,
Model \model{32} also diverges from higher resolution simulations
in many aspects, including both SFRs and ISM properties.
\footnote{
It is important to note that, with
insufficient resolution, the ISM properties fail to converge 
in unpredictable ways. When we implemented SN feedback without allowing
for overlap of SNRs at the same position and time, the result diverged
in the opposite sense from what we describe here: Models \model{32} and \model{64} had 
no hot gas within the gas scale height. SN overlapping occurs frequently in
low-resolution simulations, resulting in unrealistically highly-correlated SNe
and unphysical consequences for the ISM state. 
}
In models with higher resolution ($\Delta x \le 8\pc$),
the WNM is ubiquitous at all $z$, and the hot-ionized medium fills substantial
volume even near the midplane. 

In the forthcoming subsections, we present SFRs and
properties of the ISM measured during $(t_{11}, t_9)$ using
box-and-whisker plots to display key  statistics as simply
as possible.
In these plots, the rectangular box extends from the 25th to 75th percentile
of temporal fluctuations, and the median and mean are shown as
a horizontal bar within the box and a square symbol, respectively.
The whiskers extend to 5th and 95th percentiles with caps,
and outliers are shown as circles.

\subsection{Star Formation Rates}

\begin{figure}
\plotone{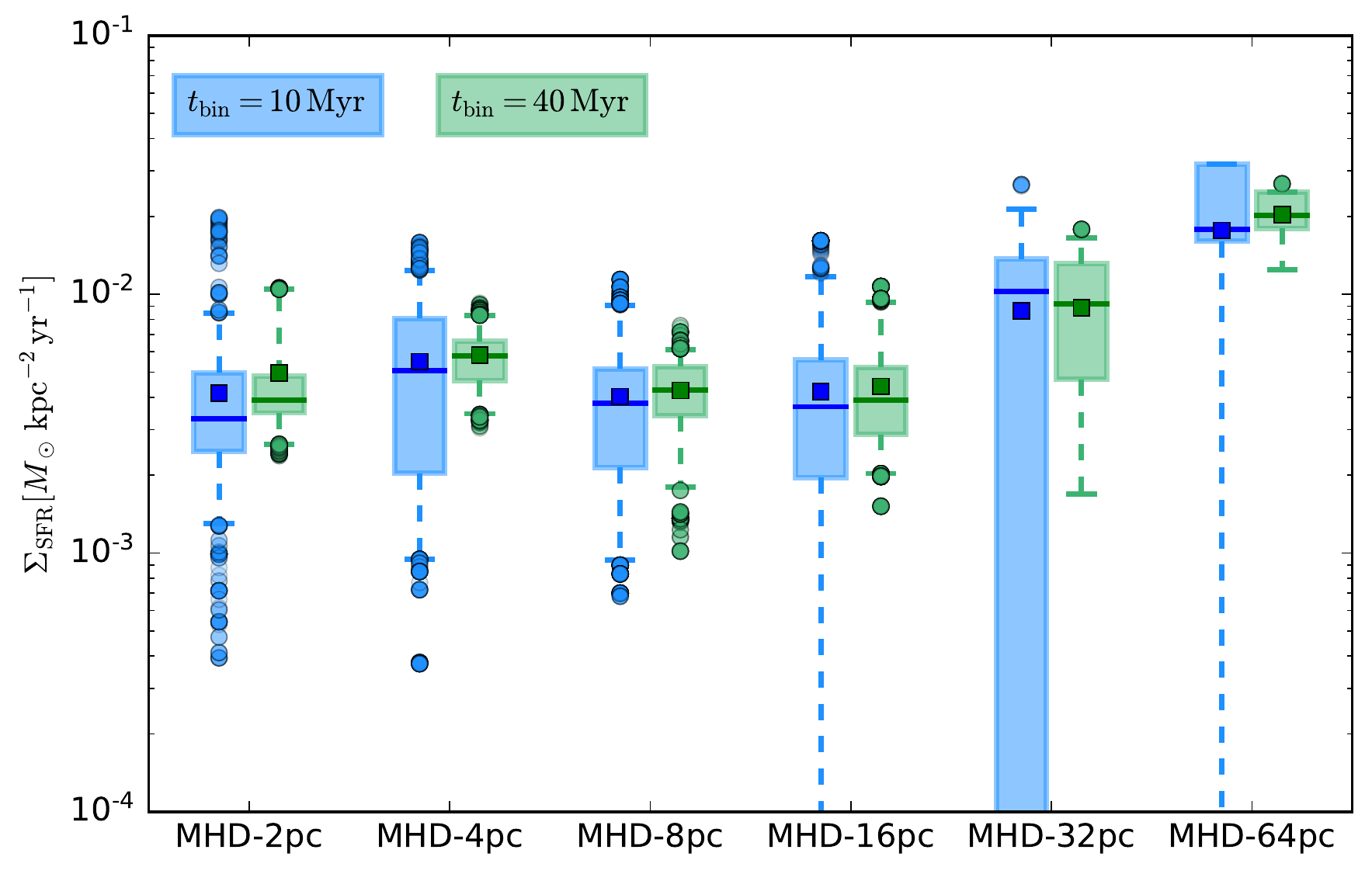}
\caption{
Dependence of SFR on numerical resolution.
    Models with resolution $\Delta x \le 16\pc$ are converged.
  Box-and-whisker plots of $\Sigma_{\rm SFR}$ statistics for simulations with
  varying resolution.   
Boxes enclose 25th to 75th percentiles with medians and means 
shown as horizontal bars and squares, respectively.
Whiskers extend to 5th and 95th percentiles (capped), with outliers shown
as circles.  For each model, we show side-by-side box-and-whisker plots
for $\Sigma_{\rm SFR}$ computed with 10Myr and 40Myr time bins.
}
\label{fig:comp_sfr}
\end{figure}

In Figure~\ref{fig:comp_sfr}, 
we first compare $\Sigma_{\rm SFR}$ using box-and-whisker plots
for time in the interval $(t_{11},t_9)$.
Between the 25th and 75th percentile (box)
with respect to the median value (horizontal bar in the box),
all results except Models \model{32} and \model{64} are in 
good agreement. While $\Sigma_{\rm SFR}$ in Model \model{32} overlaps
with the converged results, Model \model{64} has $\Sigma_{\rm SFR}$ several 
times larger than in the other models.
 As resolution gets poorer, the mass of star clusters gets larger,
increasing stochasticity of the FUV luminosity. This leads
to large temporal fluctuations in $\Sigma_{\rm SFR}(t_{\rm bin}=10\Myr)$,
extending the 5th and 25th percentiles to zero for $\Delta x \ge 16\pc$.

\subsection{Supernova treatment}

As described in Section \ref{sec:explosion}, we have
three different SN feedback treatments; the treatment that is
applied for any given SN event depends on the parameter 
$\mathcal{R}_M$ that represents the ratio of the mass that can be
resolved in the feedback region around the SN compared to the
expected SN remnant mass when it becomes radiative.  
At low numerical resolution, we expect that the Sedov-Taylor stage 
and shell formation in 
most SN explosions cannot be resolved (see KO15a), so they would
be realized in the simulation with momentum feedback (type 
{\tt MC}).  At higher resolution, we instead expect
the non-radiative early stage of evolution
to be resolved in most cases, so that the {\tt ST} type
may be applied.  In rare cases when the density is very low (mainly
for SN events in superbubbles and at high-$|z|$ from runaways), the {\tt EJ}
type would be applied.

\begin{figure}
\plotone{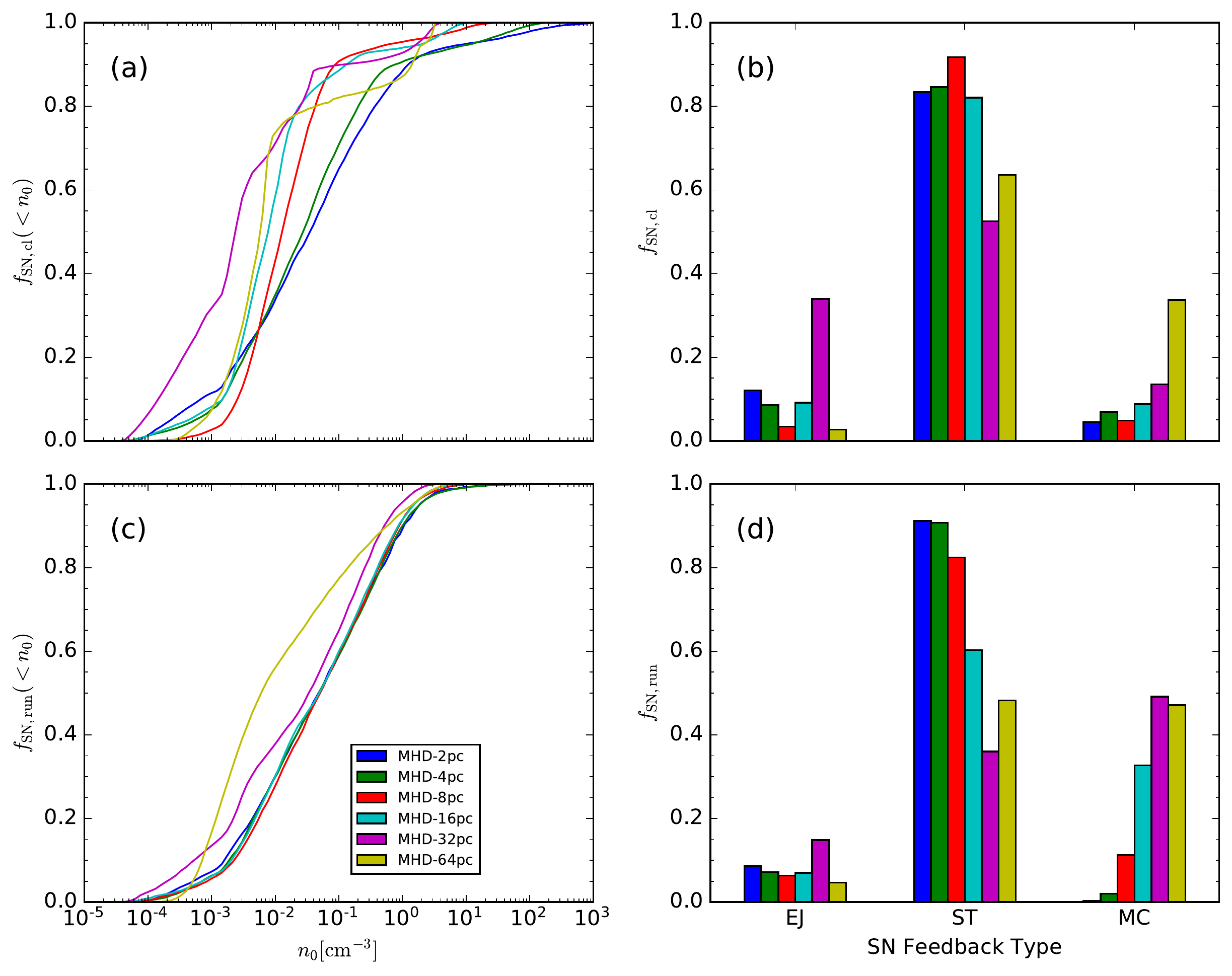}
\caption{SN environment and feedback type distribution at varying
  resolution.  
  (a) and (c): cumulative fraction of SN events that occur in regions
  where the surrounding density is smaller than $n_0$, 
for SNe in clusters and runaways, respectively.
(b) and (d): 
fraction of SN feedback type 
for SNe in clusters  and runaways, respectively.
From Section \ref{sec:explosion}, the 
type of feedback assigned depends on the ratio $\mathcal{R}_M$ between the
enclosed mass within the feedback region and expected remnant mass when
it becomes radiative.  The types correspond to different stages of
evolution for the initial resolved remnant, and different methods of
injecting energy and/or momentum.  From 
earliest to latest evolutionary stage, these
are: initial ejecta ({\tt EJ}, $\mathcal{R}_M<0.027$), Sedov-Taylor ({\tt ST},
$0.027<\mathcal{R}_M<1$), 
and momentum-conserving ({\tt MC}, $1<\mathcal{R}_M$).
Note that the ratio of runaway SN events to cluster SN events 
is about $1/3$ ($f_{\rm bin}=2/3$).
}
\label{fig:comp_SN}
\end{figure}

In Figure~\ref{fig:comp_SN}, panels (a) and (c) plot, for cluster and runaway
SNe respectively, the cumulative fraction
of SN events that occur with  surrounding number density smaller than $n_0$.
Panels (b) and (d) plot the 
fractions of SNe that are realized with each type of feedback,
separately considering SNe from cluster and runaway star particles.
Models up to $\Delta x =8\pc$ have most of their
SNe realized with the {\tt ST} prescription (panels (b) and (d)),
meaning that the
resolution is sufficient to follow evolution prior to the radiative stage.
The fraction of pure momentum feedback ({\tt MC}-type) increases
as resolution gets poorer.
In lower resolution models, SNe from cluster particles
are more correlated and more effective at forming supperbubbles  
(see Figure~\ref{fig:conv}). 
Thus, $n_0$ near SN sites is systemically lower for SNe
in clusters when resolution is poor (panel (a) of Figure~\ref{fig:comp_SN}).
However, $n_0$ near sites of runaway SNe is rather insensitive
to the resolution (panel (c)).

Not all SN events are realized with the {\tt ST} or {\tt EJ}
prescriptions even
for the highest resolution simulation.
However, if there is a cluster in a very high density region,
one or two early SNe realized by momentum injection ({\tt MC}) 
open a cavity such that subsequent SN events from the same cluster can
be realized with the {\tt ST} treatment. 
Implementation of early feedback such as stellar winds, radiation pressure, and
photoionization could also help to open cavities around clusters and to achieve
better convergence even for individual SN events at
lower resolutions (e.g., runaway SNe in Model \model{16}).\footnote{
Based on simulations including stellar winds
 and/or ionizing radiation \citep[e.g.,][]{2017MNRAS.466.1903G,2017MNRAS.466.3293P},
some have 
argued that these feedback processes, rather than just SNe, are necessary to
obtain accurate SFRs.  However,  spatial and
temporal correlation of SN clustering are at least as important to the
outcome as early feedback, so it is necessary to control for these effects
(and confirm that they match observations) before reaching any conclusions.
Our high-resolution
simulations, which have only SN feedback, achieve a realistic SFR.
}

\subsection{Turbulence and Phase Balance}

\begin{figure}
\plotone{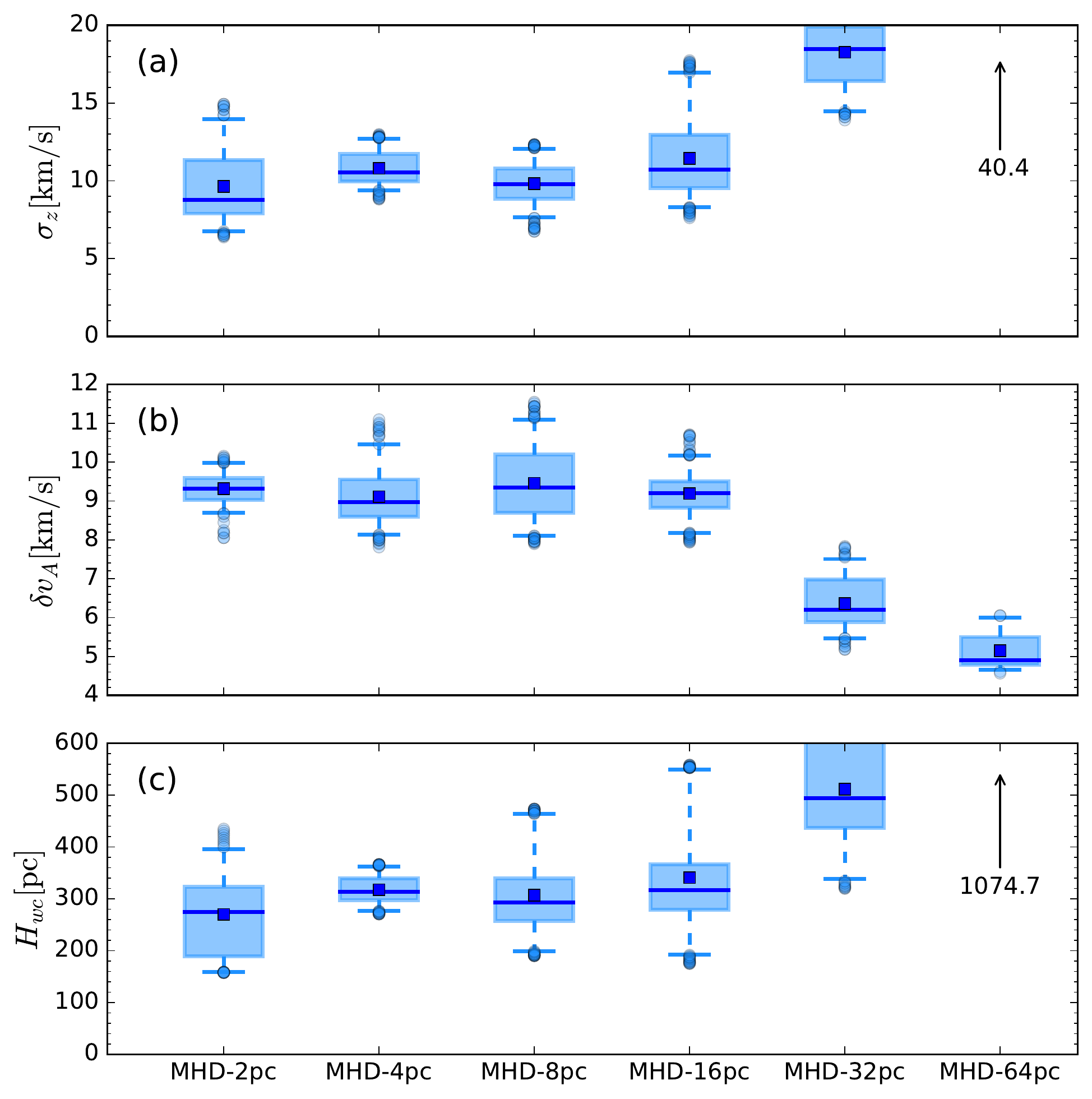}
\caption{
Dependence of ISM kinetic and magnetic properties on numerical resolution.
    Models with resolution $\Delta x \le 16\pc$ are converged.
Box-and-whisker plots of 
(a) vertical velocity dispersion,
(b) turbulent Alfv\'en velocity,
and (c) scale height of the warm-cold medium.
Boxes enclose from the 25th to 75th percentile with medians and means 
shown as horizontal bars and squares, respectively.
Whiskers extend to 5th and 95th percentiles with outliers shown as circles.
Note that in (a) and (c), Model \model{64} has extreme deviation
  from the converged
results, so we simply indicate mean values with arrows.
}
\label{fig:comp_h}
\end{figure}

In this subsection, we investigate the convergence of ISM properties,
based on statistics for time in the range $(t_{11}, t_9)$.
Figure~\ref{fig:comp_h} plots
(a) vertical velocity dispersions,
(b) turbulent Alfv\'en velocities,
and (c) scale heights of the warm-cold medium
using box-and-whisker plots.
Models with $\Delta x \le 16\pc$ are converged in these quantities,
although the temporal fluctuations are much larger in Model \model{16}
that in higher-resolution simulations.
However, Models \model{32} and \model{64} show substantially
larger velocity dispersions (panel (a)) and hence scale heights (panel (c)),
 in comparison to the resolved results.
The spatial and temporal correlations of SNe are exaggerated in these models,
with the result that overlapping
multiple SNe drive large vertical (ordered) motions.
While velocity dispersions and scale heights are large
in low resolution models, this is not due to higher amplitude
small-scale turbulence but only to very large-scale correlated
superbubble expansion.
In fact, small-scale turbulence is \textit{weaker} than in the
high-resolution models, as can be seen in the in weaker
turbulent magnetic fields (panel (b)) from a less-efficient
turbulent dynamo \citep[cf,][]{2015ApJ...815...67K}.

\begin{figure}
\plotone{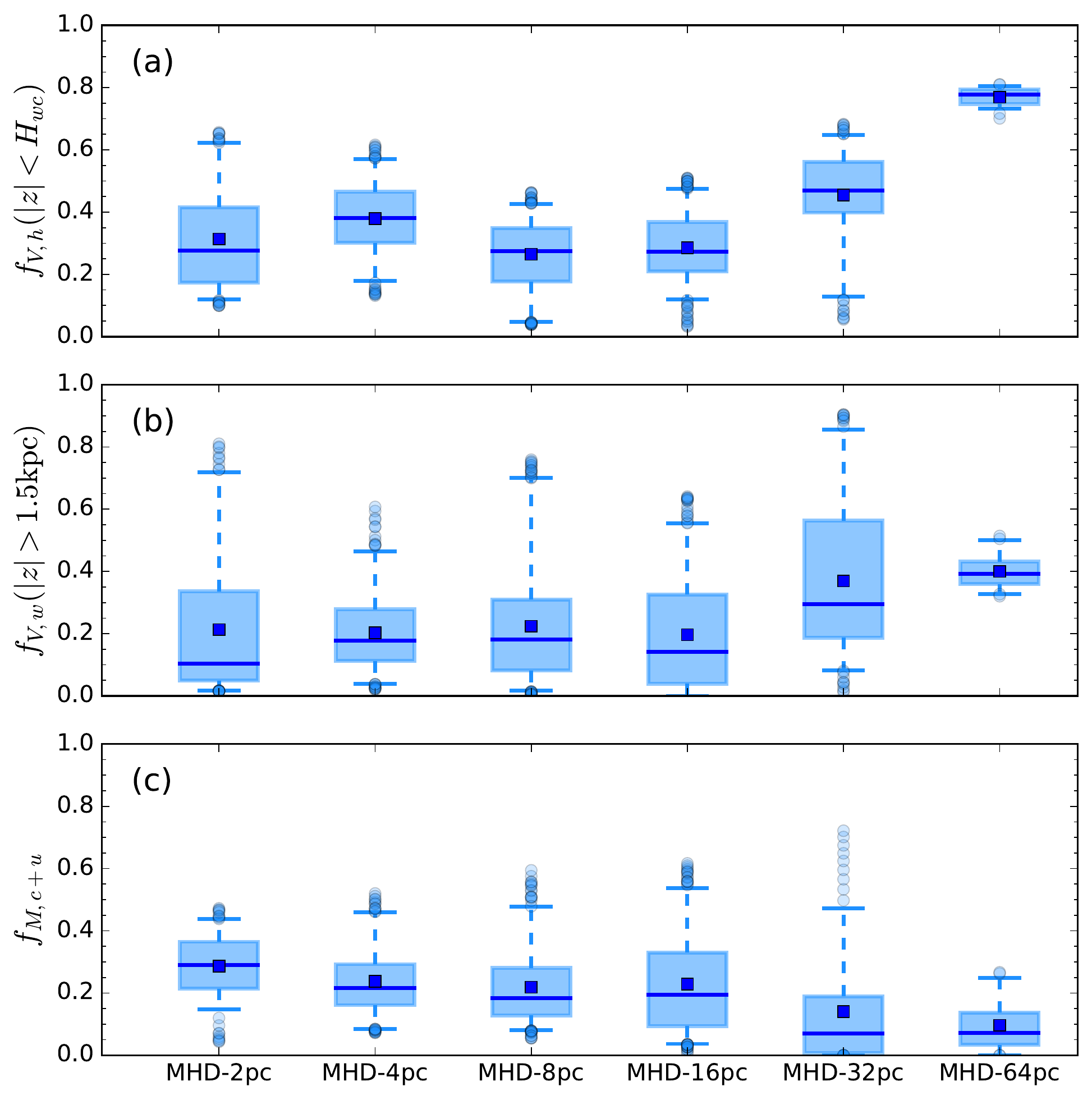}
\caption{
Dependence of ISM phase properties on numerical resolution.
    Models with resolution $\Delta x \le 16\pc$ are converged.
Box-and-whisker plots of
(a) volume fraction of the hot gas within the scale height of the
warm-cold medium,
(b) volume fraction of the WNM outside of $|z|>1.5\kpc$,
and (c) mass fraction of the CNM+UNM.
Boxes enclose from the 25th to 75th percentile with medians and means 
shown as horizontal bars and squares, respectively.
Whiskers extend to 5th and 95th percentiles with outliers shown as circles.
}
\label{fig:comp_phase}
\end{figure}

To better understand resolution dependence of phase balance,
we compare box-and-whisker plot for statistics in the time range
$(t_{11}, t_9)$.  Figure~\ref{fig:comp_phase} shows 
(a) the volume fraction of the hot gas within the scale height of the
warm-cold medium,
(b) the volume fraction of the WNM outside of $|z|>1.5\kpc$,
and (c) the mass fraction of the CNM+UNM.
These properties are essentially converged for Models
\model{8} to \model{2}.  
However, as was previously evident from Figure~\ref{fig:conv},
exaggeration of spatial and temporal correlations of SNe in 
Models \model{32} and \model{64} 
quickly blows everything away from the midplane. Model \model{64}
effectively suffers ``thermal runaway'' \citep[e.g.,][]{2015ApJ...814....4L},
so the hot gas occupies most of the volume
(see Figure~\ref{fig:conv}).

The total CNM+UNM mass fractions are more or less similar 
($f_{M,c+u}\sim 20-30\%$)
up to Model \model{16} in terms of the mean and median, but
we can see a decreasing median and increasing scatter as the resolution
gets poorer.  We note (not shown) 
that the mass fraction of the CNM alone keeps increasing
at higher resolution, presumably due to
a reduction in numerical diffusion that otherwise broadens
the phase transition layer between the CNM and WNM (artificially
turning CNM into UNM).  
We also note that in low resolution models, sink/star particles
may form at density that is insufficient to guarantee true collapse
of the CNM (see Figure~\ref{fig:threshold}).

\subsection{Mass Loss Rates}

\begin{figure}
\plotone{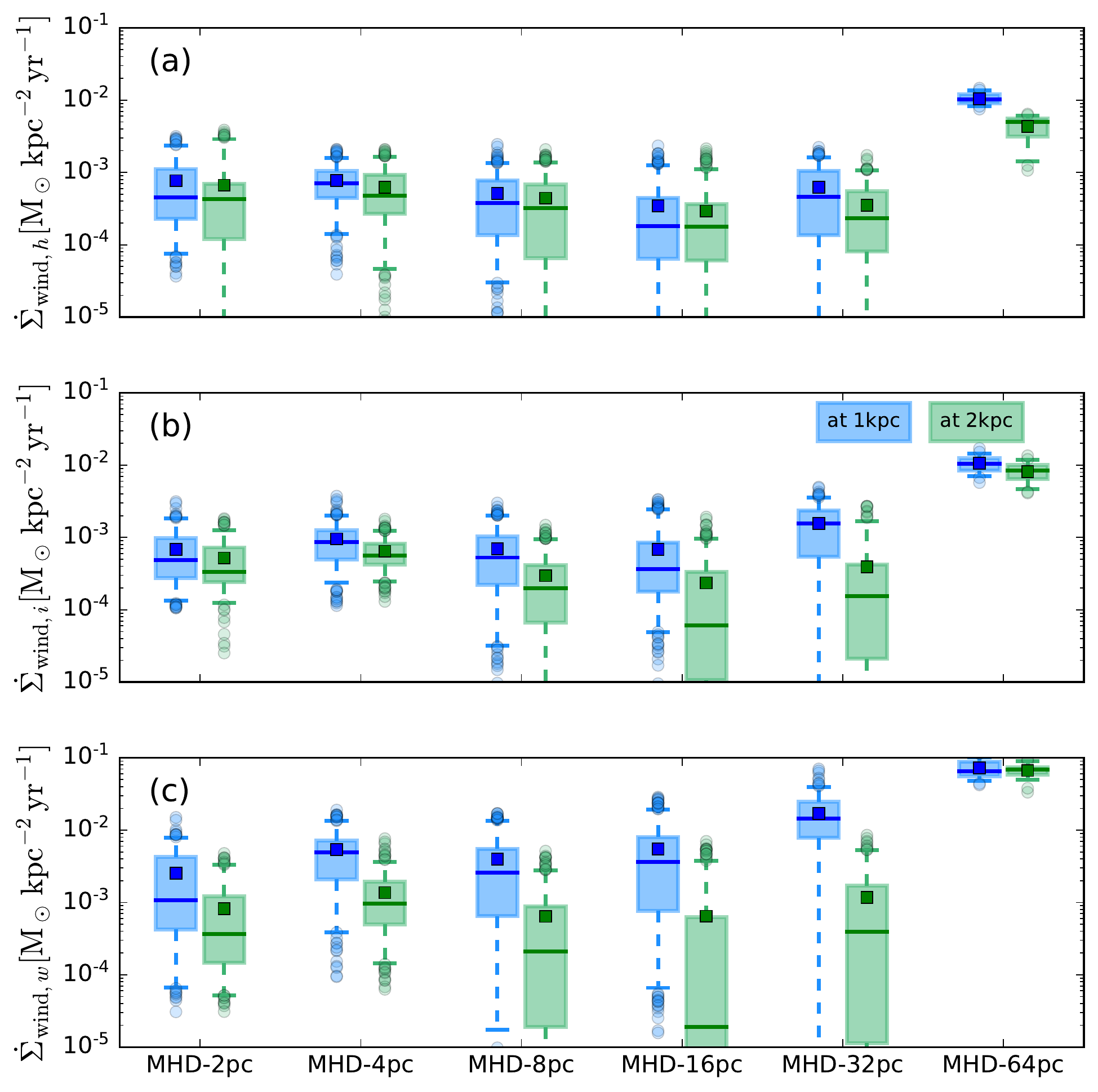}
\caption{
Dependence of wind mass-loss rates on numerical resolution.
    Models with resolution $\Delta x \le 8\pc$ are converged.
Box-and-whisker plots of
areal mass loss rate statistics in units of $\sfrunit$ for
(a) hot gas $\dot{\Sigma}_{{\rm wind},h}$,
(b) ionized gas $\dot{\Sigma}_{{\rm wind},i}$,
and (c) warm gas $\dot{\Sigma}_{{\rm wind},w}$ . 
Boxes enclose from the 25th to 75th percentile with medians and means 
shown as horizontal bars and squares, respectively.
Whiskers extend to 5th and 95th percentiles with outliers shown as circles.
For each model, we plot two mass loss rates
measured at $|z|=1\kpc$ and $|z|=2\kpc$. 
}
\label{fig:comp_massloss}
\end{figure}

We measure the areal mass loss rates of each phase at different heights
using outgoing fluxes 
\begin{equation}
\dot{\Sigma}_{{\rm wind},C}(|z|) \equiv 
\frac{\sum_{C} [(\rho v_{z,+})_{z_+}-(\rho v_{z,-})_{z_-}]\Delta V}{L_xL_y},
\end{equation}
where the outgoing velocity $v_{z,\pm}$ denotes positive or negative velocity
for positive or negative vertical coordinate $z_{\pm}$, respectively.
Figure~\ref{fig:comp_massloss} plots the areal mass loss rates
measured at $|z|=1\kpc$ and $2\kpc$ 
in blue and green boxes and whiskers, respectively. 
Overall convergence is again seen for Models \model{2} to \model{8}.
Models \model{32} and \model{64} are inconsistent with the mass fluxes
in the resolved models.  

As \citet{2016MNRAS.459.2311M} pointed out, the full evolution of
galactic winds cannot be followed in local Cartesian box simulations.  However,
local simulations are very useful for investigating the launching
condition of winds, especially because they afford very high
resolution of multiphase gas and its heating and acceleration in SN
remnants and superbubbles.  By analyzing the outflowing gas properties
for each phase, we are able to understand which material can escape and which
cannot.  In a companion paper, we shall conduct full analysis using
the Bernoulli parameter, also investigating the effect of runaways.
Here, we simply consider convergence based on the mass loss rates
measured at different $|z|$.  If some of the material passing through
$|z|=1\kpc$ eventually falls back before it reaches the vertical
boundaries at $|z|=2\kpc$, the mass loss rate is higher at lower
$|z|$. This behavior is characteristic of a fountain flow, and
is clearly evident for the WNM of the resolved models in
Figure~\ref{fig:comp_massloss}c.  However,
the hot (and ionized) gas have high enough velocity to really
escape galactic disks as winds, and mass fluxes are nearly the same at
$|z|=1\kpc$ and $|z|=2\kpc$.

From Figure~\ref{fig:comp_sfr},
the mean SFR surface density for the parameters of the present simulation is
$\Sigma_{\rm SFR}\sim 5\times10^{-3}\sfrunit$.  In comparison,  
the mass loading factors
($\beta_C\equiv \dot{\Sigma}_{{\rm wind},C}/\Sigma_{\rm SFR}$) 
of the hot and ionized gas are around
$\beta_h\sim \beta_i\sim0.1$.
This is consistent with the expected range of hot gas mass
loading factor for superbubbles that break out of the disk
after shell formation, the typical situation \citep{2017ApJ...834...25K}.
More heavily loaded winds are expected only if (1) 
the galactic gravitational potential well
is shallow enough 
for the moderate-velocity WNM
that is in cooled shells around superbubbles 
to escape, as may occur in dwarf galaxies,
or (2) a very high local SN rate creates a superbubble
that breaks out of the ISM before cooling, as may occur in some
nuclear starbursts \citep{2017ApJ...834...25K}.

\section{Summary}\label{sec:summary}

In this paper, we have presented the TIGRESS algorithms, which
we have designed and implemented in the {\tt Athena} MHD code in order
to conduct self-consistent simulations of the three-phase star-forming
ISM in a wide range of galactic disk environments.  We also
demonstrate application of these algorithms to a fiducial model
representing conditions in the Solar neighborhood, and conduct a
resolution study to assess requirements for convergence. 
In TIGRESS, the MHD equations are solved in a local frame that
includes sheared galactic rotation, and augmented with optically thin
cooling and heating, self- and external gravity, and modules to follow
star formation and the feedback it engenders.  The last two elements 
are implemented using sink/star particles
to form star clusters within gravitationally collapsing gas,
and to follow the FUV radiation and SN events that these clusters
produce based on an adopted population synthesis model. 
Our implementation allows for runaway OB stars 
that may result from a SN in a massive binary system.
 
A key aspect of TIGRESS is the detailed treatment of
sink/star particle formation and evolution. In Section \ref{sec:sp}
we delineate how we create, age, accrete onto, and move 
sink/star particles, and present tests for our implementation. 
Another key aspect is our approach to SN feedback.  Our implementation
treats SN explosions in three different ways (feedback types {\tt EJ},
{\tt ST}, and {\tt MC}), depending on the mass of
gas that is resolved in the immediate vicinity of the SN
(see \S\ref{sec:explosion}). This allows us to properly model creation of
the hot ISM when the Sedov-Taylor stage of evolution can be resolved
(which is usually possible, one of the main advantages of local
simulations).  If the Sedov-Taylor stage cannot be resolved, we
inject momentum to the warm-cold ISM surrounding a SN based on
the results of KO15a for the post-radiative stage.
Our SN feedback prescription proves to be quite robust, which has enabled
  us to run for extended evolution times ($700\Myr$, compared to only 
  $\sim 100\Myr$ in some other recent simulations with self-gravity
  and SN feedback).

By carefully treating star formation and feedback, TIGRESS
simulations yield realistic, fully self-consistent three-phase ISM models with
self-regulated star formation and galactic winds.
For the fiducial Solar neighborhood model with
gas surface density of $\sim 10\Surf$, we show that a quasi-steady 
saturated state is reached in our higher resolution simulations.
In particular, we show that the SFR, wind mass-loss rate, disk scale height,
turbulent and Alfv\'enic velocity dispersions, and volume fractions
of warm and hot phases are 
converged provided that the numerical resolution 
is at least
$\Delta x =8\pc$, such that the  
Sedov-Taylor stage is resolved for most SN events.

In numerically converged models, the SFR surface density is self-regulated to
$\Sigma_{\rm SFR}\sim 5\times10^{-3}\sfrunit$ 
with large cyclic fluctuations.
For gas at $T<2\times10^{4}\Kel$, the mass-weighted
velocity dispersions and turbulent Alfv\'en
velocity are each $\sim 10\kms$.
The scale heights of the CNM+UNM and WNM are
$\sim 80\pc$ and $\sim 360\pc$, respectively.  Hot gas fills $\sim 30-40\%$
of the volume near the midplane, while warm gas is mixed within the
mostly-hot medium at $|z|> 1 \kpc$.
High velocity hot (and ionized) gas can escape with a ratio of
hot gas outflow rate to SFR (the wind ``mass loading factor'') 
of $\sim0.1$.  The WNM cannot achieve high enough velocity to
escape the Milky Way's gravitational potential, and instead
creates a fountain flow reaching to a few $\kpc$ from the midplane.

At low resolution ($\Delta x \ge 16\pc$), the detailed
ISM properties as well as the mean SFR cannot be properly captured.
Of course, the resolution requirements we report here 
are particular to the Solar neighborhood model. One may expect more
stringent resolution requirements for galactic disk environments where the
dynamical space and time scales are shorter (e.g., Jeans length varies
inversely with the square root of density, and the SN shell formation radius
has similar dependence). Also, different prescriptions from those we describe 
for feedback and star formation (as well as other physics) may give different 
convergence trends that are difficult to predict. 
The quantitative resolution requirements
for convergence must be evaluated for any specific parameter set
and adopted physics prescriptions.

The TIGRESS algorithms are currently being applied in a suite of
numerical simulations systematically exploring SFRs,
ISM properties, and galactic winds in diverse galactic environments.
Results of these models and analyses will be presented in companion papers.

\acknowledgements

We are grateful to the referee for a detailed report, which helped us to
improve the manuscript.  This work was supported by grant AST-1312006 from the
National Science Foundation and grant NNX14AB49G from NASA.  Simulations
were performed on the computational resources supported by the PICSciE
TIGRESS High Performance Computing Center at Princeton University, and
by the NASA High-End Computing (HEC) Program through the NASA Advanced
Supercomputing (NAS) Division at Ames Research Center.

\software{Athena \citep{2008ApJS..178..137S}, STARBURST99 \citep{1999ApJS..123....3L}, yt \citep{2011ApJS..192....9T}, astropy \citep{2013A&A...558A..33A}, matplotlib \citep{Hunter:2007}, numpy \citep{vanderWalt2011}, IPython \citep{Perez2007}, pandas \citep{mckinney-proc-scipy-2010}}
\bibliography{ms_arxiv.bbl}

\begin{thebibliography}{}
\expandafter\ifx\csname natexlab\endcsname\relax\def\natexlab#1{#1}\fi
\providecommand{\url}[1]{\href{#1}{#1}}

\bibitem[{{Agertz} \& {Kravtsov}(2016)}]{2016ApJ...824...79A}
{Agertz}, O., \& {Kravtsov}, A.~V. 2016, \apj, 824, 79

\bibitem[{{Agertz} {et~al.}(2013){Agertz}, {Kravtsov}, {Leitner}, \&
  {Gnedin}}]{2013ApJ...770...25A}
{Agertz}, O., {Kravtsov}, A.~V., {Leitner}, S.~N., \& {Gnedin}, N.~Y. 2013,
  \apj, 770, 25

\bibitem[{{Astropy Collaboration} {et~al.}(2013){Astropy Collaboration},
  {Robitaille}, {Tollerud}, {Greenfield}, {Droettboom}, {Bray}, {Aldcroft},
  {Davis}, {Ginsburg}, {Price-Whelan}, {Kerzendorf}, {Conley}, {Crighton},
  {Barbary}, {Muna}, {Ferguson}, {Grollier}, {Parikh}, {Nair}, {Unther},
  {Deil}, {Woillez}, {Conseil}, {Kramer}, {Turner}, {Singer}, {Fox}, {Weaver},
  {Zabalza}, {Edwards}, {Azalee Bostroem}, {Burke}, {Casey}, {Crawford},
  {Dencheva}, {Ely}, {Jenness}, {Labrie}, {Lim}, {Pierfederici}, {Pontzen},
  {Ptak}, {Refsdal}, {Servillat}, \& {Streicher}}]{2013A&A...558A..33A}
{Astropy Collaboration}, {Robitaille}, T.~P., {Tollerud}, E.~J., {et~al.} 2013,
  \aap, 558, A33

\bibitem[{{Bai} \& {Stone}(2010)}]{2010ApJS..190..297B}
{Bai}, X.-N., \& {Stone}, J.~M. 2010, \apjs, 190, 297

\bibitem[{{Behroozi} {et~al.}(2013){Behroozi}, {Wechsler}, \&
  {Conroy}}]{2013ApJ...770...57B}
{Behroozi}, P.~S., {Wechsler}, R.~H., \& {Conroy}, C. 2013, \apj, 770, 57

\bibitem[{{Blaauw}(1961)}]{1961BAN....15..265B}
{Blaauw}, A. 1961, \bain, 15, 265

\bibitem[{{Cioffi} {et~al.}(1988){Cioffi}, {McKee}, \&
  {Bertschinger}}]{1988ApJ...334..252C}
{Cioffi}, D.~F., {McKee}, C.~F., \& {Bertschinger}, E. 1988, \apj, 334, 252

\bibitem[{{Cox} \& {Smith}(1974)}]{1974ApJ...189L.105C}
{Cox}, D.~P., \& {Smith}, B.~W. 1974, \apjl, 189, L105

\bibitem[{{Crain} {et~al.}(2015){Crain}, {Schaye}, {Bower}, {Furlong},
  {Schaller}, {Theuns}, {Dalla Vecchia}, {Frenk}, {McCarthy}, {Helly},
  {Jenkins}, {Rosas-Guevara}, {White}, \& {Trayford}}]{2015MNRAS.450.1937C}
{Crain}, R.~A., {Schaye}, J., {Bower}, R.~G., {et~al.} 2015, \mnras, 450, 1937

\bibitem[{{Dale}(2015)}]{2015NewAR..68....1D}
{Dale}, J.~E. 2015, \nar, 68, 1

\bibitem[{{Dalla Vecchia} \& {Schaye}(2012)}]{2012MNRAS.426..140D}
{Dalla Vecchia}, C., \& {Schaye}, J. 2012, \mnras, 426, 140

\bibitem[{{Dav{\'e}} {et~al.}(2016){Dav{\'e}}, {Thompson}, \&
  {Hopkins}}]{2016MNRAS.462.3265D}
{Dav{\'e}}, R., {Thompson}, R., \& {Hopkins}, P.~F. 2016, \mnras, 462, 3265

\bibitem[{{de Avillez}(2000)}]{2000MNRAS.315..479D}
{de Avillez}, M.~A. 2000, \mnras, 315, 479

\bibitem[{{de Avillez} \& {Breitschwerdt}(2004)}]{2004A&A...425..899D}
{de Avillez}, M.~A., \& {Breitschwerdt}, D. 2004, \aap, 425, 899

\bibitem[{{Draine}(1978)}]{1978ApJS...36..595D}
{Draine}, B.~T. 1978, \apjs, 36, 595

\bibitem[{{Eldridge} {et~al.}(2011){Eldridge}, {Langer}, \&
  {Tout}}]{2011MNRAS.414.3501E}
{Eldridge}, J.~J., {Langer}, N., \& {Tout}, C.~A. 2011, \mnras, 414, 3501

\bibitem[{{Federrath} {et~al.}(2011){Federrath}, {Chabrier}, {Schober},
  {Banerjee}, {Klessen}, \& {Schleicher}}]{2011PhRvL.107k4504F}
{Federrath}, C., {Chabrier}, G., {Schober}, J., {et~al.} 2011, Physical Review
  Letters, 107, 114504

\bibitem[{{Field}(1965)}]{1965ApJ...142..531F}
{Field}, G.~B. 1965, \apj, 142, 531

\bibitem[{{Field} {et~al.}(1969){Field}, {Goldsmith}, \&
  {Habing}}]{1969ApJ...155L.149F}
{Field}, G.~B., {Goldsmith}, D.~W., \& {Habing}, H.~J. 1969, \apjl, 155, L149

\bibitem[{{Foster} \& {Chevalier}(1993)}]{1993ApJ...416..303F}
{Foster}, P.~N., \& {Chevalier}, R.~A. 1993, \apj, 416, 303

\bibitem[{{Frank} {et~al.}(2014){Frank}, {Ray}, {Cabrit}, {Hartigan}, {Arce},
  {Bacciotti}, {Bally}, {Benisty}, {Eisl{\"o}ffel}, {G{\"u}del}, {Lebedev},
  {Nisini}, \& {Raga}}]{2014prpl.conf..451F}
{Frank}, A., {Ray}, T.~P., {Cabrit}, S., {et~al.} 2014, Protostars and Planets
  VI, 451

\bibitem[{{Fujii} \& {Portegies Zwart}(2011)}]{2011Sci...334.1380F}
{Fujii}, M.~S., \& {Portegies Zwart}, S. 2011, Science, 334, 1380

\bibitem[{{Gammie}(2001)}]{2001ApJ...553..174G}
{Gammie}, C.~F. 2001, \apj, 553, 174

\bibitem[{{Gatto} {et~al.}(2015){Gatto}, {Walch}, {Low}, {Naab}, {Girichidis},
  {Glover}, {W{\"u}nsch}, {Klessen}, {Clark}, {Baczynski}, {Peters},
  {Ostriker}, {Ib{\'a}{\~n}ez-Mej{\'{\i}}a}, \& {Haid}}]{2015MNRAS.449.1057G}
{Gatto}, A., {Walch}, S., {Low}, M.-M.~M., {et~al.} 2015, \mnras, 449, 1057

\bibitem[{{Gatto} {et~al.}(2017){Gatto}, {Walch}, {Naab}, {Girichidis},
  {W{\"u}nsch}, {Glover}, {Klessen}, {Clark}, {Peters}, {Derigs}, {Baczynski},
  \& {Puls}}]{2017MNRAS.466.1903G}
{Gatto}, A., {Walch}, S., {Naab}, T., {et~al.} 2017, \mnras, 466, 1903

\bibitem[{{Gent} {et~al.}(2013){Gent}, {Shukurov}, {Fletcher}, {Sarson}, \&
  {Mantere}}]{2013MNRAS.432.1396G}
{Gent}, F.~A., {Shukurov}, A., {Fletcher}, A., {Sarson}, G.~R., \& {Mantere},
  M.~J. 2013, \mnras, 432, 1396

\bibitem[{{Girichidis} {et~al.}(2016){Girichidis}, {Walch}, {Naab}, {Gatto},
  {W{\"u}nsch}, {Glover}, {Klessen}, {Clark}, {Peters}, {Derigs}, \&
  {Baczynski}}]{2016MNRAS.456.3432G}
{Girichidis}, P., {Walch}, S., {Naab}, T., {et~al.} 2016, \mnras, 456, 3432

\bibitem[{{Gong} \& {Ostriker}(2009)}]{2009ApJ...699..230G}
{Gong}, H., \& {Ostriker}, E.~C. 2009, \apj, 699, 230

\bibitem[{{Gong} \& {Ostriker}(2011)}]{2011ApJ...729..120G}
---. 2011, \apj, 729, 120

\bibitem[{{Gong} \& {Ostriker}(2013)}]{2013ApJS..204....8G}
---. 2013, \apjs, 204, 8

\bibitem[{{Gong} \& {Ostriker}(2015)}]{2015ApJ...806...31G}
{Gong}, M., \& {Ostriker}, E.~C. 2015, \apj, 806, 31

\bibitem[{{Gong} {et~al.}(2016){Gong}, {Ostriker}, \&
  {Wolfire}}]{2016arXiv161009023G}
{Gong}, M., {Ostriker}, E.~C., \& {Wolfire}, M.~G. 2016, ArXiv e-prints,
  arXiv:1610.09023

\bibitem[{{Heger} {et~al.}(2003){Heger}, {Fryer}, {Woosley}, {Langer}, \&
  {Hartmann}}]{2003ApJ...591..288H}
{Heger}, A., {Fryer}, C.~L., {Woosley}, S.~E., {Langer}, N., \& {Hartmann},
  D.~H. 2003, \apj, 591, 288

\bibitem[{{Hennebelle} \& {Iffrig}(2014)}]{2014A&A...570A..81H}
{Hennebelle}, P., \& {Iffrig}, O. 2014, \aap, 570, A81

\bibitem[{{Hill} {et~al.}(2012){Hill}, {Joung}, {Mac Low}, {Benjamin},
  {Haffner}, {Klingenberg}, \& {Waagan}}]{2012ApJ...750..104H}
{Hill}, A.~S., {Joung}, M.~R., {Mac Low}, M.-M., {et~al.} 2012, \apj, 750, 104

\bibitem[{{Hockney} \& {Eastwood}(1981)}]{1981csup.book.....H}
{Hockney}, R.~W., \& {Eastwood}, J.~W. 1981, {Computer Simulation Using
  Particles}

\bibitem[{{Hopkins} {et~al.}(2014){Hopkins}, {Kere{\v s}}, {O{\~n}orbe},
  {Faucher-Gigu{\`e}re}, {Quataert}, {Murray}, \&
  {Bullock}}]{2014MNRAS.445..581H}
{Hopkins}, P.~F., {Kere{\v s}}, D., {O{\~n}orbe}, J., {et~al.} 2014, \mnras,
  445, 581

\bibitem[{Hunter(2007)}]{Hunter:2007}
Hunter, J.~D. 2007, Computing In Science \& Engineering, 9, 90

\bibitem[{{Iffrig} \& {Hennebelle}(2015)}]{2015A&A...576A..95I}
{Iffrig}, O., \& {Hennebelle}, P. 2015, \aap, 576, A95

\bibitem[{{Joung} \& {Mac Low}(2006)}]{2006ApJ...653.1266J}
{Joung}, M.~K.~R., \& {Mac Low}, M.-M. 2006, \apj, 653, 1266

\bibitem[{{Katz}(1992)}]{1992ApJ...391..502K}
{Katz}, N. 1992, \apj, 391, 502

\bibitem[{{Keller} {et~al.}(2016){Keller}, {Wadsley}, \&
  {Couchman}}]{2016MNRAS.463.1431K}
{Keller}, B.~W., {Wadsley}, J., \& {Couchman}, H.~M.~P. 2016, \mnras, 463, 1431

\bibitem[{{Kennicutt}(1998)}]{1998ApJ...498..541K}
{Kennicutt}, Jr., R.~C. 1998, \apj, 498, 541

\bibitem[{{Kim} {et~al.}(2008){Kim}, {Kim}, \&
  {Ostriker}}]{2008ApJ...681.1148K}
{Kim}, C.-G., {Kim}, W.-T., \& {Ostriker}, E.~C. 2008, \apj, 681, 1148

\bibitem[{{Kim} {et~al.}(2010){Kim}, {Kim}, \&
  {Ostriker}}]{2010ApJ...720.1454K}
---. 2010, \apj, 720, 1454

\bibitem[{{Kim} {et~al.}(2011){Kim}, {Kim}, \&
  {Ostriker}}]{2011ApJ...743...25K}
---. 2011, \apj, 743, 25

\bibitem[{{Kim} \& {Ostriker}(2015{\natexlab{a}})}]{2015ApJ...802...99K}
{Kim}, C.-G., \& {Ostriker}, E.~C. 2015{\natexlab{a}}, \apj, 802, 99

\bibitem[{{Kim} \& {Ostriker}(2015{\natexlab{b}})}]{2015ApJ...815...67K}
---. 2015{\natexlab{b}}, \apj, 815, 67

\bibitem[{{Kim} {et~al.}(2013){Kim}, {Ostriker}, \&
  {Kim}}]{2013ApJ...776....1K}
{Kim}, C.-G., {Ostriker}, E.~C., \& {Kim}, W.-T. 2013, \apj, 776, 1

\bibitem[{{Kim} {et~al.}(2017){Kim}, {Ostriker}, \&
  {Raileanu}}]{2017ApJ...834...25K}
{Kim}, C.-G., {Ostriker}, E.~C., \& {Raileanu}, R. 2017, \apj, 834, 25

\bibitem[{{Kim} {et~al.}(2002){Kim}, {Ostriker}, \&
  {Stone}}]{2002ApJ...581.1080K}
{Kim}, W.-T., {Ostriker}, E.~C., \& {Stone}, J.~M. 2002, \apj, 581, 1080

\bibitem[{{Kimm} \& {Cen}(2014)}]{2014ApJ...788..121K}
{Kimm}, T., \& {Cen}, R. 2014, \apj, 788, 121

\bibitem[{{Korpi} {et~al.}(1999){Korpi}, {Brandenburg}, {Shukurov}, {Tuominen},
  \& {Nordlund}}]{1999ApJ...514L..99K}
{Korpi}, M.~J., {Brandenburg}, A., {Shukurov}, A., {Tuominen}, I., \&
  {Nordlund}, {\AA}. 1999, \apjl, 514, L99

\bibitem[{{Koyama} \& {Inutsuka}(2002)}]{2002ApJ...564L..97K}
{Koyama}, H., \& {Inutsuka}, S.-i. 2002, \apjl, 564, L97

\bibitem[{{Koyama} \& {Ostriker}(2009)}]{2009ApJ...693.1316K}
{Koyama}, H., \& {Ostriker}, E.~C. 2009, \apj, 693, 1316

\bibitem[{{Kroupa}(2001)}]{2001MNRAS.322..231K}
{Kroupa}, P. 2001, \mnras, 322, 231

\bibitem[{{Krumholz} {et~al.}(2014){Krumholz}, {Bate}, {Arce}, {Dale},
  {Gutermuth}, {Klein}, {Li}, {Nakamura}, \& {Zhang}}]{2014prpl.conf..243K}
{Krumholz}, M.~R., {Bate}, M.~R., {Arce}, H.~G., {et~al.} 2014, Protostars and
  Planets VI, 243

\bibitem[{{Kuijken} \& {Gilmore}(1989)}]{1989MNRAS.239..571K}
{Kuijken}, K., \& {Gilmore}, G. 1989, \mnras, 239, 571

\bibitem[{{Larson}(1969)}]{1969MNRAS.145..271L}
{Larson}, R.~B. 1969, \mnras, 145, 271

\bibitem[{{Leitherer} {et~al.}(1999){Leitherer}, {Schaerer}, {Goldader},
  {Gonz{\'a}lez Delgado}, {Robert}, {Kune}, {de Mello}, {Devost}, \&
  {Heckman}}]{1999ApJS..123....3L}
{Leitherer}, C., {Schaerer}, D., {Goldader}, J.~D., {et~al.} 1999, \apjs, 123,
  3

\bibitem[{{Lemaster} \& {Stone}(2009)}]{2009ApJ...691.1092L}
{Lemaster}, M.~N., \& {Stone}, J.~M. 2009, \apj, 691, 1092

\bibitem[{{Leroy} {et~al.}(2012){Leroy}, {Bigiel}, {de Blok}, {Boissier},
  {Bolatto}, {Brinks}, {Madore}, {Munoz-Mateos}, {Murphy}, {Sandstrom},
  {Schruba}, \& {Walter}}]{2012AJ....144....3L}
{Leroy}, A.~K., {Bigiel}, F., {de Blok}, W.~J.~G., {et~al.} 2012, \aj, 144, 3

\bibitem[{{Li} {et~al.}(2016){Li}, {Bryan}, \&
  {Ostriker}}]{2016arXiv161008971L}
{Li}, M., {Bryan}, G.~L., \& {Ostriker}, J.~P. 2016, ArXiv e-prints,
  arXiv:1610.08971

\bibitem[{{Li} {et~al.}(2015){Li}, {Ostriker}, {Cen}, {Bryan}, \&
  {Naab}}]{2015ApJ...814....4L}
{Li}, M., {Ostriker}, J.~P., {Cen}, R., {Bryan}, G.~L., \& {Naab}, T. 2015,
  \apj, 814, 4

\bibitem[{{Lopez} {et~al.}(2014){Lopez}, {Krumholz}, {Bolatto}, {Prochaska},
  {Ramirez-Ruiz}, \& {Castro}}]{2014ApJ...795..121L}
{Lopez}, L.~A., {Krumholz}, M.~R., {Bolatto}, A.~D., {et~al.} 2014, \apj, 795,
  121

\bibitem[{{Mac Low} \& {Klessen}(2004)}]{2004RvMP...76..125M}
{Mac Low}, M.-M., \& {Klessen}, R.~S. 2004, Reviews of Modern Physics, 76, 125

\bibitem[{{Martizzi} {et~al.}(2015){Martizzi}, {Faucher-Gigu{\`e}re}, \&
  {Quataert}}]{2015MNRAS.450..504M}
{Martizzi}, D., {Faucher-Gigu{\`e}re}, C.-A., \& {Quataert}, E. 2015, \mnras,
  450, 504

\bibitem[{{Martizzi} {et~al.}(2016){Martizzi}, {Fielding},
  {Faucher-Gigu{\`e}re}, \& {Quataert}}]{2016MNRAS.459.2311M}
{Martizzi}, D., {Fielding}, D., {Faucher-Gigu{\`e}re}, C.-A., \& {Quataert}, E.
  2016, \mnras, 459, 2311

\bibitem[{{McKee} \& {Ostriker}(2007)}]{2007ARA&A..45..565M}
{McKee}, C.~F., \& {Ostriker}, E.~C. 2007, \araa, 45, 565

\bibitem[{{McKee} \& {Ostriker}(1977)}]{1977ApJ...218..148M}
{McKee}, C.~F., \& {Ostriker}, J.~P. 1977, \apj, 218, 148

\bibitem[{McKinney(2010)}]{mckinney-proc-scipy-2010}
McKinney, W. 2010, in Proceedings of the 9th Python in Science Conference, ed.
  S.~van~der Walt \& J.~Millman, 51 -- 56

\bibitem[{{Moster} {et~al.}(2013){Moster}, {Naab}, \&
  {White}}]{2013MNRAS.428.3121M}
{Moster}, B.~P., {Naab}, T., \& {White}, S.~D.~M. 2013, \mnras, 428, 3121

\bibitem[{{Oh} {et~al.}(2015){Oh}, {Kroupa}, \&
  {Pflamm-Altenburg}}]{2015ApJ...805...92O}
{Oh}, S., {Kroupa}, P., \& {Pflamm-Altenburg}, J. 2015, \apj, 805, 92

\bibitem[{{Ostriker} {et~al.}(2010){Ostriker}, {McKee}, \&
  {Leroy}}]{2010ApJ...721..975O}
{Ostriker}, E.~C., {McKee}, C.~F., \& {Leroy}, A.~K. 2010, \apj, 721, 975

\bibitem[{{Ostriker} \& {Shetty}(2011)}]{2011ApJ...731...41O}
{Ostriker}, E.~C., \& {Shetty}, R. 2011, \apj, 731, 41

\bibitem[{{Penston}(1969)}]{1969MNRAS.144..425P}
{Penston}, M.~V. 1969, \mnras, 144, 425

\bibitem[{Perez \& Granger(2007)}]{Perez2007}
Perez, F., \& Granger, B.~E. 2007, Computing in Science {\&} Engineering, 9,
  21.

\bibitem[{{Peters} {et~al.}(2017){Peters}, {Naab}, {Walch}, {Glover},
  {Girichidis}, {Pellegrini}, {Klessen}, {W{\"u}nsch}, {Gatto}, \&
  {Baczynski}}]{2017MNRAS.466.3293P}
{Peters}, T., {Naab}, T., {Walch}, S., {et~al.} 2017, \mnras, 466, 3293

\bibitem[{{Portegies Zwart}(2000)}]{2000ApJ...544..437P}
{Portegies Zwart}, S.~F. 2000, \apj, 544, 437

\bibitem[{{Poveda} {et~al.}(1967){Poveda}, {Ruiz}, \&
  {Allen}}]{1967BOTT....4...86P}
{Poveda}, A., {Ruiz}, J., \& {Allen}, C. 1967, Boletin de los Observatorios
  Tonantzintla y Tacubaya, 4, 86

\bibitem[{{Quinn} {et~al.}(2010){Quinn}, {Perrine}, {Richardson}, \&
  {Barnes}}]{2010AJ....139..803Q}
{Quinn}, T., {Perrine}, R.~P., {Richardson}, D.~C., \& {Barnes}, R. 2010, \aj,
  139, 803

\bibitem[{{Raskutti} {et~al.}(2016){Raskutti}, {Ostriker}, \&
  {Skinner}}]{2016ApJ...829..130R}
{Raskutti}, S., {Ostriker}, E.~C., \& {Skinner}, M.~A. 2016, \apj, 829, 130

\bibitem[{{Rosdahl} {et~al.}(2017){Rosdahl}, {Schaye}, {Dubois}, {Kimm}, \&
  {Teyssier}}]{2017MNRAS.466...11R}
{Rosdahl}, J., {Schaye}, J., {Dubois}, Y., {Kimm}, T., \& {Teyssier}, R. 2017,
  \mnras, 466, 11

\bibitem[{{Sanders} {et~al.}(1998){Sanders}, {Morano}, \&
  {Druguet}}]{1998JCoPh.145..511S}
{Sanders}, R., {Morano}, E., \& {Druguet}, M.-C. 1998, Journal of Computational
  Physics, 145, 511

\bibitem[{{Schaye} {et~al.}(2015){Schaye}, {Crain}, {Bower}, {Furlong},
  {Schaller}, {Theuns}, {Dalla Vecchia}, {Frenk}, {McCarthy}, {Helly},
  {Jenkins}, {Rosas-Guevara}, {White}, {Baes}, {Booth}, {Camps}, {Navarro},
  {Qu}, {Rahmati}, {Sawala}, {Thomas}, \& {Trayford}}]{2015MNRAS.446..521S}
{Schaye}, J., {Crain}, R.~A., {Bower}, R.~G., {et~al.} 2015, \mnras, 446, 521

\bibitem[{{Schober} {et~al.}(2012){Schober}, {Schleicher}, {Federrath},
  {Glover}, {Klessen}, \& {Banerjee}}]{2012ApJ...754...99S}
{Schober}, J., {Schleicher}, D., {Federrath}, C., {et~al.} 2012, \apj, 754, 99

\bibitem[{{Somerville} \& {Dav{\'e}}(2015)}]{2015ARA&A..53...51S}
{Somerville}, R.~S., \& {Dav{\'e}}, R. 2015, \araa, 53, 51

\bibitem[{{Sternberg} {et~al.}(2002){Sternberg}, {McKee}, \&
  {Wolfire}}]{2002ApJS..143..419S}
{Sternberg}, A., {McKee}, C.~F., \& {Wolfire}, M.~G. 2002, \apjs, 143, 419

\bibitem[{{Stone} \& {Gardiner}(2009)}]{2009NewA...14..139S}
{Stone}, J.~M., \& {Gardiner}, T. 2009, \na, 14, 139

\bibitem[{{Stone} \& {Gardiner}(2010)}]{2010ApJS..189..142S}
{Stone}, J.~M., \& {Gardiner}, T.~A. 2010, \apjs, 189, 142

\bibitem[{{Stone} {et~al.}(2008){Stone}, {Gardiner}, {Teuben}, {Hawley}, \&
  {Simon}}]{2008ApJS..178..137S}
{Stone}, J.~M., {Gardiner}, T.~A., {Teuben}, P., {Hawley}, J.~F., \& {Simon},
  J.~B. 2008, \apjs, 178, 137

\bibitem[{{Sutherland} \& {Dopita}(1993)}]{1993ApJS...88..253S}
{Sutherland}, R.~S., \& {Dopita}, M.~A. 1993, \apjs, 88, 253

\bibitem[{{Tammann} {et~al.}(1994){Tammann}, {Loeffler}, \&
  {Schroeder}}]{1994ApJS...92..487T}
{Tammann}, G.~A., {Loeffler}, W., \& {Schroeder}, A. 1994, \apjs, 92, 487

\bibitem[{{Teyssier} {et~al.}(2013){Teyssier}, {Pontzen}, {Dubois}, \&
  {Read}}]{2013MNRAS.429.3068T}
{Teyssier}, R., {Pontzen}, A., {Dubois}, Y., \& {Read}, J.~I. 2013, \mnras,
  429, 3068

\bibitem[{{Thornton} {et~al.}(1998){Thornton}, {Gaudlitz}, {Janka}, \&
  {Steinmetz}}]{1998ApJ...500...95T}
{Thornton}, K., {Gaudlitz}, M., {Janka}, H.-T., \& {Steinmetz}, M. 1998, \apj,
  500, 95

\bibitem[{{Truelove} {et~al.}(1997){Truelove}, {Klein}, {McKee}, {Holliman},
  {Howell}, \& {Greenough}}]{1997ApJ...489L.179T}
{Truelove}, J.~K., {Klein}, R.~I., {McKee}, C.~F., {et~al.} 1997, \apjl, 489,
  L179

\bibitem[{{Turk} {et~al.}(2011){Turk}, {Smith}, {Oishi}, {Skory}, {Skillman},
  {Abel}, \& {Norman}}]{2011ApJS..192....9T}
{Turk}, M.~J., {Smith}, B.~D., {Oishi}, J.~S., {et~al.} 2011, \apjs, 192, 9

\bibitem[{van~der Walt {et~al.}(2011)van~der Walt, Colbert, \&
  Varoquaux}]{vanderWalt2011}
van~der Walt, S., Colbert, S.~C., \& Varoquaux, G. 2011, Computing in Science
  {\&} Engineering, 13, 22.

\bibitem[{{Vogelsberger} {et~al.}(2014){Vogelsberger}, {Genel}, {Springel},
  {Torrey}, {Sijacki}, {Xu}, {Snyder}, {Bird}, {Nelson}, \&
  {Hernquist}}]{2014Natur.509..177V}
{Vogelsberger}, M., {Genel}, S., {Springel}, V., {et~al.} 2014, \nat, 509, 177

\bibitem[{{Vorobyov} \& {Basu}(2005)}]{2005MNRAS.360..675V}
{Vorobyov}, E.~I., \& {Basu}, S. 2005, \mnras, 360, 675

\bibitem[{{Walch} \& {Naab}(2015)}]{2015MNRAS.451.2757W}
{Walch}, S., \& {Naab}, T. 2015, \mnras, 451, 2757

\bibitem[{{Walch} {et~al.}(2015){Walch}, {Girichidis}, {Naab}, {Gatto},
  {Glover}, {W{\"u}nsch}, {Klessen}, {Clark}, {Peters}, {Derigs}, \&
  {Baczynski}}]{2015MNRAS.454..238W}
{Walch}, S., {Girichidis}, P., {Naab}, T., {et~al.} 2015, \mnras, 454, 238

\bibitem[{{Wang} {et~al.}(2010){Wang}, {Li}, {Abel}, \&
  {Nakamura}}]{2010ApJ...709...27W}
{Wang}, P., {Li}, Z.-Y., {Abel}, T., \& {Nakamura}, F. 2010, \apj, 709, 27

\bibitem[{{Zhang} {et~al.}(2013){Zhang}, {Rix}, {van de Ven}, {Bovy}, {Liu}, \&
  {Zhao}}]{2013ApJ...772..108Z}
{Zhang}, L., {Rix}, H.-W., {van de Ven}, G., {et~al.} 2013, \apj, 772, 108

\end{thebibliography}

\end{document}